\documentclass[11pt]{article}

\usepackage{overpic}
\usepackage{array}
\usepackage{rotating}

\usepackage{pict2e}

\usepackage{longtable}
\usepackage{makeidx}
\usepackage{fullpage}
\usepackage{amsmath}
\usepackage{amssymb}
\usepackage{setspace}
\usepackage{bbm}
\usepackage{dsfont}
\usepackage{graphics}
\usepackage{epsfig}
\usepackage[font=footnotesize,labelfont=bf]{caption}
 \usepackage{multirow}
\usepackage{array,booktabs}
\usepackage[scaled=2]{CountriesOfEurope}

\usepackage{color}

\usepackage{cite}

\usepackage{hyperref}

\hypersetup{
    bookmarks=true,%
    colorlinks,%
    citecolor=blue,%
    filecolor=black,%
    linkcolor=blue,%
    urlcolor=blue
}

 \usepackage{xcolor}

\newcommand{\bqn}{\begin{eqnarray}}
\newcommand{\eqn}{\end{eqnarray}}
\newcommand{\lt}{\left}
\newcommand{\rt}{\right}

\newcommand{\Om}{\Omega}
\newcommand{\la}{\lambda}

\newcommand{\ph}{\phi}

\newcommand{\na}{\nabla}

\newcommand{\pd}{\partial}

\begin{document}
\numberwithin{equation}{section}
{
\begin{titlepage}
\begin{center}

\hfill \\
\hfill \\
\vskip 0.2in


{\LARGE \bf Hyperscaling violating black holes \\
\vspace{.3cm} with spherical and hyperbolic horizons}\\


\vskip 0.4in

{\large  Juan F. Pedraza${}^{\text{{\large \CountriesOfEuropeFamily\Netherlands}}}$, Watse Sybesma${}^{\text{\CountriesOfEuropeFamily\Iceland}}$ and Manus R. Visser${}^{\text{{\large \CountriesOfEuropeFamily\Netherlands}}}$}\\

\vskip 0.3in

${}^{
\text{{\large \CountriesOfEuropeFamily\Netherlands}}
}${\it Institute for Theoretical Physics, University of Amsterdam,\\
Science Park 904, Postbus 94485, 1090 GL Amsterdam, The Netherlands} \vskip .5mm
${}^{
\text{\CountriesOfEuropeFamily\Iceland}
}${\it University of Iceland, Science Institute, Dunhaga 3, 107 Reykjav\'{i}k, Iceland} \vskip .5mm

\texttt{jpedraza@uva.nl, watse@hi.is, m.r.visser@uva.nl}

\end{center}

\vskip 0.6in

\begin{center} {\bf ABSTRACT } \end{center}
We present a new family of charged black holes with hyperscaling violating asymptotics and non-trivial horizon topology,  for arbitrary  Lifshitz exponent $z$ and hyperscaling   violation parameter  $\theta$. They are $(d+2)$-dimensional analytic  solutions to a generalized Einstein-Maxwell-Dilaton theory  with an additional vector field supporting   spherical or hyperbolic horizons.  The hyperbolic solutions are only consistent with the null energy condition for $1  \le  z  < 2$ and $\theta =  d(z-1)$, which explains why they are absent in the pure Lifshitz case. 
We further study the (extended) thermodynamics of these black holes, both in the fixed   charge and fixed potential ensemble, and find that phase transitions only occur  for   spherical black holes with $1 \le z \le 2$ and  any  $\theta$.

%

\vfill

\noindent \today

\end{titlepage}
}

\newpage

\tableofcontents

\section{Introduction}
Through the holographic duality, one can relate a strongly coupled quantum field theory to a weakly coupled gravitational theory and extract information about the former via the latter \cite{Maldacena:1997re,Witten:1998qj,Gubser:1998bc}. In the quest of describing condensed matter systems in the vicinity of a quantum critical point, these holographic tools need to be extended beyond the usual relativistic domain. For a recent review about the holographic correspondence for condensed matter systems, see \cite{Hartnoll:2016apf}.

In close proximity of a quantum critical point, observables   typically exhibit a scaling behavior given by a plethora of critical exponents \cite{Goldenfeld:1992qy}. For example, some non-relativistic quantum critical systems exhibit an anisotropic scaling symmetry between space and time \cite{Kachru:2008yh,Balasubramanian:2009rx,Son:2008ye,Taylor:2008tg}, which can be parametrized by the dynamical critical exponent $z$: $\{ t, \vec x\} \rightarrow \{ \zeta^z t, \zeta \vec x \}$.  Theories with $z=1$ support Lorentz invariance, whereas theories with $z\neq 1$ are invariant under   non-relativistic symmetries. On the gravity side, a family of backgrounds that geometrically realizes this scaling symmetry is given by    Lifshitz spacetimes, which are a generalization of   Anti-de Sitter (AdS) space \cite{Kachru:2008yh}.  One consequence of such  an anisotropic scaling is that, at low temperature, the specific heat scales as $c_V \sim T^{d/z}$ (where $d$ is the number of boundary space dimensions).
Therefore,  asymptotically Lifshitz black holes are good candidates to holographically describe (non-relativistic) versions of Fermi liquids for $z=d$. Ideally, one would be interested in obtaining the same linear scaling behavior for Lorentz invariant theories, but this is not possible in the pure Lifshitz case.

A quantum critical point can be characterized by many other critical exponents, each of which determine the scaling of a given observable with respect to the correlation length.  These exponents can  satisfy various relations and hence depend on each other \cite{Goldenfeld:1992qy}. A class of these critical exponent relations, known as hyperscaling relations, differs from other scaling relations in that the dimensionality of space appears explicitly \cite{Widom1965}.  It is well known that  the  hyperscaling   relation known as the Josephson relation   could be violated in the presence of   irrelevant couplings \cite{wegner1973logarithmic, fisher1974renormalization}. Examples of critical theories which do not respect this relation include those above their upper critical dimension, or some even below their critical dimension, as in the random-field Ising model \cite{fisher1986scaling}.

Theories that violate the (Josephson) hyperscaling relation have a specific heat $c_V\sim T^{ (d-\theta)/z}$ \cite{Gouteraux:2011ce,Huijse:2011ef}, where $\theta$ is the so-called hyperscaling violating parameter. In a sense, $\theta$ effectively lowers the dimensionality of the theory. Interestingly, relativistic theories with $\theta=d-1$
can have the particular linear scaling of the specific heat, which is characteristic of Fermi liquid theory. The extra parameter $\theta$ can be implemented in the bulk theory by means of a hyperscaling violating geometry, which generalize the aforementioned Lifshitz spacetimes.
Specific solutions for hyperscaling violating black holes were notably found in \cite{Charmousis:2010zz,Dong:2012se,Gath:2012pg,Alishahiha:2012qu,Gouteraux:2012yr,OKeeffe:2013xdv,Ghodrati:2014spa,Cremonini:2014gia,Ganjali:2015cba,Roychowdhury:2015fxf,Li:2016rcv,Ge:2016lyn,Cremonini:2016avj,Cremonini:2018jrx}. These are solutions to gravitational theories with higher order gravitational corrections or additional matter fields, such as massive vector fields or a Maxwell field coupled to a dilation.  We stress, however, that all previously constructed hyperscaling violating solutions only support black branes with \textit{planar} topology.

In this paper, we construct the first spherical and hyperbolic black holes in asymptotically hyperscaling violating spacetimes. The motivation to study this type of black holes is that the topology introduces an extra scale in the theory, allowing for a non-trivial phase structure. Our black holes are analytic solutions to a generalized  Einstein-Maxwell-Dilaton (EMD) theory, which consists of Einstein gravity minimally coupled to a real scalar field and   three   $U(1)$ vector fields: one to support Lifshitz asymptotics, one to accommodate the non-trivial topology, and one to allow for an electric charge.  EMD theories have the advantage that they admit analytic solutions, as opposed to Einstein gravity coupled to massive vector fields, whose  black hole solutions are studied only numerically or for specific $z$  \cite{Taylor:2008tg}. Our black holes are generalizations of previously discovered solutions to a generalized EMD, namely the charged black branes with arbitrary $z$ and $\theta$   in \cite{Alishahiha:2012qu} (to other topologies), and the charged spherical Lifshitz black holes found in \cite{Tarrio:2011de} (to general $\theta$).


Hyperbolic black holes are known to exhibit exotic features as compared to their spherical and planar siblings. Hyperbolic black holes were first discussed in the context of holography in \cite{Emparan:1998he,Birmingham:1998nr,Emparan:1999gf}, in an asymptotically AdS background. The massless limit of these black holes has finite temperature and non-vanishing entropy, which contrasts the spherical and planar case. Another interesting feature is that this massless hyperbolic   black hole is isomorphic to a Rindler wedge of AdS spacetime.
Furthermore, AdS hyperbolic black holes  do not have phase transitions as function of temperature, mimicking the result for planar black holes. In contrast, hyperbolic black holes do have an extra length scale $\ell$ as compared to planar black holes.

The fact that we are able to engineer a hyperbolic black hole with $z>1$ might come as a surprise, since this solution is known to be absent for the Lifshitz ($\theta=0$, $z>1$) EMD setup. This situation can be understood from the results of the null energy conditions summarized in Table \ref{table1}. From inspection of this table it is obvious that hyperbolic black holes for $z>1$ can only exist when the hyperscaling violation parameter is nonzero and given by $\theta =d (z-1)$.
To our knowledge, we present the first hyperbolic black hole for $z\neq1$.

Another insight we present concerns tidal forces. It is known that Lifshitz spacetime exhibits tidal divergences. Upon further inspection of the $\theta=d(z-1)$ spacetime we find that such tidal divergences are absent, as was noted in \cite{Shaghoulian:2011aa} for the planar case. This suggests that this spacetime is geodesically complete. In addition, we show that the  \textit{massless} hyperbolic black hole has no singularity at the origin, suggesting that it is the Rindler wedge of some yet unknown space.
\begin{table}[t!!]
\begin{center}
  \begin{tabular}{| c | c | c | c | }
    \hline
    $z$& hyperbolic $k=-1$ & planar $k=0$ & spherical $k=1$\\ \hline\hline
    $z<1$ & no solution & no solution & no solution \\
    $1\leq z<2 $ &$\theta = d(z-1)$ & $\theta\leq d(z-1)$ &  $\theta\leq d(z-1)$  \\
    $z\geq2$ &no solution&  $\theta< d$ & $\theta< d$ \\
    \hline
  \end{tabular}
\end{center}
\vspace{-0.5cm}
\caption{Restrictions to the space of parameters coming from the null energy condition.}

\label{table1}
\end{table}

Furthermore, we   deploy an exhaustive study of the thermodynamics of the hyperscaling violating black holes. We find that phase transitions solely occur if the black hole is spherical and if $1 \le z \le 2$, which in the fixed electric potential ensemble is analogous to the Hawking-Page transition \cite{Hawking:1982dh,Witten:1998zw} and in the fixed charge ensemble  to the liquid-gas phase transition \cite{Chamblin:1999tk,Chamblin:1999hg}. Notably, the qualitative behavior of these phase transitions seems independent of the parameters $d$ or $\theta$.
Finally, we   study the extended thermodynamics of this system. We determine the thermodynamic volume and pressure, following from the Smarr relation, and show that the critical exponents are the same as those of the Van der Waals fluid.

This paper is organized in the following manner. In Section \ref{setup} we construct the   black hole solutions from an explicit action, and study the   null energy condition and tidal forces. The thermodynamics for fixed electric charge and  fixed electric potential are examined in Section \ref{secthermodynamics}. In Section~\ref{secextendedthermo} we study the extended thermodynamics. We end with a summary and discussion, followed by   Appendix \ref{euclideanaction} in which we compute the background subtracted Euclidean on-shell action.

\section{Setup and black hole solutions}
\label{setup}

In Ref. \cite{Alishahiha:2012qu}, the authors constructed electrically charged black branes with arbitrary Lifshitz exponent $z$ and hyperscaling violation parameter $\theta$, starting from a specific Einstein-Maxwell-Dilaton action. The solutions obtained there, however, were constrained to have planar horizons. In this section we consider a generalized EMD action that will allow us to construct black holes akin to the ones mentioned above, but including new solutions with spherical and hyperbolic topologies. These new solutions also include as particular cases the spherical Lifshitz black holes found in \cite{Tarrio:2011de}. In  the rest of the section we derive the restrictions on $z$ and $\theta$ coming from the null energy condition, discuss in more detail the one-parameter family of solutions with $\theta=d(z-1)$, which can support solutions with hyperbolic horizons, and study  the  tidal forces   for our black hole solutions.

\subsection{Generalized Einstein-Maxwell-Dilaton theory}
The theory that we consider is a variation of the standard EMD theory \cite{Dong:2012se} with two extra vector fields, one ($H$) supporting the non-trivial topology, and another one ($K$) supporting states with finite charge density:\footnote{The fact that one can trace these properties back to the specific vector fields arises due to specific choices made when solving the Einstein equations.}
\bqn
\label{action1}
S=
	-\frac1{16\pi G}\int d^{d+2}x\sqrt{-g}\lt[
	R-\frac12(\na_\mu\ph)^2+V(\ph)-\frac14 X(\ph)F^2-\frac14 Y(\ph)H^2-\frac14 Z(\ph)K^2
	\rt]\,,
\eqn
where $F=dA$, $H=dB,$ and $K=dC$, respectively. We consider the following potential and dilaton couplings:
\bqn
V=V_0 e^{\lambda_0 \ph},\;\;X=X_0 e^{\la_1\ph},\;\;Y=Y_0 e^{\la_2\ph},\;\;Z=Z_0 e^{\la_3\ph}\,,
\eqn
with arbitrary constants $V_0$, $X_0$, $Y_0$, $Z_0$ and $\lambda_i$. The equations of motion that follow from this action are:
\bqn
\na^2\ph
+\pd_\ph V(\ph)
-\frac{1}{4}  \partial_\phi X(\phi )  F^2
-\frac{1}{4}   \partial_\phi Y(\phi ) H^2
-\frac{1}{4}  \partial_\phi  Z(\phi )   K^2  &=&0,\\
\nabla_{\mu }\lt(X(\phi )F^{\mu \nu }\rt)=0,\qquad
\nabla _{\mu }\lt(Y(\phi )H^{\mu \nu }\rt)=0,\qquad
\nabla_{\mu }\lt(Z(\phi )K^{\mu \nu }\rt)&=&0,\\
R_{\mu \nu }+\frac{ g_{\mu \nu } }{d}V(\phi ) -\frac12\pd_\mu\ph\pd_\nu\ph
-\frac{1}{2} X(\phi ) \left(F_{\alpha \mu } F^{\alpha }{}_{\nu }-\frac{ g_{\mu \nu }}{2 d}F^2\right)\;\;&&  \nonumber \\
-\frac{1}{2} Y(\phi ) \left(H_{\alpha \mu } H^{\alpha }{}_{\nu }-\frac{g_{\mu \nu }}{2 d}H^2\right)
-\frac{1}{2} Z(\phi ) \left(K_{\alpha \mu } K^{\alpha }{}_{\nu }-\frac{g_{\mu \nu }}{2 d}K^2\right)&=&0\,.
\eqn
In order to construct explicit solutions we start with the following Ansatz for all the fields, with a single blackening factor $f(r)$ in the metric,\footnote{We work in natural units, i.e. we set $c=\hbar = k_B=1$.}
\bqn
&&ds^2=\lt(\frac{r}{r_F}\rt)^{  -2 \theta /d   }\lt(
-\lt(\frac r\ell\rt)^{2z}f(r) dt^2+\frac{\ell^2}{f(r)r^2}dr^2+r^2 d\Om^2_{k,d}
\rt) \,,  \label{ansatz}   \\
&&A=a(r) dt\,,\qquad B=b(r) dt\,,\qquad C=c(r) dt\,,\qquad \ph=\ph(r)\,.\label{ansatz2}
\eqn
Here $k=-1,0,1$ labels the hyperboloid, planar, or spherical topology for the black hole horizon, respectively, where
\bqn
\begin{split}
	d\Omega^2_{k=1,d}
	=
	d\chi_{0}^{2}
	+\sin(\chi_{0})^{2}d\chi_{1}^{2}
	+
	\dots
	+
	\sin(\chi_{0})^{2}\cdots \sin(\chi_{d-2})^{2}d\chi_{d-1}^{2}
	\,,\\
	d\Omega^2_{k=0,d}
	=
	\frac{d\vec{x}_{d}^{2}}{\ell^{2}}	
	\,, \qquad
	d\Omega^2_{k=-1,d}
	=
	d\chi_{0}^{2}
	+
	\sinh(\chi_{0})^{2}d
	\Omega^{2}_{k=1,d-1}
	\,,\qquad
\end{split}
\eqn
and $\chi_{i}$  are the standard angles. The constant $\ell$ in the metric sets the overall scale of the spatial geometry and can be thought of as a generalization of the AdS radius. The constants $z$ and $\theta$ are exponents that characterize the symmetries of the underlying theory. In particular, near the asymptotic boundary we expect $f(r)\to1$ and, in this limit, the Ansatz (\ref{ansatz}) is the most general metric that is covariant under the scale transformations
\bqn
t\to\zeta^z\,t\,,\qquad \Omega\to\zeta\, \Omega\,,\qquad r\to\zeta^{-1}\, r\,,\qquad ds\to\zeta^{\theta/d}\,ds\,.
\eqn
We refer to $z$ and $\theta$ as the Lifshitz dynamical exponent and the hyperscaling violation exponent, respectively.\footnote{Notice that in standard Lifshitz theories, the symmetry generators include time translations $P_0$, space translations $P_i$, anisotropic scale transformations $D$ and space rotations $M_{ij}$. Our background metric is invariant under space translations only in the planar case, where $k=0$. Thus, strictly speaking the spherical and hyperbolic metrics above are not Lifshitz invariant.}
As  usual in  holographic models,  the  radial  direction $r$ is mapped into an energy scale of the dual field theory.
For $\theta<d$,  in  the  coordinates  we  have  chosen  above, $r\to\infty$ and $r\to0$ describe the UV  and  IR  of  the  theory, respectively.
We assume this condition throughout this paper, because otherwise the UV and IR are not well behaved.
We emphasize that our scaling metrics provide a good description of the theory up to the scale $r\sim r_F$, but they generally require a UV completion. For example, if the dual field theory under consideration flows from a UV fixed point to a critical point with non-trivial $z$ and $\theta$ we expect that our metric should be matched to an AdS geometry at large $r$. The radius $r_F$ here, which is set by UV physics, marks the location where the IR solution fails, and is indeed responsible for restoring the canonical dimensions in the UV.\footnote{For example, in models with a Fermi surface, $r_{\rm F}$ is set by the Fermi momentum \cite{Ogawa:2011bz}. Examples of UV completions of these kind of models were studied in \cite{Dong:2012se,Gath:2012pg,Perlmutter:2012he} and, notably, in \cite{Giataganas:2017koz} for spatially anisotropic cases.} On the other hand, the theory may flow to some other fixed point in the deep IR, or develop a mass gap, etc. In such cases the metric (\ref{ansatz}) ceases to be valid for very small $r$ as well \cite{Dong:2012se}. Incidentally, theories with arbitrary $z$ and $\theta$ are known to develop genuine IR singularities which  may  require  stringy  effects  to  be  resolved \cite{Shaghoulian:2011aa,Harrison:2012vy,Bao:2012yt,Copsey:2012gw}. We will ignore these issues in the rest of the paper, and assume that the gravity background is valid within a certain range of energy scales.

\subsection{Hyperscaling violating  black hole solutions}
As mentioned above, the three gauge fields of our gravitational system are introduced in order to obtain solutions with the desired geometric properties and symmetries. To reiterate, $F$ is introduced to support the Lifshitz asymptotics of the geometry, $H$ to support the topology of the internal space, and $K$ to allow for solutions with electric charge. The scalar potential $V(\phi)$ facilitates the hyperscaling violating factor of the solution.

The black hole solutions with arbitrary $z$, $\theta$ and $k$ can be compactly written as follows:\footnote{In the
appropriate ``infinite volume'' limit the $k=\pm1$ classes of solutions approach the planar black brane with $k=0$ \cite{Emparan:1999gf,Chamblin:1999tk,Tarrio:2011de}.
Such a limit can be reached by rescaling $r\to\eta r$, $t\to\eta^{-z}t$, $\ell^2 d \Omega^2_{\pm 1, d} \to \eta^{-2} d \Omega_{0,d}^2$, $r_F\to \eta r_F$, $m\to \eta^{d-\theta+z}m$, $q\to\eta^{d-\theta+z-1} q$ and taking $\eta\to\infty$, making the radius of the $S^d$ or $H^d$   much larger than the thermal wavelength of the system \cite{Witten:1998zw}.}
 \bqn
 \ph&=&\ph_0+\gamma\log r,   \label{phispherical}\\
F&=&-\rho _1 e^{-\lambda _1 \phi(r) }r^{-\frac{2 \theta }{d}-d+\theta +z-1}dtdr,\\
H&=&-\rho_2e^{-\lambda _2 \phi(r) }r^{-\frac{2 \theta }{d}-d+\theta +z-1}dtdr,\\
K&=&-\rho_3e^{-\lambda _3 \phi(r) }r^{-\frac{2 \theta }{d}-d+\theta +z-1}dtdr,\\
f &=&1     +k\frac{(d-1)^2  }{  (d-\theta +z-2)^2} \frac{\ell^2}{r^2}       -  \frac{m}{r^{ d-\theta +z} } +  \frac{ q^2}{ r^{2 (d- \theta + z-1)}}  \label{blackeningspherical}
\eqn
where we have defined $\gamma\equiv\sqrt{ 2 \left(d- \theta \right) \left( z-1 -  \theta/d \right)}$, and with the following parameters:
\bqn
\lambda_0 &=& \frac{2 \theta  }{\gamma d } , \;\;
\lambda _1 = -\frac{2 \left(d- \theta +\theta /d \right)}{ \gamma  },\;\;  \lambda _2= -\frac{2 (d-1) (d-\theta )}{ \gamma  d}, \;\;    \lambda _3= \frac{\gamma }{d-\theta },  \\
V_0&=& (d-\theta +z-1) (d-\theta +z) \ell^{-2} r_F^{- 2 \theta  /d} e^{-\lambda_0 \phi_0 },\\
\rho _1^2&=&  2       (z-1) (d-\theta +z)  X_0^{-1}   \ell ^{-2 z} r_F^{ 2 \theta /d}   e^{  \lambda_1  \phi_0 }    ,\\
\rho _2^2&=&2  k \frac{ (d-1)   (d (z-1)-\theta )    }{d-\theta +z-2} Y_0^{-1}  \ell ^{2(1-z)} r_F^{ 2 \theta /d} e^{\lambda_2 \phi_0 }   , \label{rho2}\\
\rho_3^2&=&  2  q^2 (d-\theta) (d-\theta + z-2) Z_0^{-1}  \ell^{-2z}r_F^{2 \theta/d}e^{ \lambda_3 \phi_0 } .
\eqn
In the blackening factor $f(r)$ we have, in addition, the mass and charge parameters $m$ and $q$, which can be arbitrary (provided we do not have a naked singularity). The parameters $X_{0}$, $Y_{0}$, $Z_{0}$ are positive and represent the strength of the coupling of the gauge fields with gravity. Furthermore, in order to arrive at this solution we have assumed that $d- \theta +z -2>0$ and $\gamma\in\mathbb{R}$.

As mentioned above, gravity solutions with generic $z$ and $\theta$ are known to develop generic curvature singularities, even in the vacuum case, where $m=0$ and $q=0$. This can be corroborated by computing scalar quantities such as $R$, $R_{\mu\nu}R^{\mu\nu}$, $R_{\mu\nu\sigma\rho}R^{\mu\nu\sigma\rho}$, etc. For example, a quick calculation shows that for our black hole solutions:
\bqn\label{ricciscalar}
R=\frac{1}{\ell^2}\left(\frac{r}{r_F}\right)^{\frac{2 \theta }{d}}\left[a_1+a_2\frac{k\ell^2}{r^2}+a_3\frac{m}{r^{d-\theta +z}}+a_4\frac{q^2}{r^{2 (d - \theta + z - 1)}}\right],
\eqn
where
\bqn
a_1=-2 z (d+z)-\frac{(d+1)(d-\theta )^2}{d}+\frac{2(d+1) z\theta}{d},\quad a_2=\frac{(d-1)(d-dz+\theta )[4-3d+z(d-2)+\theta] }{d(d-\theta +z-2)^2},\nonumber&&\\
a_3=d-z(d-\theta )-\frac{\theta^2}{d},\quad a_4=-\frac{(d-\theta ) [2 (2-z)+(d-3) (d-\theta )]}{d}.\qquad\qquad\qquad\quad\nonumber&&
\eqn
Depending on the various parameters, $R$ can blow up at either $r\to\infty$, $r\to0$, or both. The same is also true for $R_{\mu\nu}R^{\mu\nu}$, $R_{\mu\nu\sigma\rho}R^{\mu\nu\sigma\rho}$, which have similar but more longwinded expression that we will not
transcribe here. The fact that one or various of these scalar invariants blow up signals a genuine physical singularity. There are a few methods available in the literature to deal with these kind of singularities which involve the addition of stringy corrections into the gravitational description \cite{Shaghoulian:2011aa,Harrison:2012vy,Bao:2012yt,Copsey:2012gw}. These corrections will change drastically the deep IR of the theory, effectively dressing up the singularity. We will not be concerned with these corrections here, but rather, we will assume that the gravity description is valid within an intermediate range of energies.

\subsection{Null energy condition and allowed values of $z$ and $\theta$}
Let us study the restrictions in the space of parameters coming from the null energy condition. Letting $\xi^{\mu}$ be a null vector, i.e. $\xi^{2}=0$, the null energy condition is formulated as
\begin{equation}\label{NECdef}
	T_{\mu\nu}\xi^{\mu}\xi^{\nu}
	\geq
	0
	\,,
\end{equation}
where $T_{\mu\nu}$ is the energy momentum tensor. The null energy condition is generally assumed to provide a sufficient condition to have a physically sensible holographic dual in the semiclassical limit. This condition has been argued for from general properties of RG flows, causality and unitarity, and various entropic-related properties of quantum field theories, see e.g. \cite{Allais:2011ys,Myers:2012ed,Callan:2012ip,Caceres:2013dma,Headrick:2014cta}.\footnote{However there are known examples where this condition is too strong, e.g. RG flows with a well-behaved and monotonous $c$-function that `average out' small violations of null energy condition\cite{Cremonini:2013ipa}.} In the following we will therefore assume that (\ref{NECdef}) must hold in order to have a consistent duality.

Following \cite{Dong:2012se}, we use the Einstein equations to recast the above requirement as,
\begin{equation}\label{eq:null2}
	R_{\mu\nu}\xi^{\mu}\xi^{\nu}
	\geq
	0
	\,,
\end{equation}
where $R_{\mu\nu}$ is the Ricci tensor (notice that terms such as the scalar potential and the Ricci scalar, which are proportional to the metric tensor, vanish when contracted with null vectors). Considering two orthogonal null vectors one derives the following inequalities:
\begin{equation}
	(d-\theta)(d(z-1)-\theta)\geq0
	\,,
\end{equation}
\begin{equation}
\frac{r^{2}}{\ell^{2}}(z-1)(d-\theta+z)+	k\frac{(d-1)(d(z-1)-\theta)}{d-\theta+z-2}
+q^{2} \frac{(d-\theta)(d(z-1)-\theta)}{\ell^{2}r^{2(d-\theta+z-2)}}
	\geq0\label{NEC2}
	\,.
\end{equation}
The second inequality can be simplified further. Let us first consider the neutral case, where $q=0$. In this case, the last term in (\ref{NEC2}) drops out and we end with two terms one of which   scales with $r^{2}$. Since the inequality should hold for all values of $r$, these terms have to be greater or equal to zero independently (one can reach this conclusion by analyzing the limits $r\to0$ and $r\to\infty$). Keeping in mind that for the derivation of the black hole solutions we have assumed $d-\theta +z-2>0$ and $\theta<d$, we conclude that at zero charge the inequalities   following from the null energy condition reduce to:
\begin{equation}\label{NECsummary}
	d(z-1)-\theta\geq0
	\,,
	\quad
	\quad
	(z-1)(d-\theta+z)
	\geq0
	\,,
	\quad\quad
	k(d(z-1)-\theta)
	\geq0
	\,.
\end{equation}
For finite charge $q\neq0$   the third term in (\ref{NEC2}) dominates over the second term at small $r$. The coefficient of the third term is real and positive definite provided that the first inequality in (\ref{NECsummary}) is satisfied, so we do not get any additional constraints at finite charge.\footnote{At finite charge the third inequality in (\ref{NECsummary}) is not strictly necessary. However, we assume it is satisfied for all $q$, by continuity in the limit $q\to0$. This extra inequality does not add any information for $k=\{0,1\}$. For $k=-1$, however, it implies that $\theta=d(z-1)$.} Further, for $k=0$ we notice that the third inequality is identically satisfied, and we recover the results from, e.g., Ref. \cite{Chemissany:2014xsa}.

In Table \ref{table1}, placed in the introduction, we summarize the constraints on $\theta$ and $z$ for hyperbolic, planar and spherical black holes. For the spherical case, the third constraint in \eqref{NECsummary} does not contain extra information, which is consistent with previous findings for $\theta=0$\cite{Tarrio:2011de}.
We highlight the fact that, opposite to the results from Ref. \cite{Tarrio:2011de} which studied the $\theta=0$ case, in our space of solutions it is possible to find $z\neq1$ hyperbolic black holes for the specific value $\theta = d(z-1)$. However, notice that precisely for $\theta=d(z-1)$ we have $\gamma=0$, implying that $\lambda_i\to\infty$ for $i=\{0,1,2\}$ and $\lambda_3=0$. This looks like a singular limit. In the next section we will show that, under a specific rescaling of the scalar field, this one-parameter family of solutions is actually physically sensible.

\subsection{The case $\theta=d(z-1)$}
For $\theta = d(z-1)$  the null energy condition is satisfied as long as the dynamical exponent is in the range $1 \le z < 2$, but the solution itself seems singular. In order to properly analyze this limit we redefine the scalar as follows:
\begin{equation}
	\phi(r)
	\to
	\gamma
	\tilde{\phi}(r)
	\,,
\end{equation}
so that the action becomes:
\bqn
\label{actionredefined}
S= \!
	-\frac1{16\pi G} \! \int \! d^{d+2}x\sqrt{-g}\lt[
	R-\gamma^{2}\frac12(\na_\mu\tilde{\ph})^2+V(\tilde{\ph})-\frac14 X(\tilde{\ph})F^2-\frac14 Y(\tilde{\ph})H^2-\frac14 Z(\tilde{\ph})K^2
	\rt].
\eqn
The potentials and couplings are now given by:
\bqn
V=V_0 e^{\tilde{\lambda}_0 \tilde{\ph}},\;\;X=X_0 e^{\tilde{\la}_1\tilde{\ph}},\;\;Y=Y_0 e^{\tilde{\la}_2\tilde{\ph}},\;\;Z=Z_0 e^{\tilde{\la}_3\tilde{\ph}}\,,
\eqn
where the constants $\tilde{\lambda}_i$ are:
\bqn
\tilde{\lambda}_0 = \frac{2 \theta  }{d } , \;\;
\tilde{\lambda} _1 = -2 \left(d- \theta +\theta /d \right),\;\;  \tilde{\lambda} _2= -\frac{2 (d-1) (d-\theta )}{d}, \;\;    \tilde{\lambda} _3= \frac{\gamma^2 }{d-\theta }.
\eqn
In the limit $\gamma\to0$ a number of simplifications take place. First, already from (\ref{rho2}) we can see that the charge associated to $H$ vanishes,
$\rho_2\to0$. This means that for this family of solutions we do not require a vector field to support the non-trivial topologies $k=\{-1,1\}$. At the level of the action we get two additional simplifications $i)$ $\tilde{\lambda}_{3}=0$ so the coupling between the dilaton and $K$ is trivial $Z(\tilde{\phi})=Z_0$, and $ii)$ the kinetic term for the dilaton vanishes so the action becomes:
\begin{equation}\label{action2}
S=
	-\frac1{16\pi G}\int d^{d+2}x\sqrt{-g}\lt[
	R
	+
	V(\tilde{\ph})
	-
	\frac14 X(\tilde{\ph})F^2
	-
	\frac14 Z_{0}K^2
	\rt]\,.
\end{equation}
In these kind of actions the dilaton can be interpreted as a strongly coupled scalar, see e.g. \cite{Balasubramanian:2009rx}.

Black hole solutions for this theory take the following form:\footnote{Remarkably, there is an exact map between the hyperbolic black hole with $\theta=d(z-1)$ and a planar black brane with axion charge \cite{Ge:2016lyn,Cremonini:2016avj}. For instance, by setting $\beta_0^2\to 2(d-1)$ and $d\vec{x}^2_d\to d\Omega^2_{k=-1,d-2}$ in \cite{Ge:2016lyn} one recovers our hyperbolic solution. We thank  Ioannis Papadimitriou for making this observation.}
 \bqn
\tilde{\ph}&=&\tilde{\ph}_0+\log r,    \label{phihyper}  \\
F&=&-\rho _1 e^{-\tilde{\lambda}_1 \tilde{\phi} (r)}r^{-d (2-z)-z+1}dtdr,\\
K&=&-\rho_3r^{-d (2-z)-z+1}dtdr,\\
f &=&1     +\frac{k}{(2-z)^2} \frac{\ell^2}{r^2}       -  \frac{m}{r^{d (2-z)+z} } +  \frac{ q^2}{ r^{2 [d (2 - z) + z - 1]}}   \label{blackeninghyper}
\eqn
where
\bqn
\tilde{\lambda}_0 &=& 2(z-1) , \;\;
\tilde{\lambda}_1 = -2 [d(2-z)+z-1 ], \\
V_0&=& [d (2 - z) + z] [d (2 - z) + z - 1] \ell^{-2} r_F^{- 2 (z-1)} e^{-\tilde{\lambda}_0 \tilde{\phi}_0 },\\
\rho _1^2&=&  2       (z-1) [d (2 - z) + z]  X_0^{-1}   \ell ^{-2 z} r_F^{ 2(z-1)}   e^{  \tilde{\lambda}_1  \tilde{\phi}_0 }    ,\\
\rho_3^2&=&  2  q^2 d(d-1) (2-z)^2  Z_0 ^{-1} \ell^{-2z}r_F^{2 (z-1)}.
\eqn
As advertized above, these solutions are now well behaved and absent of any singular limit.

It is interesting to analyze   the curvature invariants for this family of solutions. Plugging the value   $\theta=d(z-1)$ into (\ref{ricciscalar}) we obtain the following expression for the Ricci scalar:
\bqn\label{scalar1}
R=-\frac{1}{\ell^2}\left(\frac{r}{r_F}\right)^{2(z-1)}\left[(2+d (2-z)) (z+d(2-z))+\frac{(d-1) (d-2) (2-z)^2q^2}{r^{2[d(2-z)+z-1]}}\right].
\eqn
We notice that the terms depending on $k$ and $m$ vanish for this family, in complete analogy with   standard AdS black holes (in fact, they are part of this family of solutions and correspond to the $z=1$ and $\theta=0$  case). Moreover, since $1\leq z<2$, the vacuum solution is regular at $r\to0$. This is intriguing, at least for the $k=-1$ case, because it implies that the hyperbolic black hole is regular at the origin from the point of view of the Ricci scalar. In order to gain more intuition we compute the other two curvature scalars for this family. A brief calculation leads to the following structure:
\bqn
R_{\mu\nu}R^{\mu\nu}&=&\frac{1}{\ell^4}\left(\frac{r}{r_F}\right)^{4(z-1)}\left[b_1+b_2\frac{q^2}{r^{2[d(2-z)+z-1]}}+b_3\frac{q^4}{r^{4[d(2-z)+z-1]}}\right],\\
R_{\mu\nu\sigma\rho}R^{\mu\nu\sigma\rho}&=&\frac{1}{\ell^4}\left(\frac{r}{r_F}\right)^{4(z-1)}\bigg[c_1+c_2\frac{m}{r^{d(2-z)+z}}+c_3\frac{q^2}{r^{2[d(2-z)+z-1]}}\\
&&\qquad\qquad\qquad\quad+c_4\frac{m^2}{r^{2[d(2-z)+z]}}+c_5\frac{q^4}{r^{4[d(2-z)+z-1]}}+c_6\frac{mq^2}{r^{3d(2-z)+3z-2}}\bigg],\label{scalar3} \nonumber
\eqn
with constants $b_i$ and $c_i$ that depend on $z$ and $d$. The specific values of these constants are not important. What is interesting here is that the vacuum solution (i.e. for $m=0$ and $q=0$) is regular at $r\to0$ regardless the value of $k$. This means that the hyperbolic black hole with zero mass and charge does not have a true curvature singularity, in complete parallel with the standard AdS case. We recall that the massless AdS black hole with $k=-1$ is merely a Rindler wedge of   $k=0$ (or $k=1$) empty AdS, and therefore, the horizon at $r=r_h$ is actually not a black hole event horizon but an acceleration horizon. The temperature in this case is associated to the Unruh temperature of an observer in pure AdS with constant proper acceleration. Such an observer  has access to a restricted portion of the full manifold, specifically, to its Rindler wedge, and this is precisely the region of the spacetime covered by the zero-mass hyperbolic black hole.

One might wonder if the above is also true for other members of the family with $z\neq1$. However, after a close inspection, one reaches a negative conclusion. The reason is the following: in empty AdS, the diffeomorphism that maps between the $k=-1$ and $k=0$ (or $k=1$) solution mixes \emph{all} bulk coordinates, including the holographic coordinate $r$. Such a diffeomorphism must induce a conformal transformation in the boundary that maps the plane (or the sphere) to an hyperboloid, and hence the mixing \cite{Caceres:2010rm}. This is possible in AdS because the isometry group allows for such conformal mapping; however this is no longer the case for other values of $z\neq1$. From the point of view of the curvature invariants
(\ref{scalar1})-(\ref{scalar3}) the reason is also clear: the $z=1$ (or AdS) case is the only solution of the family where all the scalars are constant. Therefore, any coordinate transformation would leave all these invariants intact. On the other hand, other values of $z$ lead to curvature scalars with a non-trivial radial dependence, e.g., $R\sim r^{2(z-1)}$, and so on. Hence, a general coordinate transformation that mixes $r$ with the boundary coordinates would generally lead to a different expression for the curvature invariants. Therefore, we conclude that for any $z\neq1$, vacuum solutions with different $k$ are in fact \emph{different} solutions (from the same action) and physically inequivalent.

\subsection{Tidal forces}
From inspection of curvature invariants such as the Ricci scalar   \eqref{ricciscalar} it can be seen that, for specific choices of the parameters such as $k=q=m=0$,   there are no curvature singularities. This observation does not depend on the values of $\theta$, $z$ and $d$. However, as was argued in \cite{Copsey:2010ya,Horowitz:2011gh,Pang:2009ad}, when traveling on a timelike geodesic, one can nevertheless experience diverging tidal forces at certain points in the $k=q=m=0$ spacetime, if $z>1$ and $\theta=0$. This property renders the spacetime geodesically incomplete. Such tidal divergences are absent when $z=1$.

In this section we investigate tidal forces for the generic hyperscaling violating spacetime \eqref{ansatz}. We especially focus on the massless and chargeless $\theta=d(z-1)$ case, in which case we show that diverging tidal forces are    absent at the origin. We can detect the presence of tidal divergences from divergences in the Riemann tensor, when it is boosted in an orthonormal frame along a timelike radial geodesic. In order to perform this computation we introduce a timelike geodesic $u^{\mu}u_{\mu}=-1$ with $u_{\mu}=\dot{x}_{\mu}$, where the dot denotes a derivative with respect to proper time. Furthermore, we establish
\begin{equation}
	E
	=
	g_{tt}\dot{t}
	\,,
	\quad
	P_{\hat{\i}}=g_{\hat{\i}\hat{\i}}\dot{x}_{\hat{\i}}
	\,,
	\quad
	\dot{r}^{2}
	=
	-g^{rr}(1+g_{tt}\dot{t}^{2})
	\,.
\end{equation}
Here $E$ stands for energy and $P_{\hat{\i}}$ for momentum along the geodesic. We adopt the notation that there is no summation implied over hatted indices such as $\hat{\i}$. Since we will be considering a radial geodesic, we put $P_{\hat{\i}}=0$. The orthonormal frame, before being boosted, is given by the following Vielbeins
\begin{equation}
	(e_{0})_{\mu}
	=
	-\sqrt{-g_{tt}}\partial_{\mu}t
	\,,\quad
	(e_{1})_{\mu}
	=
	\sqrt{g_{rr}}\partial_{\mu}r
	\,,\quad
	(e_{\hat{\i}})_{\mu}
	=
	\sqrt{g_{\hat{\i}\hat{\i}}}\partial_{\mu}x_{\hat{\i}}
	\,.
\end{equation}
We now perform a boost on the Vielbeins such that $(\tilde{e}_{0})_{\mu}=u_{\mu}$, which is the desired frame choice for probing tidal divergences. The boosted frame, i.e. the orthonormal frame parallel propagated along a geodesic, denoted by a tilde, is given by
\begin{equation}\begin{aligned}
	(\tilde{e}_{0})_{\mu}
	=&\;
	-E\partial_{\mu}t
	+
	E\sqrt{-g^{tt}g_{rr}}\sqrt{1-\frac{-g_{tt}}{E^{2}}}\partial_{\mu}r
	\,,
	\\
	(\tilde{e}_{1})_{\mu}
	=&\;
	-E\sqrt{1-\frac{-g_{tt}}{E^{2}}}\partial_{\mu}t
	+
	E\sqrt{-g^{tt}g_{rr}}\partial_{\mu}r
	\,,
	\\
	(\tilde{e}_{\hat{\i}})_{\mu}
	=&\;
	\sqrt{g_{\hat{\i}\hat{\i}}}\partial_{\mu}\hat{\i}
	\,.
\end{aligned}\end{equation}
We can now express the Riemann tensor in a frame boosted along a timelike radial geodesic

\begin{equation}
	\tilde{R}_{mnab}
	\equiv
	R^{\mu\nu\alpha\beta}
	(\tilde{e}_{m})^{\mu}(\tilde{e}_{n})^{\nu}(\tilde{e}_{a})^{\alpha}(\tilde{e}_{b})^{\beta}
	\,.
\end{equation}
The corresponding non-zero entries of the Riemann tensor in the boosted frame are
\begin{equation}\begin{aligned}
		R_{0101}
		=&
		\frac{1}{2dr^{2}}
		\left(
			\frac{r}{r_{F}}
		\right)^{2\theta/d}
		\left[
		k\frac{2(d-1)^{2}(z-2)(d(z-1)-\theta)}{(d- \theta + z-2)^{2}}
		+
		\frac{
			2r^{2(d-1)+2z}z(dz-\theta)
		}{\ell^{2} r^{2(d+z-2)}}
		\right.
		\\&
		+
		\left.
		(d-\theta)
		\frac{
			2q^{2}r^{2\theta}(d-1)(2(d-1)-2\theta+z)
			-
			mr^{d+\theta+z-2}(d (d-z)-(d-2)\theta)
		}{\ell^{2} r^{2(d+z-2)}}
		\right]
		\,,
		\\
		R_{0\hat{\i}0\hat{\i}}=&
		\frac{d-\theta}{d^{2}\ell^{2}}r^{-2(d+z-1)}
		\left(
			\frac{r}{r_{F}}
		\right)^{2\theta/d}
		\left[
			r^{2(d+z-1)}
			\left(
				d
				+
				E^{2}
				\left(
					\frac{r}{\ell}
				\right)^{-2z}
				\left(
					\frac{r}{r_{F}}
				\right)^{2\theta/d}
				(d(z-1)-\theta)
			\right)
		\right.
		\\&
		\left.	
			+
			(d-\theta+z-2)
			\left(
				m\frac{d}{2}r^{d-\theta +z-2}
				-
				q^{2}dr^{2\theta}
			\right)
		\right]
		,
		\\
		R_{1\hat{\i}1\hat{\i}}=&
		-\frac{d-\theta}{d^{2}r^{2}}\left(\frac{r}{r_{F}}\right)^{2\theta/d}
		\left[
			k\frac{(d-1)^{2}(d(z-1)-\theta)}{(d-\theta+z-2)^{2}}
			-
			\frac{q^{2}(d-1)r^{-2(d-\theta +z-2)}(d-\theta)}{\ell^{2}}
		\right.
		\\
		&
		\left.
		+
		\frac{mr^{-(d-\theta+z-2)}(d^{2}+2\theta-d(z+\theta))}{2\ell^{2}}
		+
			\frac{r^{2}	}{\ell^{2}}
				\left(dz-\theta
				-
				E^{2}\left(
						\frac{r}{\ell}
					\right)^{-2z}
					\left(
						\frac{r}{r_{F}}
					\right)^{2\theta/d}
					(d(z-1)-\theta)
				\right)
		\right]
		,
		\\
		R_{0\hat{\i}1\hat{\i}}=&
		\frac{(d-\theta)(d(z-1)-\theta)E^{2}}{d^{2}\ell^{2}}
		\left(
			\frac{r}{\ell}
		\right)^{-2z}
		\left(
			\frac{r}{r_{F}}
		\right)^{4\theta/d}
		\sqrt{
			1
			-
			\left(
				\frac{r}{\ell}
			\right)^{2z}
			\left(
				\frac{r}{r_{F}}
			\right)^{-2\theta/d}
			E^{-2}
			f(r)
		}
		\,,
		\\
		R_{\hat{\i}\hat{\j}\hat{\i}\hat{\j}}=&
		\frac{1}{d^{2}r^{2}}\left(\frac{r}{r_{F}}\right)^{2\theta/d}
		\left[
			k\frac{(d(z-1)-\theta)(d( 2d - 2 \theta  +z-3)+\theta)}{(d-\theta+z-2)^{2}}
			+
			(d-\theta)^{2}\frac{r^{2}}{\ell^{2}}
		\right.
		\\&\left.+
			(d-\theta)^{2}\frac{q^{2}r^{2+2\theta}-mr^{d+z-\theta}}{r^{2(d+z-1)}\ell^{2}}
		\right]
		\Theta_{k}
		\,.
		\nonumber
\end{aligned}\end{equation}
Here $f(r)$ is the blackening factor as defined in \eqref{blackeningspherical} and $\hat{\i}\neq\hat{\j}$. Also, $\Theta_{k}=\text{Csc}(\chi_{0})^{2},1,\text{Csch}(\chi_{0})^{2}$, depending on whether $k=+1,0,-1$, respectively.

We can consider ``pure'' hyperscaling violating spacetime if we put $m$ and $q$ to zero. If we furthermore require $\theta=0$, we recover the results of \cite{Copsey:2010ya,Horowitz:2011gh,Pang:2009ad}. We observe that for any $z>1$ we encounter divergences in the components of the Riemann tensor as $r\to0$, which signals diverging tidal forces. However, there is one exception, namely   the specific value $\theta=d(z-1)$. In the case of $\theta=d(z-1)$, when $1\leq z <2$, the expressions for the Riemann tensor simplify to
\begin{equation}
		R_{0101}
		=
		\frac{z}{2-z}R_{0\hat{\i}0\hat{\i}}
		=
		-\frac{z}{2-z}R_{1\hat{\i}1\hat{\i}}
		=
		-\frac{z}{(2-z)^{2}\Theta_{k}}
		R_{\hat{\i}\hat{\j}\hat{\i}\hat{\j}}
		=
		\frac{z}{\ell^{2}}
		\left(
			\frac{r}{r_{F}}
		\right)^{2(z-1)}
		\!\!\!\!\!,
		\quad
		R_{0\hat{\i}1\hat{\i}}=
		0\,.
\end{equation}
This solution holds for any value of $k$. The $k=0$ case was studied in \cite{Shaghoulian:2011aa}. When $z=1$ we return to the regular AdS case. For $1< z <2$, we notice the absence of diverging tidal forces as $r\to 0$, in contrast to all other $z\neq1$ cases. This result suggests that the $m=q=0$,   $\theta=d(z-1)$ spacetime is an example of a geodesically complete spacetime with non-relativistic symmetries on the boundary.

\section{Thermodynamics and phase structure}
\label{secthermodynamics}
In the previous section we have presented two new families of charged black hole solutions in a generalized EMD theory: the first one  (\ref{phispherical})-(\ref{blackeningspherical}) with spherical horizons and the second one (\ref{phihyper})-(\ref{blackeninghyper}) with hyperbolic horizons. The first family has two independent parameters, $z$ and $\theta$, while the second one, with $\theta=d(z-1)$, is restricted to just one free parameter. In the present section we analyze the thermodynamics   and phase structure associated to these two families of black holes.
We   closely follow the work \cite{Tarrio:2011de} of Tarrio and Vandoren, since many qualitative features of the thermodynamics of charged Lifshitz black holes ($\theta=0$) carry over to the hyperscaling violating case ($\theta \neq 0$).
In most instances we keep the topology parameter $k$ general, but sometimes we discuss the hyperbolic black hole separately since its thermodynamic variables are  slightly  different from those of the spherical and planar black hole. Only the $k=1$ case is found to have a non-trivial phase structure, so most of the explanations and figures will be specific for     the spherical case.




In appendix \ref{euclideanaction} we also compute the (background subtracted) on-shell Euclidean action, and derive the free energy from the action both in the canonical and grand canonical ensemble.\footnote{A more rigourous treatment would be to consider the full-fledged holographic renormalization of the theory \cite{Papadimitriou:2005ii}. The additional counterterms needed to regularize the on-shell action in the various cases would possibly give rise to  different expressions for the (Casimir) energies, depending on the number of dimensions. However, the background subtracted quantities that we compute here are sufficient for our purposes, namely, for studying the thermodynamics and phase structure of our solutions.} From the free energy and the temperature one can obtain the entropy and   energy of a thermodynamic system. We check in the appendix that the entropy and mass of the black hole obtained in this way agree with the expressions presented in the present section, which are derived in a more pedestrian way using the Bekenstein-Hawking entropy  formula and the ADM  mass expression.


\subsection{Thermodynamic quantities and first law}
\label{firstlaw}

Charged black holes can have several inner horizons and an outer horizon. We denote the position of the  outer horizon   by $r_h$,  i.e.   the  largest positive root of $f(r_h) = 0$. Apart from the horizon radius $r_h$, our black hole systems are controlled by several other   length scales: the  curvature radius $\ell$, the UV scale $r_F$, the scalar amplitude $\phi_0$, and the charge parameter $q$. We will  express most thermodynamic quantities in terms of these parameters. \\


\noindent \textbf{Temperature and entropy} From the condition $f(r_h) = 0$ we can solve for the mass parameter $m$ and express it in terms of the horizon radius $r_h$ as follows
\begin{equation}\label{massparameter}
m = r_h^{d+z-\theta} \left [ 1+ k \frac{ (d-1)^2  }{ (d-\theta +z-2)^2} \frac{\ell^{2}}{r_h^2}
+ \frac{q^2} { r_h^{2 (d-\theta  + z - 1)} } \right] \, .
\end{equation}
For planar and spherical black holes $m$ is non-negative, whereas for hyperbolic black holes it can become negative (but not arbitrarily negative as we will see below).
By using the standard Euclidean trick, one can find  the following temperature for the metric Ansatz (\ref{ansatz})
\begin{equation}\label{HawkT}
T = \frac{1}{4\pi} \left ( \frac{r_h}{\ell} \right)^{z+1} \big | f ' (r_h) \big |
\,.
\end{equation}
Note that the conformal factor (which includes the radius $r_F$) does not feature in this expression, since the Hawking temperature is conformally invariant \cite{Jacobson:1993pf}.
By inserting     formula (\ref{blackeningspherical}) for the blackening factor and     (\ref{massparameter}) for the mass parameter we find
\begin{equation}   \label{temperature}
T = \frac{r_h^{z}}{4 \pi \ell^{z+1}}  \left [(d-\theta +z)
 +k \frac{(d-1)^2  }{  (d-\theta +z-2)}  \frac{\ell^2}{r_h^2}
 -  \frac{ (d-\theta + z -2)  q^2 }{  r_h^{2 ( d- \theta + z-1)} }
   \right] \, .
\end{equation}
The entropy is given by the Bekenstein-Hawking formula, which in our case  takes the form
\begin{equation}
S = \frac{\omega_{k,d}}{4 G} r_h^{d-\theta} r_F^\theta \,
\,.
\end{equation}
Here $\omega_{k,d}$ is the volume of the space described by the unit metric $d \Omega_{k,d}^2$, e.g. $\omega_{1,d}$ is the volume of the unit $d$-sphere.
In this form, the entropy is independent of the Lifshitz dynamical exponent $z$, but it depends explicitly on the hyperscaling violation exponent $\theta$.

Furthermore, the black hole becomes extremal when the temperature vanishes. This happens when the charge parameter is given by
 \begin{equation} \label{extremalcharge}
 q^2_{ext} =      r_{ext}^{2(d+z-\theta -1)} \left [\frac{d-\theta+z}{d-\theta + z- 2 } + k \frac{\ell^2}{r_{ext}^2} \frac{(d-1)^2}{(d-\theta + z - 2 )^2} \right] \, ,
 \end{equation}
 where   $r_{ext}$ denotes the extremal horizon radius, defined by  $f(r_{ext}) = f'(r_{ext}) = 0$.
 The extremal value of the mass parameter can be expressed in terms of $r_{ext}$ as follows
  \begin{equation} \label{extremalmass}
 m_{ext}= 2  r_{ext}^{d-\theta +z}  \left [       \frac{d-\theta +z-1}{d-\theta +z-2} + k  \frac{\ell^2}{r_{ext}^2} \frac{  (d-1)^2    }{(d-\theta +z-2)^2}   \right]
 \,.
 \end{equation}
The extremal black hole solution corresponds to the ground state of the theory in the canonical ensemble. Its finite entropy implies that the ground state is highly degenerate, which is a well-known feature of charged AdS black holes \cite{Chamblin:1999tk}. The special case of vanishing temperature and  zero charge ($q_{ext}=0$) corresponds to the ground state in the grand canonical ensemble. For planar and spherical black holes this solution is obtained by setting $m=0$ and $r_h=0$. The latter condition implies that it has zero entropy and therefore the ground state is non-degenerate.  For the hyperbolic black hole, however, the ground state in the grand canonical ensemble has a negative mass parameter and a finite horizon radius
  \begin{eqnarray}  \label{groundhyperbolic}
 k=-1:  \quad m_{ground}  =  - \frac{2   \ell^2 r_{h,ground}^{(d-1)(2-z)}  }{(2-z)^2(z +d (2-z))}    \, , \quad  r_{h,ground}   =\ell \sqrt{  \frac{(d-1) }{(2-z)(z +d(2-z))}}
 \,,
 \end{eqnarray}
where we inserted the  only physical value for the hyperscaling violating parameter $\theta = d(z-1)$. Hence, the ground state of the hyperbolic black hole has  finite entropy, even in the grand canonical ensemble. A similar observation holds for AdS black holes with a hyperbolic horizon, and indeed the minimal values of $m$ and $r_h$ above agree with those found in \cite{Birmingham:1998nr,Emparan:1999gf} for $z=1$.    \\

\noindent \textbf{Mass} The mass of the black hole solutions can be found by evaluating the ADM expression
\begin{equation}
M_T = - \frac{1}{8 \pi G} \int_{S_{k,d}}  d^d x   \sqrt{\sigma} \, N  \Theta \Big |_{r=R}\,.
\end{equation}
Here $\sigma$ is the determinant of the induced metric on the codimension-two surface $S_{k,d}$ (a radial slice $r=R$ of a constant-$t$ surface), $N$ is the lapse function and $\Theta$ is the  trace of the extrinsic curvature of  $S_{k,d}$ as embedded in the constant-$t$ surface. For our metric Ansatz  (\ref{ansatz}) we have
\begin{equation}
N = \sqrt{ |g_{tt} |}  = \left  ( \frac{r}{r_F} \right)^{\!\!-\theta/d} \!\! \left (  \frac{r}{\ell} \right)^z   \sqrt{f(r)} \, , 
 \qquad \Theta = \frac{1}{\sqrt{g_{rr}}} \partial_r \log \sqrt{\sigma} =    \left ( \frac{r}{r_F}\right)^{ \!\!\theta/d}      \frac{d-\theta}{\ell}  \sqrt{f(r)}    
 \,.
\end{equation}
The regularized mass can be obtained by subtracting the result for the thermal case (for $k=0,1$ this is given by the $m=q=0$ solution with a periodic Euclidean time circle) and taking the limit where the surface $S_{k,d}$ goes to spatial infinity:
\begin{equation} \label{sphericalmass}
k=0,1: \qquad M = \lim_{R \rightarrow \infty}  \left ( M_T - \frac{\sqrt{f(R)} }{\sqrt{f_0(R)}}M_0 \right) = \frac{\omega_{k,d}}{16 \pi G} (d- \theta) m \ell^{-z-1} r_F^\theta
\,.
\end{equation}
The factor $\sqrt{f(R)}/\sqrt{f_0(R)}$ is included to ensure  that the intrinsic geometry of the hypersurface $r=R$ is the same for the black hole and for the vacuum hyperscaling violating spacetime with Euclidean thermal circle. Namely, this factor effectively rescales the lapse function of the vacuum spacetime  so that it becomes identical to $N$ for the black hole at $r=R$.

The mass indeed has the correct dimensions $[M] = [L]^{d-1}$, since the mass parameter $m$ is a dimensionful quantity that scales with length as $[m] = [L]^{d+z-\theta}$. Furthermore, we see clearly from this expression that the hyperscaling violation parameter $\theta$ effectively reduces the number of dimensions. For the pure Lifshitz black hole ($\theta =0$) our result agrees with the Komar mass  calculated in  \cite{Tarrio:2011de} and with the mass expression in \cite{Brenna:2015pqa} that follows from imposing the Smarr formula. For the hyperscaling violating black brane ($k=0$) it is consistent with the energy term in the renormalized stress-tensor  found   in \cite{Kiritsis:2016rcb} and with the ADM mass computed in \cite{Tarrio:2013tta}.



For the hyperbolic black hole it is more natural to take the extremal solution  with zero charge  (\ref{groundhyperbolic})   as the background spacetime,  instead of  the $m=q=0$ solution, since the former corresponds to the ground state in the grand canonical ensemble. With this choice of background the regularized mass becomes
\begin{equation} \label{hypermass}
 k=-1: \qquad M =   \frac{\omega_{k,d}}{16 \pi G} d (2-z )  (m - m_{ground}) \ell^{-z-1} r_F^{d(z-1)}\,.
\end{equation}
Notice that with such a subtraction we are only measuring energies above the ground state. As a consequence, from the background subtraction method one cannot determine for example  whether the ground state (\ref{groundhyperbolic})  has negative  energy or zero energy.  This issue could be resolved with the method of counterterm subtraction (or holographic renormalization) \cite{Henningson:1998gx,Balasubramanian:1999re}, which has indeed been employed for the case of hyperbolic AdS black holes in \cite{Emparan:1999pm,Emparan:1999gf}, but we leave that analysis for future work. 
We stress though that such an analysis can only change the absolute value of the mass by a constant term, but it does not modify the background subtracted results (\ref{sphericalmass}) and (\ref{hypermass}), and hence does not affect the thermodynamics and phase structure of the   black holes.\\ 


\noindent \textbf{Conserved charges and  potentials}
The total electric charge of the black hole is given by the conserved charge of the field strength $K$
\begin{eqnarray} \label{electriccharge}
Q \equiv Q_K = \frac{1}{16\pi G} \int Z(\phi ) * K = \frac{\omega_{k,d}}{16\pi G} Z_0  \rho_3 \ell^{z-1} r_F^{\theta- 2 \theta/d} \, .
\end{eqnarray}
 Similar expressions exist for the other two conserved charges $Q_F$ and $Q_H$ in terms of $\rho_1$ and $\rho_2$. However, as we will see below, these two charges do not have a thermodynamic interpretation.
We can express all three conserved charges in terms of the parameters that characterize the solution
\begin{eqnarray}
Q_F &=&  \frac{\omega_{k,d}}{16\pi G}  \sqrt{2 X_0 (z-1)(d-\theta + z)} \ell^{-1} r_F^{\theta-  \theta/d} e^{\lambda_1 \phi_0/2} \,,  \label{chargeF}\\
Q_H &=&  \frac{\omega_{k,d}}{16\pi G} \sqrt{k Y_0 \frac{2(d-1)(d(z-1)-\theta)}{d-\theta + z-2}}   r_F^{\theta-   \theta/d} e^{ \lambda_2 \phi_0 /2}  \,, \label{chargeH}\\
Q_K &=& \frac{\omega_{k,d}}{16 \pi G}  \sqrt{2 Z_0 (d-\theta)(d-\theta + z-2)} \, q  \, \ell^{-1} r_F^{\theta -  \theta/d} e^{\lambda_3 \phi_0 /2}\,.\label{chargeq}
\end{eqnarray}
Furthermore, the gauge field potentials are given by
\begin{eqnarray}
A&=&\frac{\rho _1}{d-\theta +z} e^{-\lambda _1 \phi_0}   \left( r^{d-\theta +z} -  r_h^{d-\theta +z} \right)dt\,,   \label{gaugefield1} \\
B&=&\frac{\rho_2}{d-\theta +z-2}   e^{-\lambda _2 \phi_0}  \left ( r^{d-\theta +z-2}  - r_h^{d-\theta +z-2} \right) dt\,,   \label{gaugefield2} \\
C&=& - \frac{\rho_3}{d-\theta +z-2 }   e^{-\lambda _3 \phi_0}  \left ( r^{-(d-\theta +z-2)  } - r_h^{-(d-\theta +z-2)  } \right) dt\,,   \label{gaugefield3}
\end{eqnarray}
 where we have chosen the integration constant such that the gauge field vanishes at the horizon.
The gauge fields $A$ and $B$ are   needed  to support the asymptotics of the spacetime and the topology of the internal space, respectively. The charges associated to these gauge fields must be kept fixed, otherwise the boundary theory would not be properly defined. Hence, they do not feature in the thermodynamic first law. On the other hand, the charge associated to $C$ can be kept fixed or can be varied, depending on the ensemble. Different values of $Q$ give rise to states in the boundary theory at different charge density. The potential associated to the electric charge $Q$ is given by  the asymptotic value of gauge field $C$,
\begin{equation} \label{gaugepotential}
\Phi \equiv C_{t} (\infty) = \frac{q}{c \, r_h^{d- \theta +z-2}} \, , \qquad \text{with} \qquad c = \sqrt{\frac{Z_0 (d- \theta + z-2)}{2 (d-\theta)}} \ell^z r_F^{-\theta/d} e^{\lambda_3 \phi_0 /2} \, .
\end{equation}
As a remark, we notice that   the temperature (\ref{temperature}) can be expressed purely as a function  of  $\Phi$ and  $q$ as follows
   \bqn\label{eqntemp}
    T   =   \frac{   (d-\theta +z)   + (d-\theta + z-2) \left (  k \ell^2  \frac{(d-1)^2}{(d-\theta+z-2)^2}  -    c^2 \Phi^2        \right)  \left ( \frac{c \Phi}{q   }         \right)^{\frac{2}{d-\theta + z-2}    }  }{   4 \pi \ell^{z+1} \left ( \frac{c \Phi}{q   }        \right)^{\frac{z}{d-\theta + z-2}}} \, .
\eqn
Finally,  we emphasize that for the hyperbolic black hole the   field strength $H$ is absent, so $Q_H$ does not exist. However, the   expressions for the electric charge (\ref{electriccharge}) and potential (\ref{gaugepotential}) do hold for $k=-1$, with the specific values $\theta = d(z-1)$, $\lambda_3 =0$ and $\lambda_{i}\to\tilde{\lambda}_{i}$ for $i=0,1,2$.
  \\

 \noindent \textbf{First law of thermodynamics} With these definitions of the thermodynamic quantities, it is now straightforward to check that the first law of thermodynamics holds. In the grand canonical ensemble we have
\begin{equation}\label{first_law}
dM = T dS + \Phi dQ \, .
\end{equation}
In the canonical ensemble we need to compare to the extremal case instead of the thermal case, and hence the first law takes the form
\begin{equation}
d \tilde{M} = T dS + \tilde{\Phi} dQ \, ,\label{first_lawB}
\end{equation}
where the mass and electric potential are now given by
 \begin{eqnarray}
&& \tilde{M} = M - M_{ext} = \frac{\omega_{k,d}}{16 \pi G} (d-\theta) (m - m_{ext}) \ell^{-z-1} r_F^\theta   \, , \label{masscanonical} \\
&& \tilde{\Phi} \,\, = \Phi - \Phi_{ext} = \frac{q}{c} \left (  \frac{1}{r_h^{d- \theta + z-2}} -  \frac{1}{r_{ext}^{d- \theta + z-2}}   \right) \, . \label{potentialcanonical}
   \end{eqnarray}
As mentioned above, the gauge potentials $A$ and $B$ do not play any role in the thermodynamics. The corresponding charges of these fields are needed just to support the structure of the asymptotic spacetime and the geometry of the internal space, so they are kept fixed.\footnote{Varying the charges $Q_F$ and $Q_H$ would imply changing the symmetries and geometry of the dual field theory and this would  drastically affect the holographic interpretation.}

 We emphasize that all the above thermodynamic quantities, as well as the explicit form of the first laws, can also be derived by working out the free energy from the on-shell Euclidean action. The explicit calculations are presented in appendix \ref{euclideanaction}. Needless to say, the two methods give exactly the same results, so the computations shown in the appendix can serve as a non-trivial consistency check of the results presented in this section.


\subsection{Grand canonical ensemble (fixed potential)}

In the following sections we will study the phase structure of the hyperscaling violating black holes, both in the grand canonical and canonical ensemble.
In the grand canonical ensemble the electric charge $Q$ is free to vary, but its potential $\Phi$ and   temperature $T$ are held fixed. 
We will only find non-trivial phase structure in the spherical case $k=1$, but for generality we keep the explicit dependence on $k$ in most of the expressions.

We start by  defining a critical value of the potential
\begin{equation}
\Phi_c^2 = k \frac{(d-1)^2}{(d- \theta + z - 2)^2}\frac{\ell^{2}}{c^{2}}\,.
\end{equation}
The physical meaning of this quantity will become   clear below. With this definition we can rewrite the temperature (\ref{temperature}) in a   compact form
\begin{eqnarray}
T
   =  \frac{r_h^{z}}{4 \pi \ell^{z+1}}  \left [(d-\theta +z)
 +    (d-\theta +z-2)    \left ( \Phi_c^2 - \Phi^2  \right)  \frac{c^2}{ r_h^{2}}   \right].
    \end{eqnarray}
Similarly,   the thermodynamic potential or (Gibbs) free energy, $W = M - TS -  \Phi Q  $,  
  for the different black hole solutions is given by
\begin{eqnarray}  \label{free_energy}
&&k=0,1: \quad W     =  \frac{ \omega_{k,d}}{16 \pi G} \frac{r_F^{\theta } r_h^{d-\theta +z}}{\ell^{z+1}}
    \left[ -z + (2-z) (\Phi_c^2 - \Phi^2)\frac{c^{2} }{r_h^{2}} \right]\,, \\
   && k=-1 \, : \quad W =  \frac{ \omega_{k,d}}{16 \pi G} \frac{r_F^{d(z-1) } }{\ell^{z+1}}
    \left[ -r_h^{d (2 - z)+z} \left (  z     + \frac{1}{2-z} \frac{\ell^2}{r_h^2}  \right)   - (2-z)  q^2    -  d(2-z) m_{ground} \right]\,, \nonumber
\end{eqnarray}

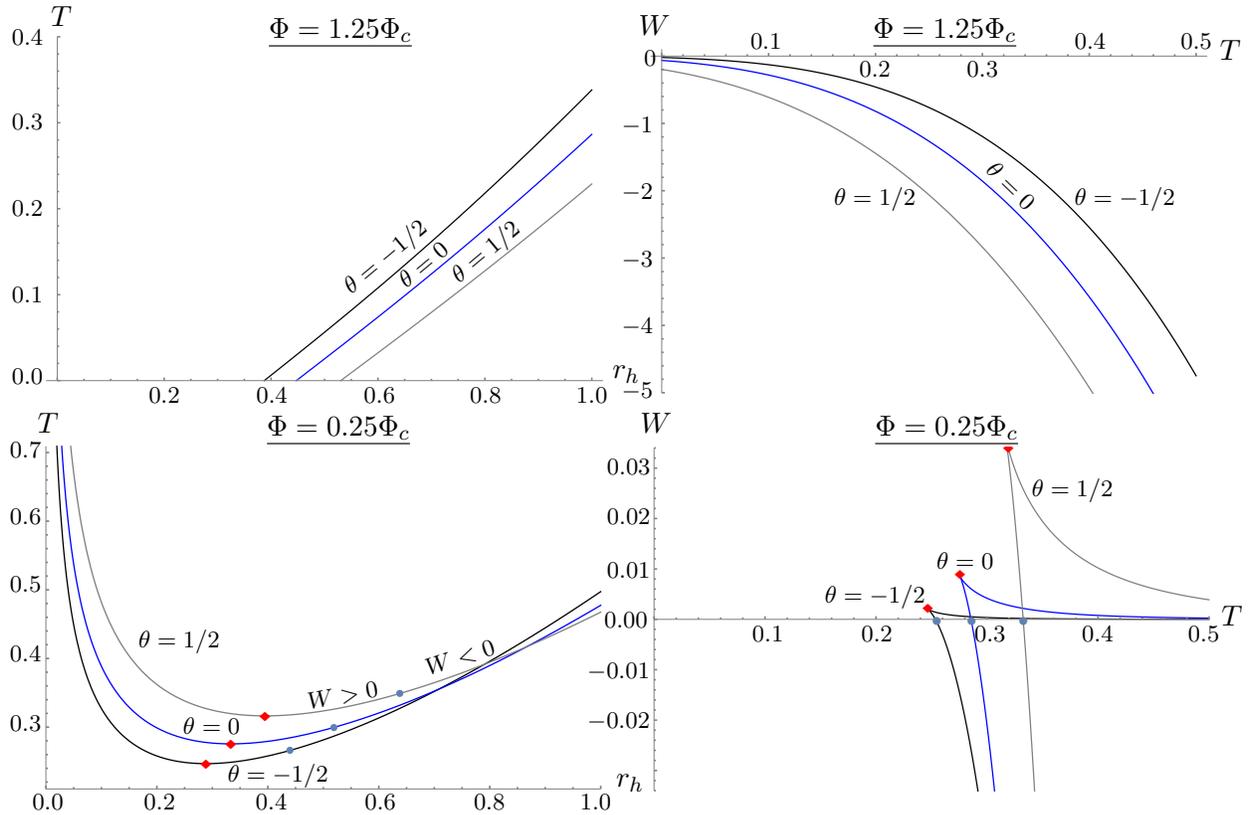
\begin{figure}[h!!]
\begin{center}
\begin{overpic}[width=.45\textwidth]{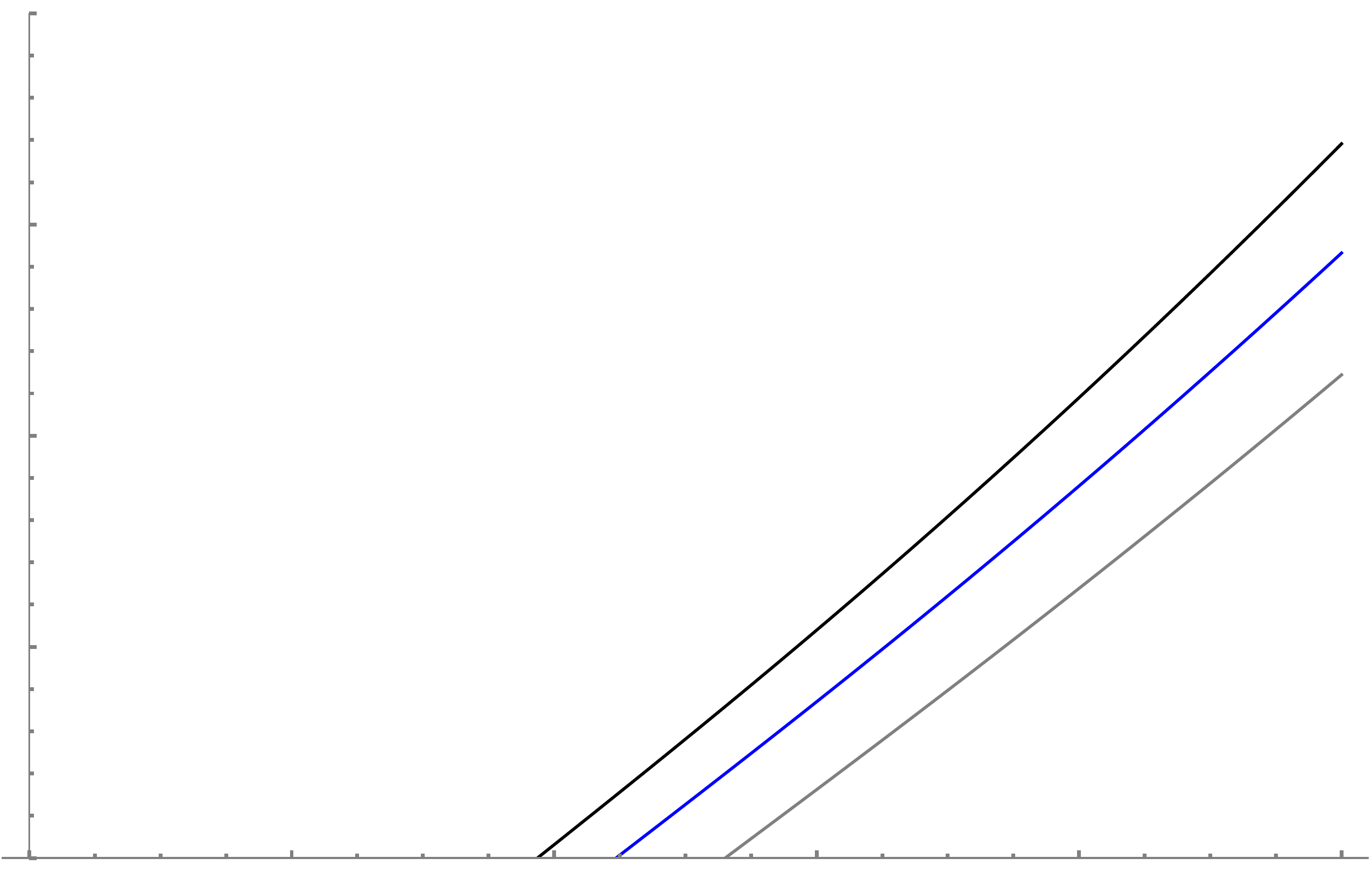}
		\put (-6,2) {\footnotesize{$0.0$}}
		\put (-6,16.5) {\footnotesize{$0.1$}}
		\put (-6,32) {\footnotesize{$0.2$}}
		\put (-6,47.5) {\footnotesize{$0.3$}}
		\put (-6,62.75) {\footnotesize{$0.4$}}
		
		\put (18.5,-1.5) {\footnotesize{$0.2$}}
		\put (37.75,-1.5) {\footnotesize{$0.4$}}
		\put (57,-1.5) {\footnotesize{$0.6$}}
		\put (76,-1.5) {\footnotesize{$0.8$}}
		\put (95,-1.5) {\footnotesize{$1.0$}}
		

		\put (72,20) {\rotatebox{40}{\footnotesize{$\theta=1/2$}}}
		\put (62.75,19) {\rotatebox{40}{\footnotesize{$\theta=0$}}}
		\put (52.75,18) {\rotatebox{40}{\footnotesize{$\theta=-1/2$}}}
		
		\put (40,64) {$\underline{\Phi=1.25\Phi_{c}}$}
		
		\put (1,66) {$T$}
		\put (101.5,3) {$r_{h}$}
\end{overpic}
\hspace{0.35cm}
\vspace{0.65cm}
\begin{overpic}[width=.45\textwidth
]{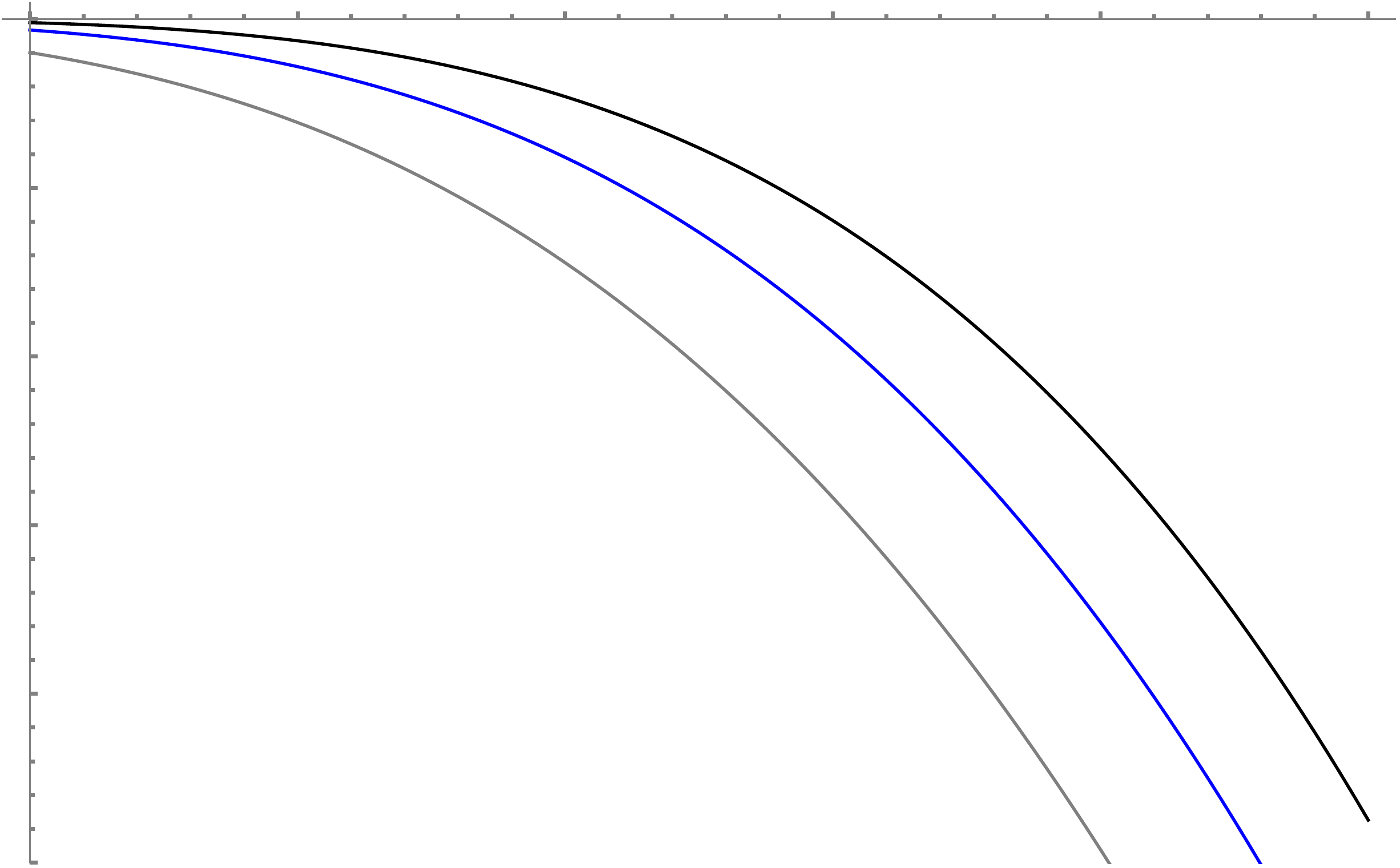}

		\put (-1,59) {\footnotesize{$0$}}
		\put (-5,47) {\footnotesize{$-1$}}
		\put (-5,35) {\footnotesize{$-2$}}
		\put (-5,23) {\footnotesize{$-3$}}
		\put (-5,11) {\footnotesize{$-4$}}	
		\put (-5,-0.5) {\footnotesize{$-5$}}

		\put (40,64) {$\underline{\Phi=1.25\Phi_{c}}$}

		\put (18,62) {\footnotesize{$0.1$}}
		\put (37.5,57) {\footnotesize{$0.2$}}
		\put (56.75,57) {\footnotesize{$0.3$}}
		\put (76,62) {\footnotesize{$0.4$}}
		\put (95,62) {\footnotesize{$0.5$}}

		\put (33,34) {\footnotesize{$\theta=1/2$}}
		\put (60,40) {\rotatebox{-44}{\footnotesize{$\theta=0$}}}
		\put (76,34) {\footnotesize{$\theta=-1/2$}}

		\put (-2,64.5) {$W$}
		\put (102,60) {$T$}
		

\end{overpic}
\begin{overpic}[width=.45\textwidth]{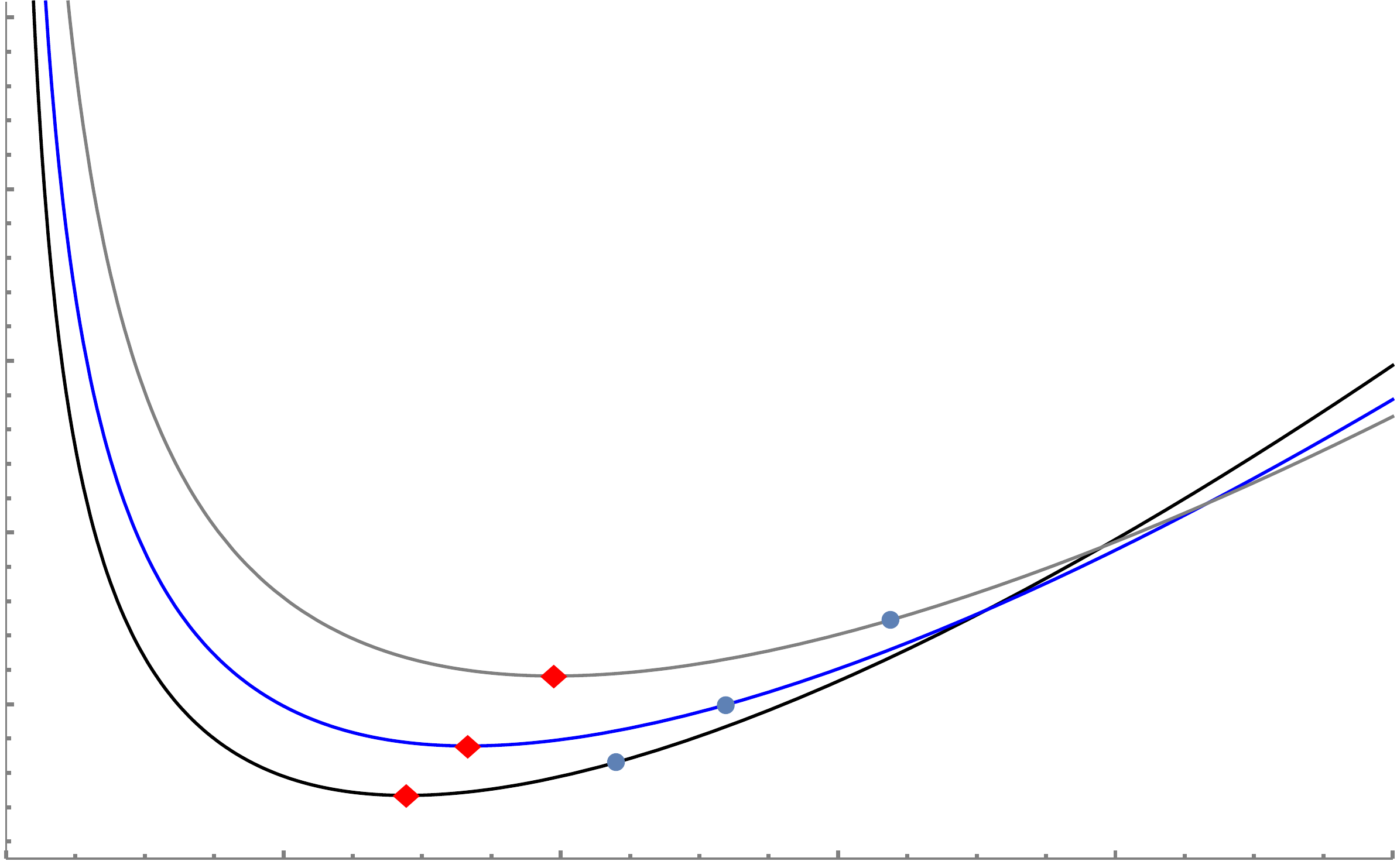}
		\put (-6,11) {\footnotesize{$0.3$}}
		\put (-6,23.5) {\footnotesize{$0.4$}}
		\put (-6,36) {\footnotesize{$0.5$}}
		\put (-6,48) {\footnotesize{$0.6$}}
		\put (-6,60) {\footnotesize{$0.7$}}

		\put (-2,-3) {\footnotesize{$0.0$}}		
		\put (18,-3) {\footnotesize{$0.2$}}
		\put (38,-3) {\footnotesize{$0.4$}}
		\put (57.5,-3) {\footnotesize{$0.6$}}
		\put (77,-3) {\footnotesize{$0.8$}}
		\put (96,-3) {\footnotesize{$1.0$}}
		
		\put (40,64) {$\underline{\Phi=0.25\Phi_{c}}$}

		\put (47,15) {\rotatebox{10.5}{\footnotesize{$W>0$}}}
		\put (68,21) {\rotatebox{18.5}{\footnotesize{$W<0$}}}
		
		\put (17,26) {\footnotesize{$\theta=1/2$}}
		\put (25,10.5) {\footnotesize{$\theta=0$}}
		\put (33,2) {\footnotesize{$\theta=-1/2$}}

		\put (-1,64.5) {$T$}
		\put (102.5,1) {$r_{h}$}
\end{overpic}
\hspace{0.4cm}
\begin{overpic}[width=.45\textwidth]{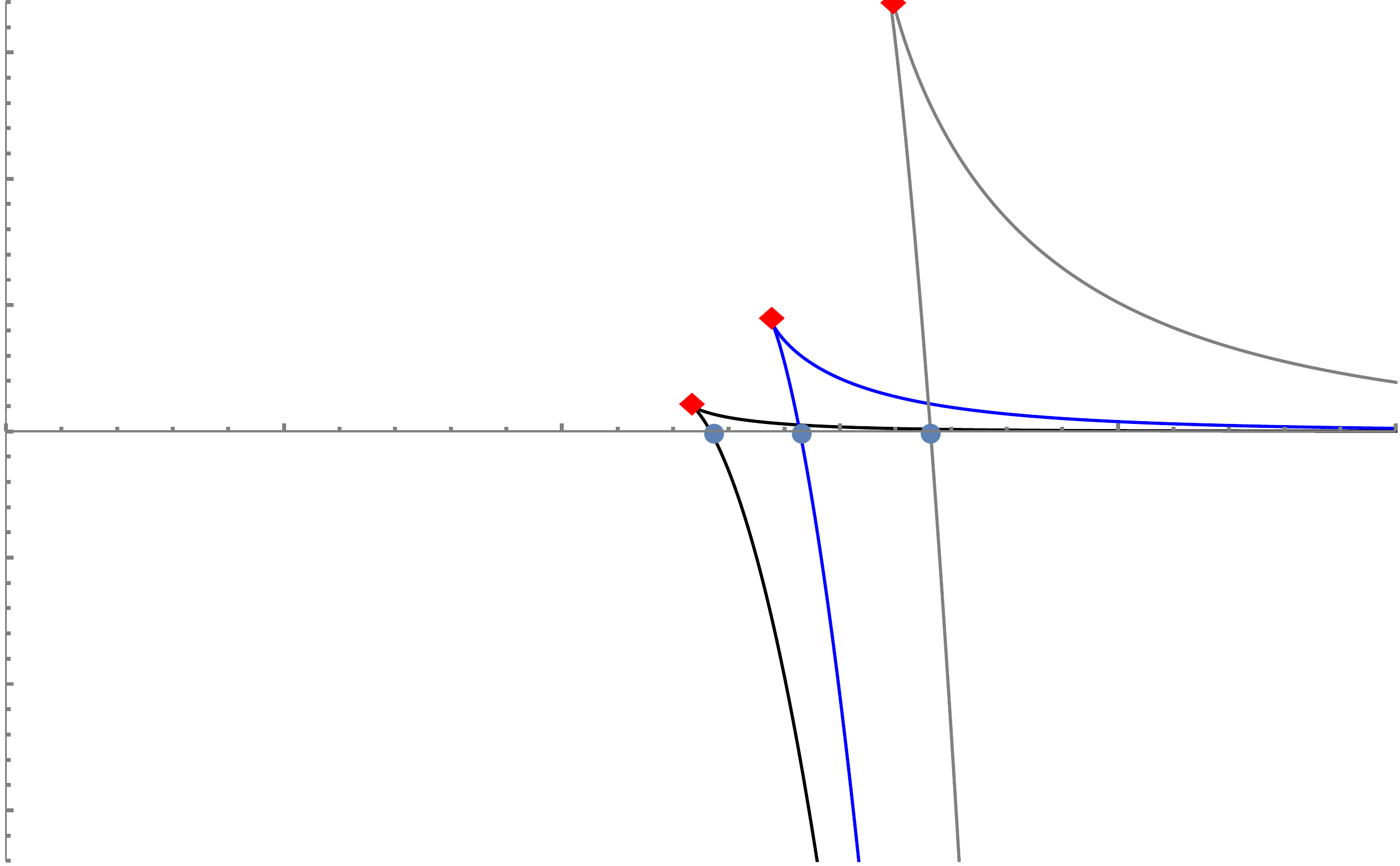}

		\put (-8,57) {\footnotesize{$0.03$}}
		\put (-8,48) {\footnotesize{$0.02$}}
		\put (-8,38.5) {\footnotesize{$0.01$}}	
		\put (-8.5,29.5) {\footnotesize{$0.00$}}
		\put (-11.5,20.5) {\footnotesize{$-0.01$}}
		\put (-11.5,11.5) {\footnotesize{$-0.02$}}

		\put (18,27) {\footnotesize{$0.1$}}
		\put (37.5,27) {\footnotesize{$0.2$}}
		\put (57.75,27) {\footnotesize{$0.3$}}
		\put (77,27) {\footnotesize{$0.4$}}
		\put (96,27) {\footnotesize{$0.5$}}

		\put (68,53) {\footnotesize{$\theta=1/2$}}
		\put (51,40) {\footnotesize{$\theta=0$}}
		\put (31,34) {\footnotesize{$\theta=-1/2$}}

		\put (-2,64.5) {$W$}
		\put (102,30) {$T$}
		
		\put (40,64) {$\underline{\Phi=0.25\Phi_{c}}$}

\end{overpic}
\end{center}
\vspace{-0.2cm}
\caption{
Plots of temperature vs. horizon radius and free energy vs. temperature  for two fixed values of the electric potential.
The upper two plots correspond to  $\Phi>\Phi_{c}$, whereas the lower two correspond to $\Phi<\Phi_{c}$. Three different values of $\theta$ are shown in each plot and other parameters are fixed at: $k=\phi_{0}=Z_{0}=r_{F}=16\pi G/\omega_{k,d}=1$, $z=3/2$ and $d=3$. It is shown that  different values of $\theta$ only yield quantitative differences.   In the case $\Phi < \Phi_c$, at the blue dot a Hawking-Page phase transition occurs between a thermal spacetime ($W > 0$) and the black hole solution ($W < 0$). The upper branch starting from the red dot in the free energy plot is thermodynamically unstable. In the case   $\Phi>\Phi_{c}$,   the temperature $T$ becomes an injective function of  the horizon radius $r_{h}$, and the free energy $W$ has a single and strictly negative branch for every value of the temperature, which rules out any phase transition. The same happens when $z>2$.
}\label{fig:grand_canonical_free}
\end{figure}

\noindent where $m_{ground}$ is given by (\ref{groundhyperbolic}). The free energy is measured with respect to the thermal spacetime, which is given by the $(m=q=0)$ solution for $k=0,1$, and    the $(m=m_{ground}, q=0)$ solution for $k=-1$. One can verify  that for the planar and hyperbolic black hole the free energy is always negative,  hence the black hole solutions  dominates the  entire phase diagram (except at $\Phi=T=0$ where the ground state dominates).   However, for the spherical black hole   the free energy can switch sign, which occurs precisely  at  the Hawking-Page phase transition \cite{Hawking:1982dh}   between the black hole solution and the thermal spacetime. 
  The   black hole dominates the ensemble in the regime where its free energy
is negative, while the thermal spacetime is thermodynamically preferred in cases where the   free energy is positive. In Figure \ref{fig:grand_canonical_free} we show some illustrative examples of this transition.

For AdS-Schwarzschild black holes the Hawking-Page transition occurs for all dimensions $d>0$ \cite{Witten:1998zw} and it only depends on the horizon size (or equivalently on the temperature): for $r_h <\ell$ the free energy is positive and for $r_h > \ell$ it is negative. For more complicated black holes, however, the   transition is found to depend on other parameters as well. For example, for charged AdS black holes the transition depends on the value of the electric potential  \cite{Chamblin:1999tk}, and hence the $(\Phi, T)$   diagram contains an entire line of first order phase transitions.  Moreover, the transition only applies to spherical AdS black holes, and not to planar  \cite{Witten:1998zw} or hyperbolic black holes \cite{Birmingham:1998nr, Emparan:1999gf}, and therefore also depends on the horizon topology parameter $k$. Finally, for Lifshitz black holes the transition also depends  on the dynamical exponent $z$, i.e. it only exists  for the values $1 \le z \le 2$  \cite{Tarrio:2011de}.

It is interesting to ask whether the Hawking-Page phase transition for hyperscaling violating black holes depends on   $\theta$. After a close inspection, we  find that the transition indeed holds for all physical values of $\theta$ for which the black hole solution is valid. 
This agrees with the (somewhat naive) intuition that the hyperscaling violating parameter   effectively only reduces the number of dimensions,  and it is consistent with the fact that the  transition holds for all dimensions.

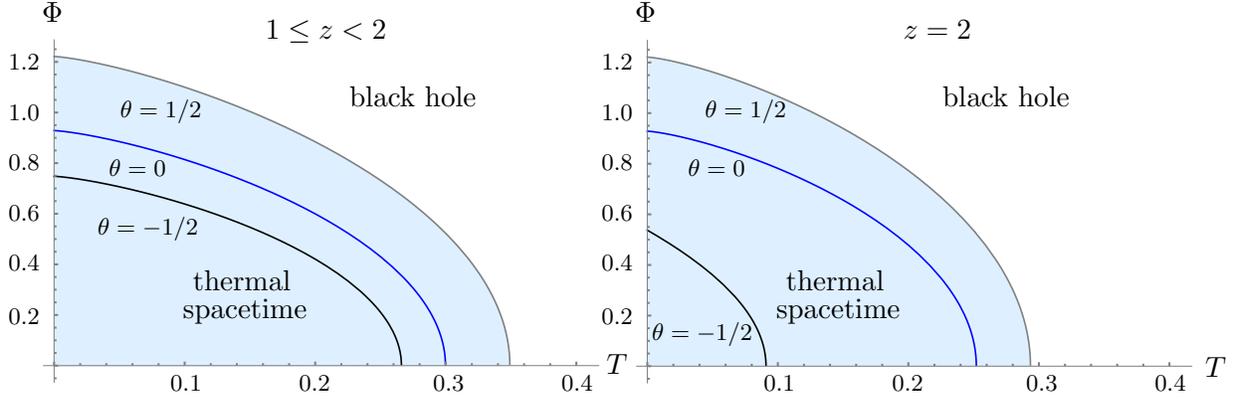
\begin{figure}[t!!]
\vspace{0.4cm}
\begin{center}
\begin{overpic}[width=.45\textwidth]{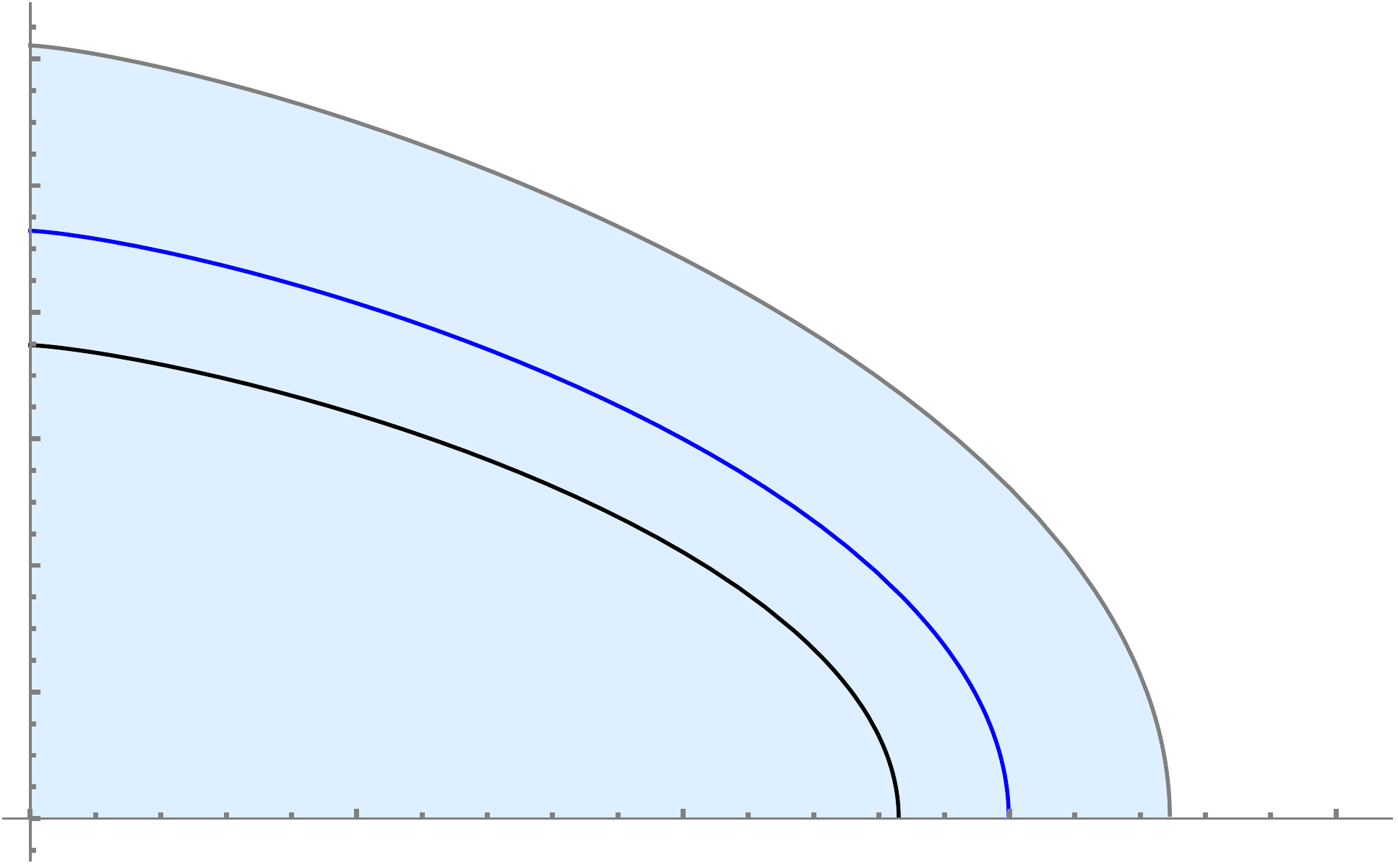}

		\put (14,48) {\footnotesize{$\theta=1/2$}}
		\put (12,37.5) {\footnotesize{$\theta=0$}}
		\put (10,27) {\footnotesize{$\theta=-1/2$}}

		\put (-6,11) {\footnotesize{0.2}}
		\put (-6,20.5) {\footnotesize{0.4}}
		\put (-6,29.5) {\footnotesize{0.6}}
		\put (-6,38.5) {\footnotesize{0.8}}
		\put (-6,47.5) {\footnotesize{1.0}}
		\put (-6,56.5) {\footnotesize{1.2}}
		
		\put (23,-1) {\footnotesize{0.1}}
		\put (46,-1) {\footnotesize{0.2}}
		\put (70,-1) {\footnotesize{0.3}}
		\put (93,-1) {\footnotesize{0.4}}
		
		\put (0,65) {$\Phi$}
		\put (101,2) {$T$}
		\put (40,62) {$1\leq z<2$}

		\put (55,50) {black hole}
		\put (27,17) {thermal}
		\put (25.25,12) {spacetime}
\end{overpic}
\hspace{0.2cm}
\begin{overpic}[width=.45\textwidth]{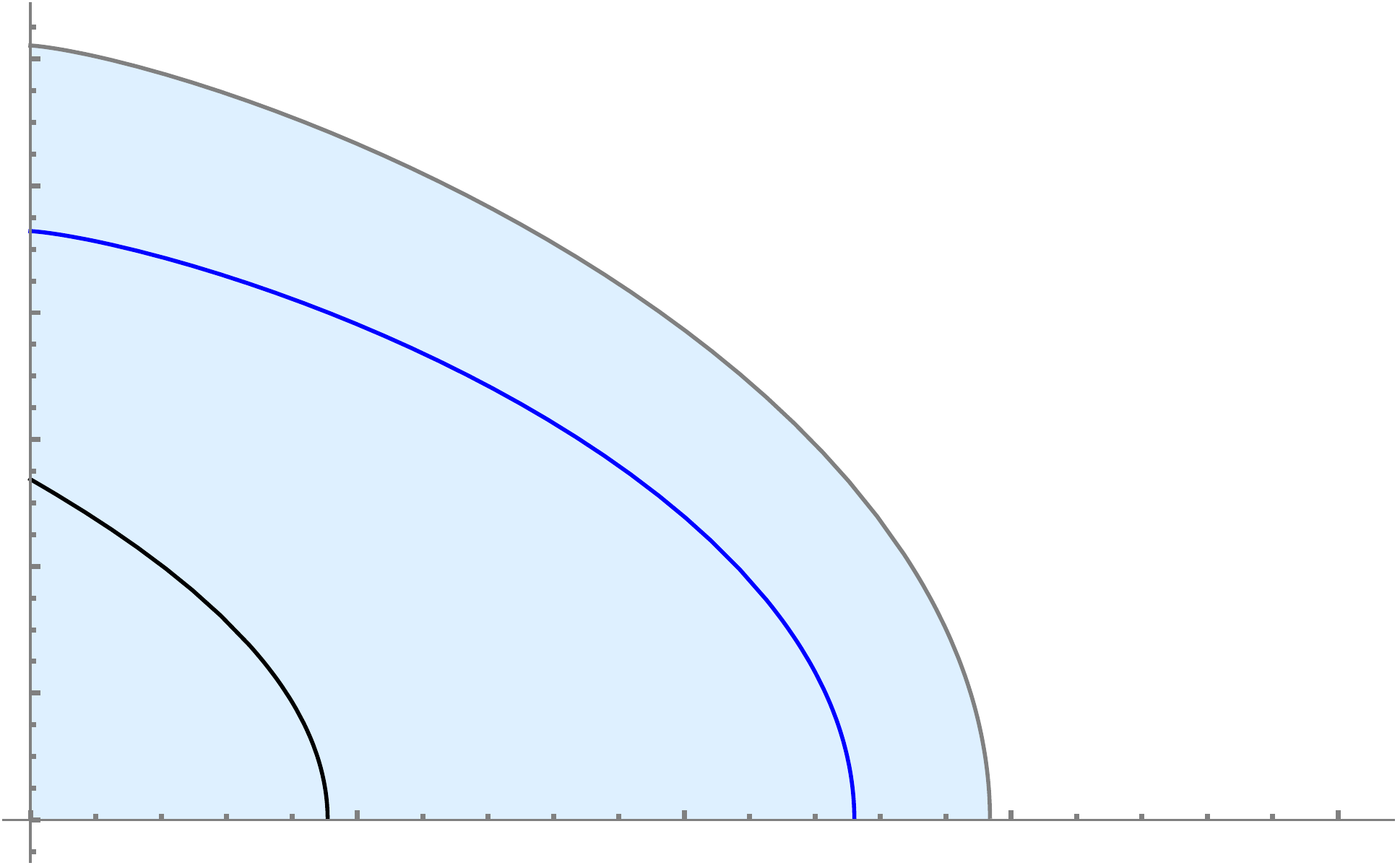}

		\put (-6,11) {\footnotesize{0.2}}
		\put (-6,20.5) {\footnotesize{0.4}}
		\put (-6,29.5) {\footnotesize{0.6}}
		\put (-6,38.5) {\footnotesize{0.8}}
		\put (-6,47.5) {\footnotesize{1.0}}
		\put (-6,56.5) {\footnotesize{1.2}}
		
		\put (23,-1) {\footnotesize{0.1}}
		\put (46,-1) {\footnotesize{0.2}}
		\put (70,-1) {\footnotesize{0.3}}
		\put (93,-1) {\footnotesize{0.4}}

		\put (12.5,48) {\footnotesize{$\theta=1/2$}}
		\put (9.5,37.5) {\footnotesize{$\theta=0$}}
		\put (3,8) {\footnotesize{$\theta=-1/2$}}

		\put (0,65) {$\Phi$}
		\put (48,62) {$z=2$}

		\put (102,1.5) {$T$}
		
		\put (55,50) {black hole}
		\put (27,17) {thermal}
		\put (25.25,12) {spacetime}
	\end{overpic}

\end{center}
\vspace{-0.2cm}
\caption{Two phase diagrams in the grand canonical ensemble for $z=3/2$ and $z=2$, plotted for three different values of $\theta$. Other parameters are fixed at: $k=\ell=\phi_{0}=Z_{0}=r_{F}=16\pi G/\omega_{k,d}=1$,  and $d=3$. In both diagrams there is a Hawking-Page phase transition between a thermal spacetime solution and a black hole solution. The phase of the thermal solution, represented by the blue shaded area,  has a different boundary depending on the value of $\theta$. For $z>2$ the black hole solution dominates everywhere, except for $T=0$ with $\Phi=0$ and $\Phi=\Phi_{c}$.}\label{fig:grand_canonical_phase_diagram}
	\label{fig:grand}
\end{figure}

Let us now consider the $\theta$ dependence of the phase diagram, depicted in Figure \ref{fig:grand_canonical_phase_diagram}. The line along which the first order phase transitions occur in the $(\Phi, T)$ plane can be computed by equating the free energy to zero and eliminating the horizon radius   in favour of the temperature. We find an analytic expression for the transition
\begin{equation}  \label{grandcanonicaltransition}
	 \Phi
	=
	\frac{\Phi_{c}}{T_{c}^{1/z}}
	\sqrt{T^{2/z}_{c}-T^{2/z}}  \, ,
	\end{equation}
where the critical value of the temperature $T_c$ is given by
\begin{equation}	\label{HPtemperature}
    T_c  = \frac{d-\theta}{2 \pi \ell   (2-z) } \left ( k \frac{ (2-z)(d-1)^2   }{z(d-\theta+z-2)^2}  \right)^{z/2}
	\,.
\end{equation}
The endpoints of the transition are precisely given by the critical values defined above: at $T=0$ the transition occurs at $\Phi = \Phi_c$, and  at $\Phi = 0 $ the Hawking-Page temperature is given by $T=T_c$.  That means that for large potentials $\Phi > \Phi_c$ and for  large temperatures $T>T_c$ the charged black hole is always thermodynamically preferred.  At $T=0$ the ensemble is dominated by the thermal hyperscaling violating spacetime for $\Phi < \Phi_c$, and by extremal black holes for $\Phi > \Phi_c$.

  By studying the expression for $T_{c}$, we see that only for $k=1$ and $z\leq 2$   phase transitions are allowed.  Beyond this bound, for $z>2$, the line of first order phase transitions disappears and reduces to a point: $\Phi= \Phi_c$ at $T=0$. At this point and at the origin $\Phi =T=0$ the phase diagram is dominated by the hyperscaling violating spacetime, whereas the black hole dominates everywhere else. Therefore, combining this result   with the restriction $z \ge 1$, coming from the null energy condition, we conclude that the Hawking-Page transition for hyperscaling violating spacetimes only occurs when $1 \le z \le 2$, i.e. the same condition found for pure Lifshitz spacetimes in \cite{Tarrio:2011de}.

  In conclusion, there  are no extra restrictions on $\theta$ coming from the expression for $T_c$, since we already assumed $d> \theta$ and $d - \theta + z-2>0$ when constructing the solutions. Thus, the phase transition  occurs for all  physical values of $\theta$. The phase transition line  in the $(\Phi, T)$ diagram does change location when varying $\theta$, since the expressions for $T_c$ and $\Phi_c$ depend on $\theta$. In Figure \ref{fig:grand} we have summarized the resulting thermodynamic phase structure for the grand canonical ensemble, and plotted the different phase transition lines for different values of $\theta$. We observe  that negative values for $\theta$ lower  the phase transition line, whereas positive values lift the line in the phase diagram.


%
%
%



\subsection{Canonical ensemble (fixed charge)}
\label{sec:canonical}

In the canonical ensemble the charge $Q$ is kept fixed, whereas the potential at infinity $\Phi$ is allowed to vary.
Since the charge takes a fixed value, we compare the black hole solution in this ensemble
with an extremal black hole of finite charge $Q$.
The thermodynamic potential or (Helmholtz) free energy  in the canonical ensemble is   defined as
\begin{equation}
\begin{aligned}
 F & \quad = \quad      \tilde M - T S   \\
 & \quad= \quad  \frac{\omega_{k,d}}{16 \pi G} \ell^{-z-1} r_F^\theta \Big [    - m_{ext} (d-\theta) - z r_h^{d-\theta +z}  + k  \frac{(d-1)^2   (2-z)}{  (d-\theta +z-2)^2} \ell^2 r_h^{d-\theta+z-2}   \\
 &\qquad \quad + (2 d -2\theta+z -2 ) q^2    r_h^{ -(   d -\theta + z-2) } \Big] \, ,
 \end{aligned}
 \end{equation}
 where $m_{ext}$ is given by (\ref{extremalmass}).
 This result holds for all values of $k$, since the extremal black hole is always the correct ground state in the canonical ensemble.
Here $\tilde M$  measures the energy difference between  the black hole and the extremal black hole. In the canonical ensemble these solutions share the   same charge parameter $q=q_{ext}$, given by (\ref{extremalcharge}). Although the free energy is expressed in terms of the horizon radius $r_h$ and the charge parameter $q$, it can also be viewed as a function of the temperature and the charge, since $r_h$    depends  implicitly on $T$ and $q$ through   (\ref{temperature}).


From (numerical) inspection it follows that for planar and hyperbolic black holes the free energy is always negative, and hence the black hole   dominates the entire phase diagram.
 Moreover, for all physical values of $z$ and $\theta$ there is only one black hole solution for a given temperature, so there is no phase transition for these black holes. For spherical black holes, however, the phase structure of the canonical ensemble is non-trivial and is very similar to that of charged AdS black holes,  discovered  in \cite{Chamblin:1999tk,Chamblin:1999hg}. As we will see, the phase diagram of spherical black holes is qualitatively the same for all $\theta$, but it differs for various values of   $z$. We will therefore restrict ourselves  to spherical hyperscaling violating solutions in the rest of this section.

\begin{figure}[t!!]
\vspace{0.4cm}
\begin{center}

\begin{overpic}[width=.45\textwidth]{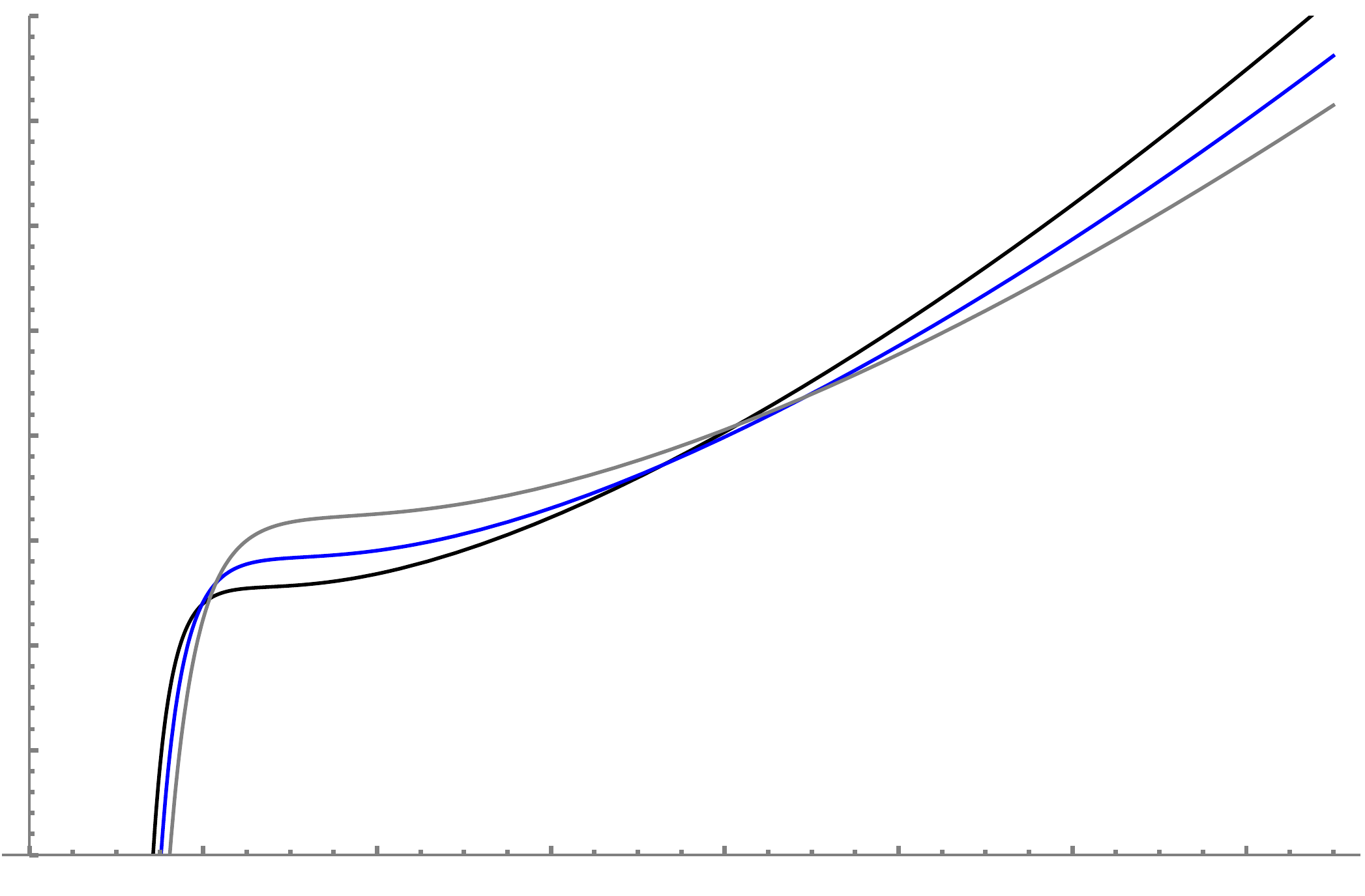}
		\put (-6,2) {\footnotesize{$0.0$}}
		\put (-6,9.5) {\footnotesize{$0.1$}}
		\put (-6,17) {\footnotesize{$0.2$}}
		\put (-6,24.5) {\footnotesize{$0.3$}}
		\put (-6,32.25) {\footnotesize{$0.4$}}
		\put (-6,39.75) {\footnotesize{$0.5$}}
		\put (-6,47.5) {\footnotesize{$0.6$}}
		\put (-6,55.25) {\footnotesize{$0.7$}}
		\put (-6,63) {\footnotesize{$0.8$}}
		
		\put (12.5,-1) {\footnotesize{$0.2$}}
		\put (25,-1) {\footnotesize{$0.4$}}
		\put (37.5,-1) {\footnotesize{$0.6$}}
		\put (50,-1) {\footnotesize{$0.8$}}
		\put (62.75,-1) {\footnotesize{$1.0$}}
		\put (75.75,-1) {\footnotesize{$1.2$}}
		\put (88.5,-1) {\footnotesize{$1.4$}}
		

		\put (18,30) {\footnotesize{$\theta=1/2$}}
		\put (20,19) {\footnotesize{$\theta=-1/2$}}
		\put (0,65.5) {$T$}
		\put (101,3) {$r_{h}$}

		\put (40,63) {$\underline{Q=1.25Q_{crit}}$}		

\end{overpic}
\hspace{0.4cm}
\vspace{0.65cm}
\begin{overpic}[width=.45\textwidth]{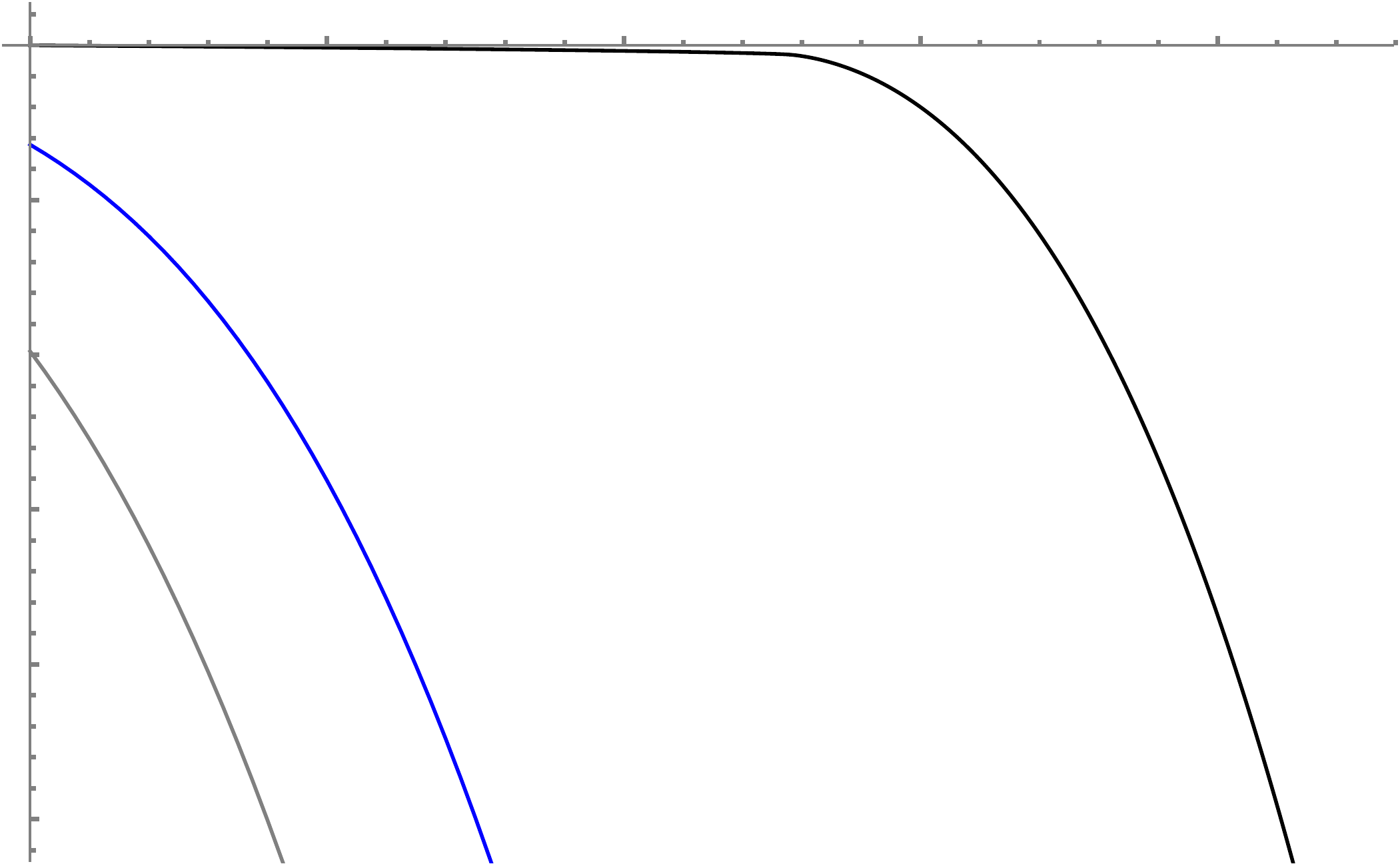}

		\put (-8,13) {\footnotesize{$-0.4$}}
		\put (-8,24) {\footnotesize{$-0.3$}}
		\put (-8,35) {\footnotesize{$-0.2$}}
		\put (-8,46) {\footnotesize{$-0.1$}}
		\put (-4.5,57.5) {\footnotesize{$0.0$}}
		
		\put (0,64) {$F$}
		\put (102,58) {$T$}

		\put (20.75,55.5) {\footnotesize{$0.1$}}
		\put (42.25,55.5) {\footnotesize{$0.2$}}
		\put (63.9,55.5) {\footnotesize{$0.3$}}
		\put (84.5,55.5) {\footnotesize{$0.4$}}

		\put (40,63) {$\underline{Q=1.25Q_{crit}}$}		
		
		\put (17,10) {\footnotesize{$\theta\!=\!1/2$}}
		\put (26,25) {\footnotesize{$\theta=0$}}
		\put (53,45) {\footnotesize{$\theta=-1/2$}}

\end{overpic}
\begin{overpic}[width=.45\textwidth]{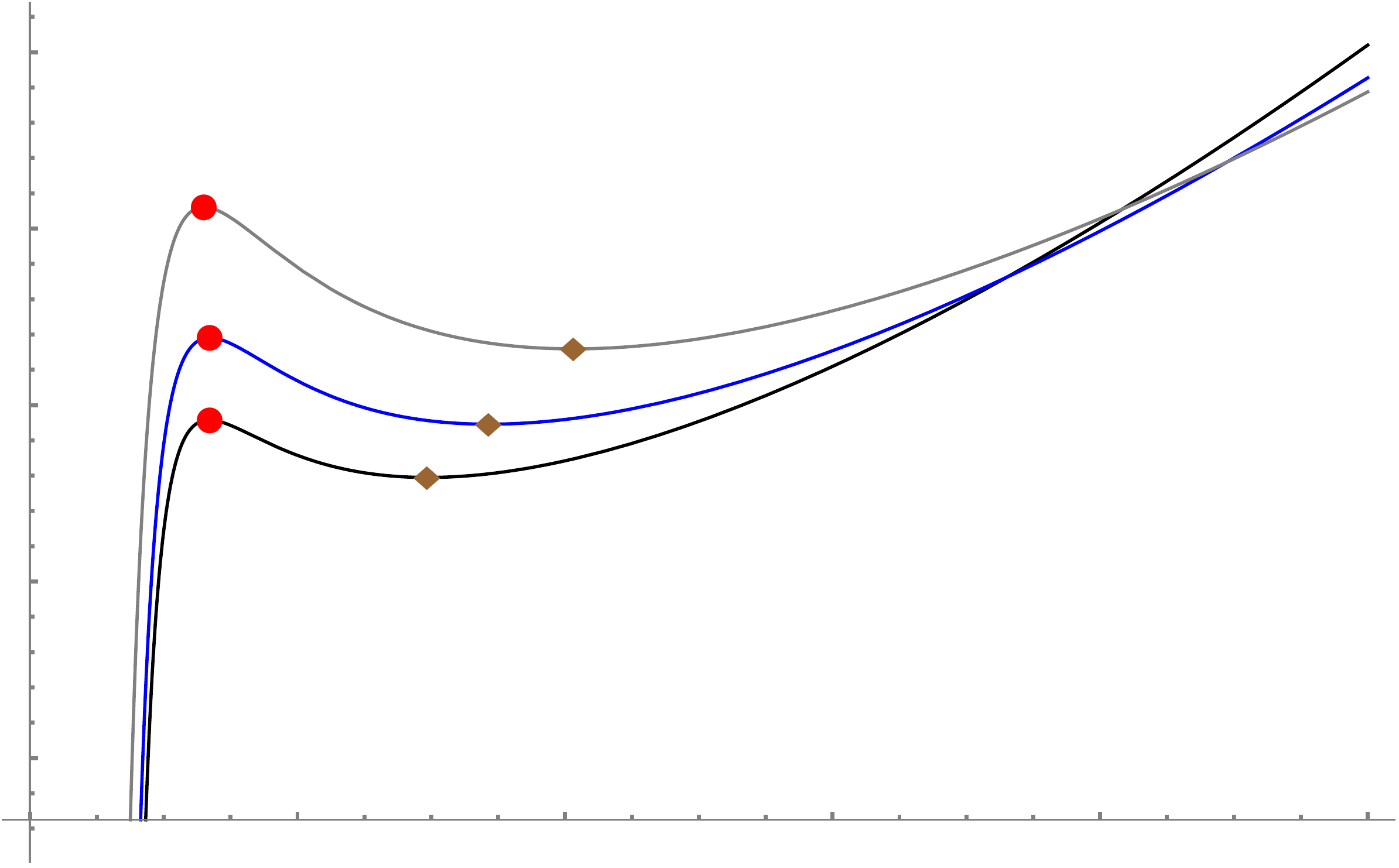}
		\put (-6,7) {\footnotesize{$0.1$}}
		\put (-6,19) {\footnotesize{$0.2$}}
		\put (-6,31.5) {\footnotesize{$0.3$}}
		\put (-6,44) {\footnotesize{$0.4$}}
		\put (-6,56) {\footnotesize{$0.5$}}
		
		\put (18,-2) {\footnotesize{$0.2$}}
		\put (38,-2) {\footnotesize{$0.4$}}
		\put (57,-2) {\footnotesize{$0.6$}}
		\put (76,-2) {\footnotesize{$0.8$}}
		\put (95,-2) {\footnotesize{$1.0$}}

		\put (40,60) {$\underline{Q=0.25Q_{crit}}$}
		

		\put (10,49) {\footnotesize{$\theta=1/2$}}
		\put (12,39) {\footnotesize{$\theta=0$}}
		\put (23,24) {\footnotesize{$\theta=-1/2$}}
		\put (0,64) {$T$}
		\put (101,3.5) {$r_{h}$}
\end{overpic}
\hspace{0.4cm}
\begin{overpic}[width=.45\textwidth]{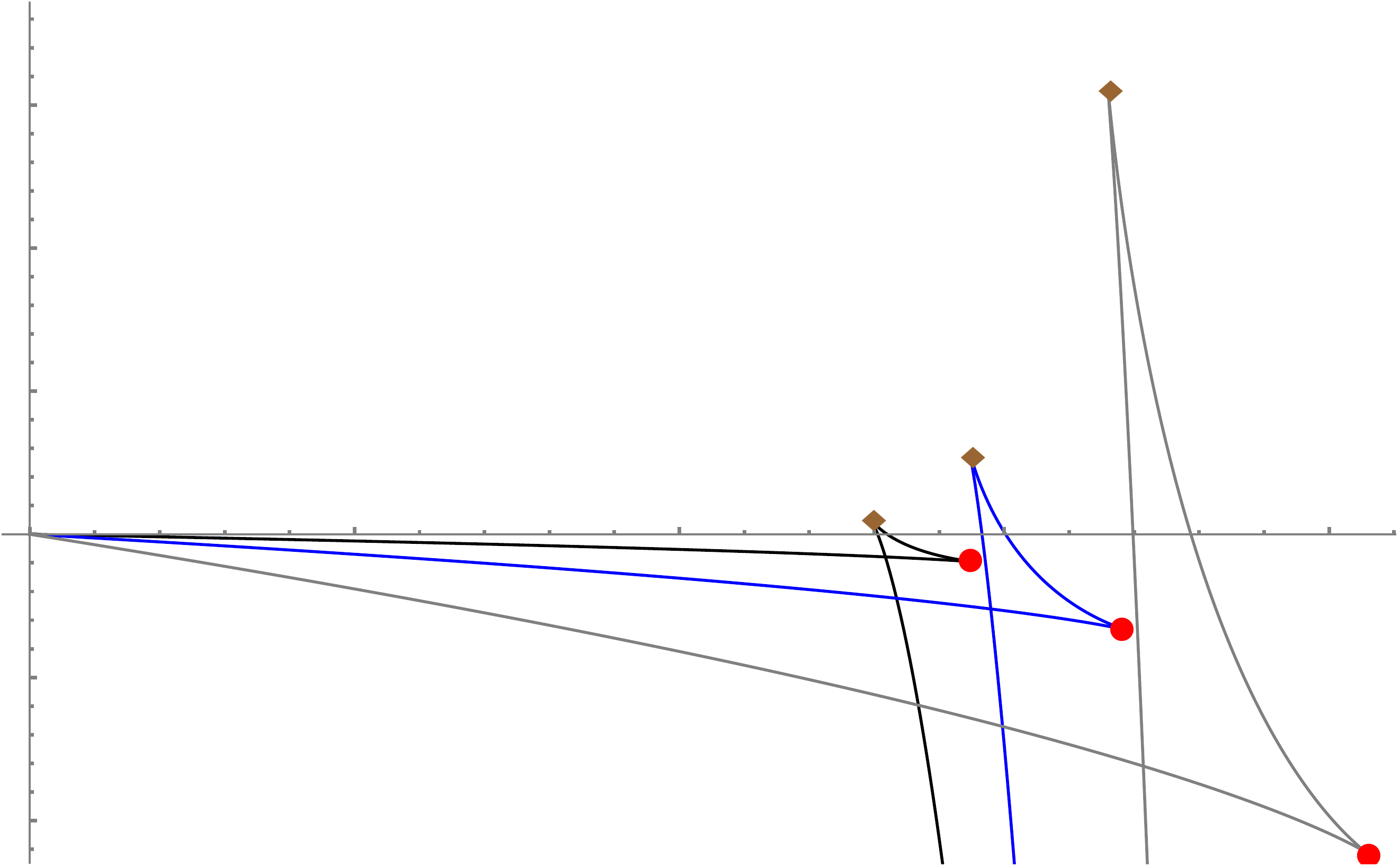}

		\put (-12.5,12) {\footnotesize{$-0.005$}}
		\put (-9,32.5) {\footnotesize{$0.005$}}
		\put (-9,43) {\footnotesize{$0.010$}}
		\put (-9,53.5) {\footnotesize{$0.015$}}
		
		\put (40,60) {$\underline{Q=0.25Q_{crit}}$}
		
		\put (0,64) {$F$}
		\put (102,23) {$T$}
		
		\put (3,20) {\footnotesize{$0.0$}}
		\put (92,20) {\footnotesize{$0.4$}}
		
		
		\put (81,52) {\footnotesize{$\theta=1/2$}}
		\put (65,30) {\footnotesize{$\theta=0$}}
		\put (43,25) {\footnotesize{$\theta=-1/2$}}

\end{overpic}
\end{center}
\vspace{-0.2cm}
\caption{Plots of temperature vs. horizon radius and free energy vs. temperature  for two fixed values of the electric charge. The upper row corresponds to $Q>Q_{crit}$ and the lower row shows $Q<Q_{crit}$. Three different values of $\theta$ are shown in each plot and other paramaters are fixed at:   $k=\ell=\phi_{0}=Z_{0}=r_{F}=16\pi G/\omega_{k,d}=1$, $z=3/2$ and $d=3$.  In the case $Q<Q_{crit}$,   there is a region for which a single value of $T$  can correspond to three values of $r_h$. The three branches of black hole solutions correspond to the region left of the red dot, in between the red dot and the brown diamond and right of the brown diamond. In the lower right panel the same three branches are depicted. A first order phase transition  occurs  where the first and third branch intersect in the free energy plot. In the  case   $Q>Q_{crit}$,   $T$ becomes an injective function of   $r_h$ and the ``swallowtail'' behaviour in the free energy plot disappears, which rules out any phase transition. This case also qualitatively reflects what happens when $z>2$.}\label{fig:canonical_temperature}
\end{figure}
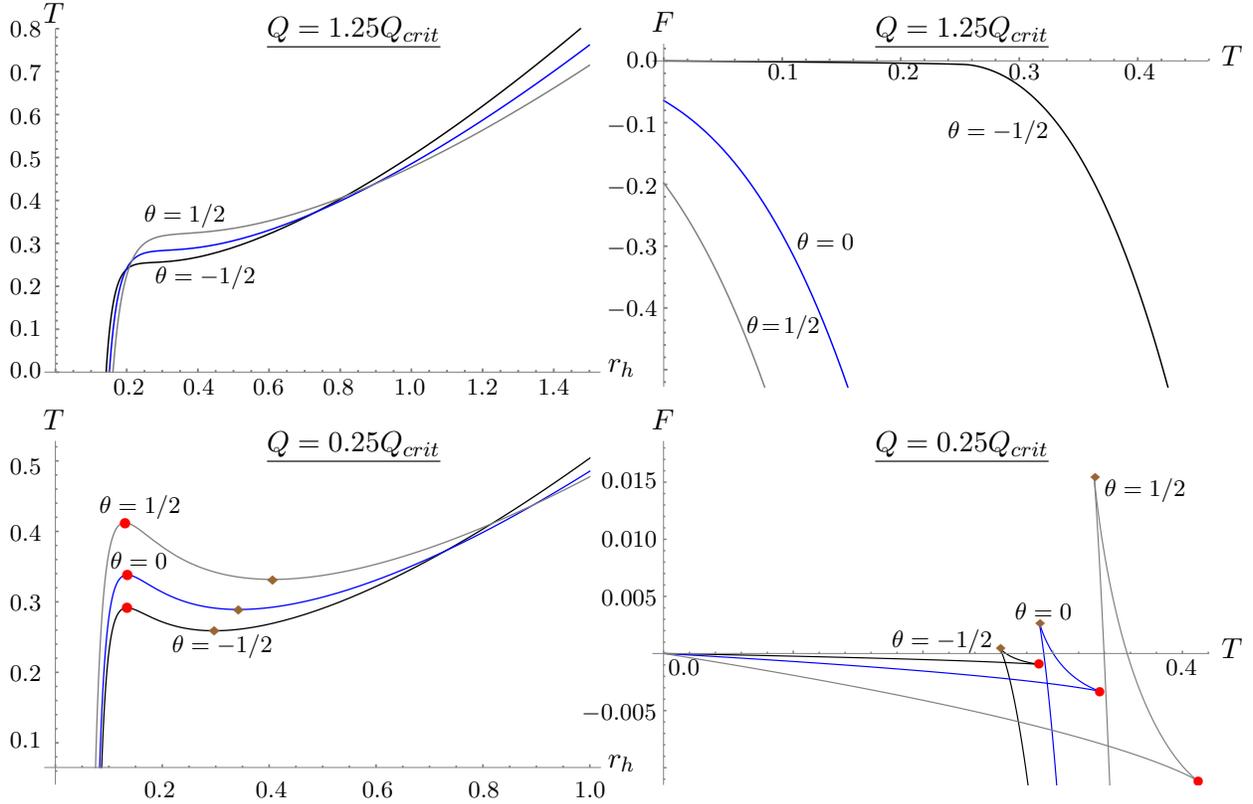

In Figure \ref{fig:canonical_temperature} we plot the temperature as a function of the horizon radius, and the free energy as a function of the temperature, both for two particular values of the charge $Q$. The plots exhibit qualitatively  different behavior for values below and above a   critical value of the charge  $Q_{crit}$ (which we compute in a moment). On the one hand, for large charges $Q>Q_{crit}$ the temperature is an injective function of the horizon radius, and the free energy is strictly negative (similar to the upper plots  in Figure \ref{fig:grand_canonical_free} for a large fixed potential). The charged black hole solutions hence dominate the phase diagram for large electric charges. On the other hand, for small charges $Q<Q_{crit}$ there is a temperature range  for which there exist three branches of black hole solutions (separated by the red dot and brown diamond).  In this case the free energy displays the characteristic ``swallowtail'' behaviour of charges AdS black holes connecting the three different branches   \cite{Chamblin:1999tk,Chamblin:1999hg}.

In the free energy plot for $Q<Q_{crit}$ the first branch starts at the origin and ends at the red dot. The second branch runs between the red dot and the brown diamond, and the third branch forms a cusp with the second branch at the brown diamond and continues downwards for large temperatures.
Precisely when the first and the third branch cross in the $(F,T)$ diagram a first order phase transition occurs between a small and large black hole solution. This phase transition depends on the   value of $Q$, so there actually exists  an entire line of first order phase transitions in the $(Q,T)$ plane. The second branch is never dominant, and hence does not play a role in the phase diagram. As we will see in the next section, the heat capacity at constant charge $C_Q$ turns out to be negative for the second branch, so it is   thermodynamically unstable.
When the electric charge approaches its   critical value $Q_{crit}$, the second branch starts to disappear, and the first and third branch merge. At $Q=Q_{crit}$ there still exists a phase transition between small and large black holes, but it is   second order and hence this point in the phase diagram is a genuine critical point.
 Finally, for $Q>Q_{crit}$ there are no phase transitions anymore.

 The value of the critical charge can   be obtained as follows.
  By analyzing the temperature as a function of the horizon radius, one can  observe that it exhibits  two turning points for $Q<Q_{crit}$   and no turning points for $Q>Q_{crit}$.
  Precisely at $Q=Q_{crit}$ the temperature has an inflection point when plotted against $r_h$, i.e. the following relations hold
 \begin{equation}
 \frac{\partial T}{\partial r_h} = 0 \quad \text{and} \quad \frac{\partial^2 T}{\partial r_h^2}=0   \qquad \text{at} \qquad r_h = r_{crit}\,,\quad q= q_{crit} \, .
 \end{equation}
 Solving these equations yields
 \begin{equation}  \label{critical}
 r_{crit}^2 = k \frac{   (d-1)^2  (2-z)\ell^2 }{z (d-\theta +z-1) (d-\theta
   +z)} \quad \text{and} \, , \quad q_{crit}^2 = \frac{z  (d-\theta +z) r_{ crit}^{2(d-\theta +z-1)}  }{(d-\theta +z-2)^2 (2 d -2\theta+z -2)}\,.
 \end{equation}
For $(\theta =0,z=1)$ this is consistent with the  results  in \cite{Chamblin:1999tk}, and for general values of  $z$ it agrees with the expressions in \cite{Tarrio:2011de}. The temperature at this critical point is 
 \begin{eqnarray} \label{criticaltemp}
 T_{crit} =
   \frac{ (d-\theta +z-1) (d-\theta +z)}{\pi  (2-z) (2 d-2\theta+z -2)} \frac{r_{crit}^z}{\ell^{z+1}}
   \,.
 \end{eqnarray}
From considering these critical quantities, it is clear that for $z>2$ or $k\neq1$ there is no critical temperature. This is consistent with our previous finding that the hyperbolic and planar hyperscaling violating black holes do not exhibit phase transitions in the canonical ensemble. Moreover, we note that the critical point exists for all physical values of $\theta$ that are consistent with the null energy condition. The positivity of the critical temperature namely follows directly from the restrictions presented in Table \ref{table1}. Thus, we find that, although the particular value of the critical point does depend on $\theta$, the phase structure in the canonical ensemble is qualitatively the same for all $\theta$.

Notice that for $z=2$ the critical horizon radius and     charge vanish, whereas the critical temperature stays finite. Since any charge is greater than $q_{crit}=0$, there are no phase transitions  in this case for finite charges. At $q=0$ we are left with the   Hawking-Page phase transition, discussed in the previous section, which   occurs at the temperature $T=T_c$.
A nice consistency check shows that the critical temperature (\ref{criticaltemp}) and the Hawking-Page   temperature (\ref{HPtemperature}) coincide for $z=2$:
\begin{equation}
T_{crit} = T_c =  \frac{ k(d-1)^2  }{4 \pi \ell  ( d-\theta)}   \qquad \text{for} \qquad z=2 \, .
\end{equation}
We emphasize that this  is a standard first order  phase transition between thermal spacetime and the black hole solution, so it is not   a critical point.


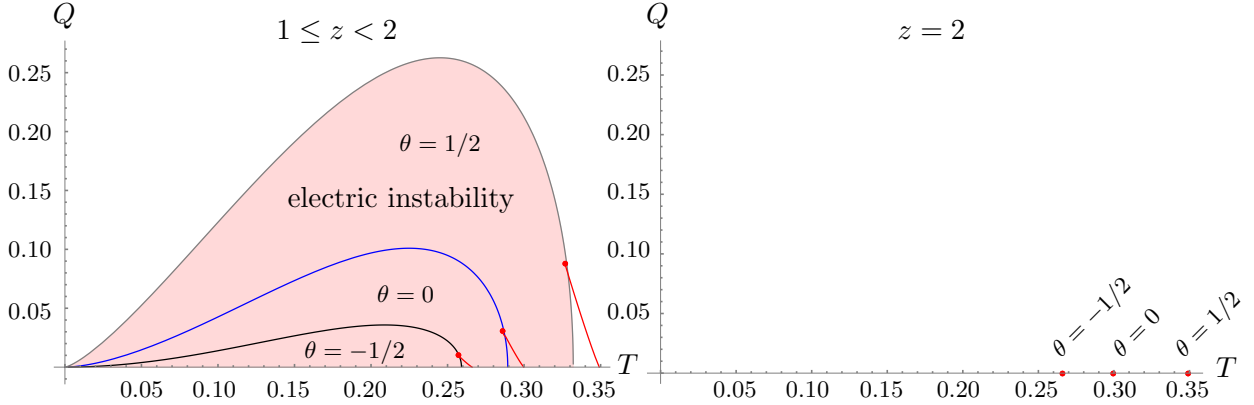
\begin{figure}[t!!]
\vspace{0.4cm}
\begin{center}
\begin{overpic}[width=.45\textwidth]{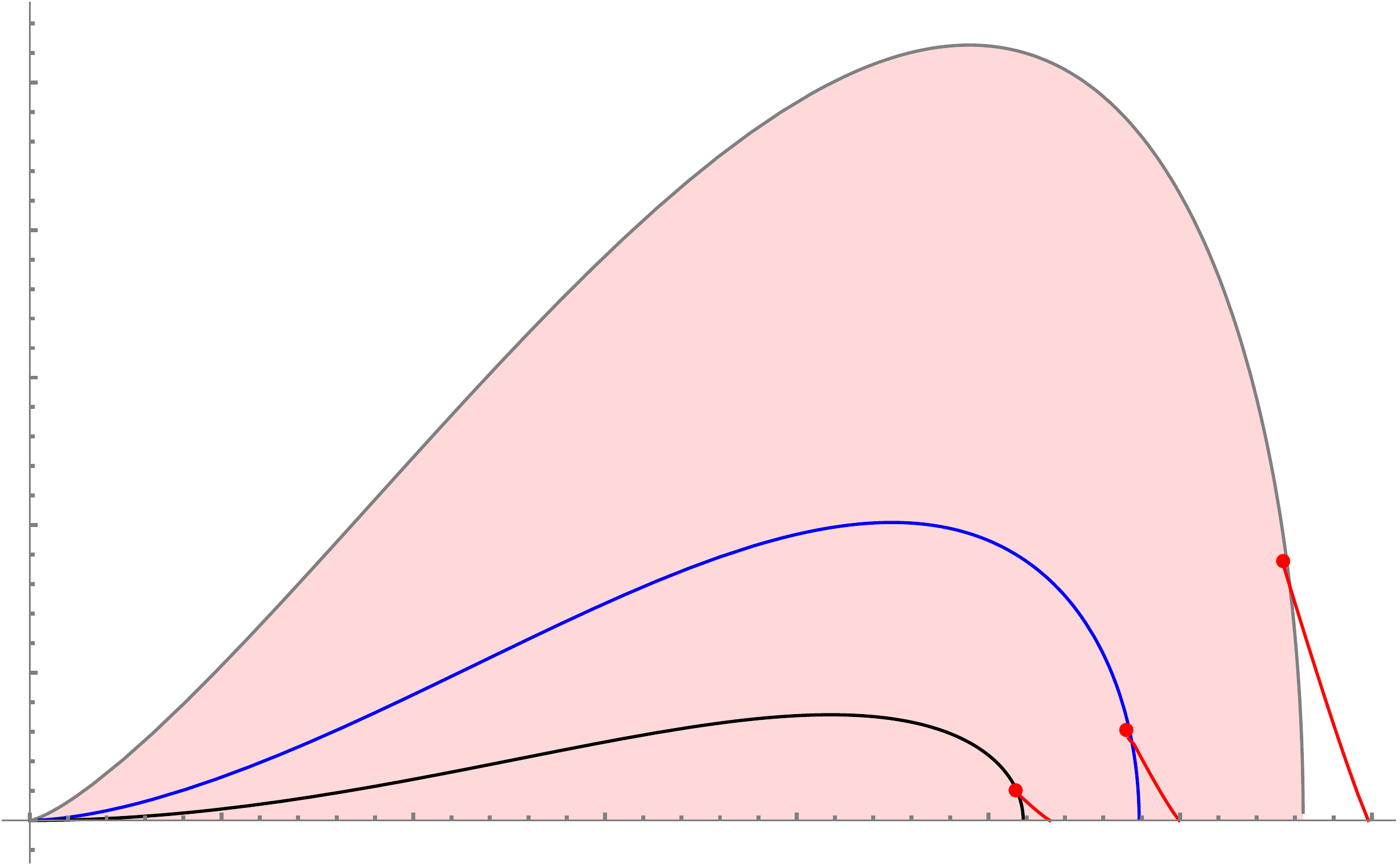}
		\put (-8,12) {\footnotesize{$0.05$}}
		\put (-8,23) {\footnotesize{$0.10$}}
		\put (-8,33.5) {\footnotesize{$0.15$}}
		\put (-8,44) {\footnotesize{$0.20$}}
		\put (-8,54.5) {\footnotesize{$0.25$}}
		
		\put (12,-2) {\footnotesize{$0.05$}}
		\put (25,-2) {\footnotesize{$0.10$}}
		\put (38.25,-2) {\footnotesize{$0.15$}}
		\put (52,-2) {\footnotesize{$0.20$}}
		\put (66,-2) {\footnotesize{$0.25$}}
		\put (80,-2) {\footnotesize{$0.30$}}
		\put (93,-2) {\footnotesize{$0.35$}}
		
		\put (42,32) {electric instability}

		\put (62,42) {\footnotesize{$\theta=1/2$}}
		\put (58,15) {\footnotesize{$\theta=0$}}
		\put (45,5) {\footnotesize{$\theta=-1/2$}}
		\put (0,65) {$Q$}
		\put (101,2) {$T$}
		\put (40,62) {$1\leq z<2$}
\end{overpic}
\hspace{0.2cm}
\begin{overpic}[width=.45\textwidth]{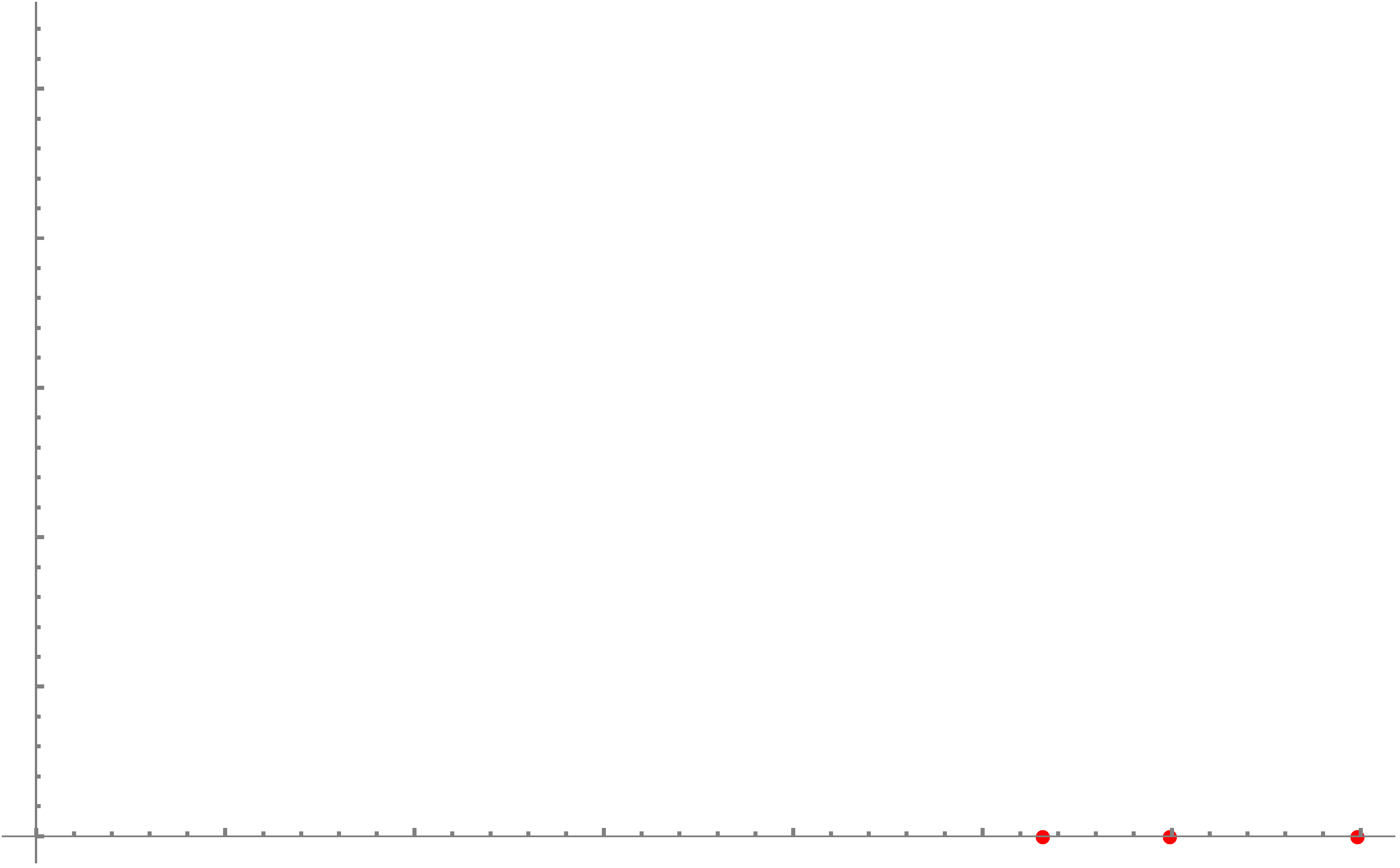}
		\put (-7,12) {\footnotesize{$0.05$}}
		\put (-7,23) {\footnotesize{$0.10$}}
		\put (-7,33.5) {\footnotesize{$0.15$}}
		\put (-7,44) {\footnotesize{$0.20$}}
		\put (-7,54.5) {\footnotesize{$0.25$}}
		
		\put (12,-2) {\footnotesize{$0.05$}}
		\put (25,-2) {\footnotesize{$0.10$}}
		\put (38.25,-2) {\footnotesize{$0.15$}}
		\put (52,-2) {\footnotesize{$0.20$}}
		\put (66,-2) {\footnotesize{$0.25$}}
		\put (80,-2) {\footnotesize{$0.30$}}
		\put (93,-2) {\footnotesize{$0.35$}}

		\put (0,65) {$Q$}
		\put (45,62) {$z=2$}
		\put (102,1.5) {$T$}
		\put (72,5) {\rotatebox{45}{\footnotesize{$\theta=-1/2$}}}
		\put (83,5) {\rotatebox{45}{\footnotesize{$\theta=0$}}}
		\put (94,5) {\rotatebox{45}{\footnotesize{$\theta=1/2$}}}
	\end{overpic}
\end{center}
\vspace{-0.2cm}
\caption{Two phase diagrams in the canonical ensemble for $z=3/2$ and $z=2$, plotted for three different values of~$\theta$. Other parameters are fixed at: $k=\ell=\phi_{0}=Z_{0}=r_{F}=16\pi G/\omega_{k,d}=1$,  and $d=3$.  In the left panel the red lines represent first order phase transitions between small and large black holes. The red dots correspond  to critical points at which a second order phase transition occurs. In the right panel there are no such critical points, but the first order phase transitions at the red dots remains (below the red dots at $Q=0$ the thermal gas dominates). The red shaded region in the left panel, of which the border depends on the value of $\theta$, is the region for which the system is unstable under electric perturbations. This electrically  unstable region is absent for $z \ge 2$. For $z>2$ the  black hole dominates the phase diagram everywhere, except at $T=Q=0$. }
\label{fig:canonical_phase_diagram}
\end{figure}

In Figure \ref{fig:canonical_phase_diagram} we show examples of the $(Q,T)$ diagram in cases where the phase structure is non-trivial. For $1 \le z < 2$ the red line indicates the line of first order phase transitions: it starts at the Hawking-Page transition $T_c$ and   ends at the critical point $T_{crit}$ (the red dot). The plot shows especially   how the line of  phase transitions depends on the hyperscaling violating parameter. We observe, for example, that the critical temperature and charge decrease with decreasing values of~$\theta$. Moreover, the phase transitions crucially depend on the dynamical exponent: the critical point only exists for $1 \le z <2$; for $z=2$   the Hawking-Page phase transition remains at $Q=0$;  and for $z>2$   the black hole spacetime dominates the entire phase diagram  except at the origin. This agrees perfectly with the results for pure Lifshitz black holes in \cite{Tarrio:2011de}.

In conclusion, the phase structure of charged hyperscaling violating black holes in the canonical ensemble is qualitatively the same as the phase structure of charged AdS black holes.  In \cite{Chamblin:1999tk,Chamblin:1999hg} it was pointed out that the   phase diagram for charged AdS black holes is strikingly similar to the  phase diagram of the Van der Waals liquid-gas system. In the latter system there is also a line of first order phase transitions between the liquid and the gas, ending at  a critical point above which a liquid can be continuously converted into a gas.   In Section \ref{criticalexp} we will explore the similarity between charged black holes and the liquid-gas system further, and compute the critical exponents for  hyperscaling violating black holes.






\subsection{Thermodynamic and electric stability}
\label{secstability}
We have discussed the phase diagrams of the grand canonical and canonical ensemble, by studying the respective free energies. In this section, we take into account the following   stability conditions
\begin{equation}\label{heatquant}
	C_{\Phi}\equiv T\left( \frac{\partial S}{\partial T}\right)_{\Phi}\geq 0\,,
	\qquad
	C_{Q}\equiv T\left( \frac{\partial S}{\partial T}\right)_{Q}\geq 0\,,
	\qquad
	\chi_{T}\equiv  \left( \frac{\partial Q}{\partial \Phi}\right)_{T}\geq 0\,.
\end{equation}
The first two quantities are heat capacities at constant potential and constant charge, respectively. When they are greater than zero, it implies that increasing the temperature of the black hole increases its entropy.
The last quantity is called the isothermal susceptibility. The positivity of this quantity implies that increasing the electric potential increases   the charge. These requirements are equivalent to   thermodynamic or electric (in)stability, respectively, when the inequalities are (not) satisfied  \cite{Chamblin:1999hg}.

We first consider the heat capacities. Using thermodynamic identities we can recast the heat capacity conditions into the following form
\begin{equation}
	C_{\Phi}
	= T
	\left(
		\frac{\partial^{2}W}{\partial T^{2}}
	\right)_{\Phi}
	\leq
	0
	\qquad
	\text{and}
	\qquad
	C_{Q}
	= T
	\left(
		\frac{\partial^{2}F}{\partial T^{2}}
	\right)_{Q}
	\leq 0
	\,.
\end{equation}
A branch in the plots of the free energies versus the temperature is thus thermodynamically (un)stable if it is concave downwards (upwards). From inspection of the lower right panel of Figure \ref{fig:grand_canonical_free} it is clear that the lower branch satisfies the first condition above, while the upper branch violates the condition and is thus   unstable. These branches correspond to large and small black holes, respectively. Turning to the lower right panel of Figure~\ref{fig:canonical_temperature} we conclude that the two branches that intersect (the first and third branch)   satisfy the second condition above. These two branches correspond, respectively, to small and large black holes. The second branch, on the other hand, corresponding to middle sized black holes, violates the condition and is thus unstable.

Another feature of the heat capacities is that they are smooth functions, except  at  the cusps in   the plots  of the free energies versus the temperature where they blow up. These cusps can be clearly distinguished in the lower right panels  of Figures \ref{fig:grand_canonical_free} and  \ref{fig:canonical_temperature}. A special case   is when the two cusps in the ($F,T$) diagram coincide, which corresponds to the critical point. Thus, we conclude that the heat capacity $C_Q$ should diverge at the critical  point.
These conclusions do not put extra constraints, though, on the allowed values of the   parameters $z$ and $\theta$, but should rather be consistent with our previous findings. As a check we will compute below  all the quantities defined in (\ref{heatquant}).\footnote{In order to compute the heat capacities and isothermal susceptibility, we use the following expressions\begin{equation}
	C_{\Phi}=\frac{\left ( \partial \left ( M - \Phi Q \right) / \partial r_h \right)_\Phi}{ \left ( \partial T/ \partial r_h \right)_\Phi}   \,,
\qquad
C_{Q}=\frac{ \left ( \partial M / \partial r_h \right)_{Q}}{\left ( \partial T / \partial r_h \right)_{Q}}\,,
	\qquad
	\chi_{T}=- \left (  \frac{\partial Q}{\partial T} \right)_{\!\Phi} \!\! \left (  \frac{\partial T}{\partial \Phi}\right)_{Q}\,.\nonumber
\end{equation}
The last equation follows from Maxwell's relation, and can be computed   using the temperature $T(\Phi, Q)$   in (\ref{eqntemp}).}

First, we compute the heat capacity at fixed potential
\begin{equation}
C_\Phi  = \frac{\pi \omega_{k,d}T }{G}    \frac{ (d-\theta)r_F^{\theta } r_h^{d-\theta-z} \ell^{z+1} }{        z (d-\theta +z)     -      k \frac{(d-1)^2(2-z)}{(d-\theta +z-2)} \frac{\ell^2}{r_h^2} + (2-z)(d - \theta +z-2)  q^2  r_h^{-2 (d- \theta + z-1)}  }
\,.
\end{equation}
This quantity is positive for large black holes, corresponding to the lower branch in the lower right plot of Figure \ref{fig:grand_canonical_free}, and it diverges at the cusp (marked by a red diamond) in the same plot. The cusp corresponds to the minimal value  of the temperature as a function of the horizon radius in the fixed potential ensemble.
%

Second, we study the heat capacity at constant charge
\begin{equation}  \label{heatcapQ}
C_Q =  \frac{\pi \omega_{k,d}T }{G}   \frac{   (d-\theta)r_F^{\theta } r_h^{d-\theta-z} \ell^{z+1}}{ z (d-\theta +z) -k \frac{ (d-1)^2
   (2-z)}{   d-\theta +z-2} \frac{\ell^2}{r_h^2} + (d-\theta +z-2)  (2 d-2   \theta +z-2)  q^2 r_h^{-2( d- \theta+ z-1)} }
   \,.
\end{equation}
As discussed above, this quantity is always positive for the first and third branch in the lower right plot of Figure  \ref{fig:canonical_temperature}. However, for $1\le z <2$ in the spherical case, there exists a value for which the heat capacity diverges. This coincides, as it should, with the critical values   $r_h=r_{crit}$ and $q=q_{crit}$ given in (\ref{critical}).

Finally, we turn to the isothermal susceptibility. It is known that for purely Lifshitz systems with $\theta=0$ an electric instability is present \cite{Tarrio:2011de}. In order to find out how this generalizes to arbitrary~$\theta$, we use the   explicit expression for the isothermal susceptibility
\begin{equation}\label{chiT}
\chi_T =  \frac{Q}{\Phi} \frac{z (d-\theta +z)  -    k  \frac{(d-1)^2 (2- z)}{(d-\theta+z-2)} \frac{\ell^2 }{r_h^2} + (d-\theta + z-2) (2 d-2 \theta +z-2) q^2 r_h^{-2(d- \theta + z-1)}   }{ z(d-\theta +z)  -  k \frac{(d-1)^2 (2-z)}{(d-\theta+z-2)} \frac{ \ell^2  }{r_h^2} + (2-z) (d-\theta + z-2) q^2 r_h^{-2(d- \theta + z-1)}   }
\,.
\end{equation}
We notice that this expression     vanishes at the critical point and diverges at the same point as where $C_\Phi$ blows up. This indicates that the `compressibilities' are not independent from each other.
When the isothermal susceptibility is negative, the system has an electric instability.  It was observed though in \cite{Chamblin:1999hg} that the border between the unstable and stable regions is actually given by the value $\chi = \infty$. Setting the denominator to zero in the expression above yields
\begin{equation}  \label{qinst}
	q_{inst}^{2}
	=
	\frac{r_{inst}^{2(d-\theta+z-1)}}{d-\theta+z-2}
	\left[
		\frac{k (d-1)^{2}}{d-\theta+z-2}\frac{\ell^{2}}{r_{inst}^2}
		-
		\frac{z}{2-z}(d-\theta+z)
	\right]
	\,.
\end{equation}
Inserting this into equation \eqref{temperature} for the temperature we find
\begin{equation} \label{tempelinst}
	T_{inst}
	=
	\frac{r_{inst}^{z}}{2\pi \ell^{z+1}}\frac{d-\theta+z}{2-z}
	\,.
\end{equation}
From these expressions we conclude that only for $k=1$ and $z< 2$ there are electric instabilities.
 The line $q_{inst}(T)$ splitting the phase diagram into electrically stable and unstable regions is found by combining (\ref{qinst}) with (\ref{tempelinst}) and is shown in Figure \ref{fig:canonical_phase_diagram} for three different values of $\theta$. The critical point always lies inside the   unstable region (colored in red), while the Hawking-Page transition (at $Q=0$ and $T=T_c$) lies  in the stable region.


\subsection{Critical exponents}
\label{criticalexp}

  \begin{table}[t!!]
        \centering
\begin{tabular}{|c||c|c|c |c|c|  }
 \hline
Van der Waals fluid& analogy 1 & analogy 2 & analogy 3 \\
 \hline
 temperature   & $Q$    &   $\beta$&   $T$\\
 pressure  &   $\beta$  & $Q$  & $P$\\
 volume  &   $r_h$  & $\Phi$  & $V$\\
 \hline
\end{tabular}
         \caption{Three analogies between the thermodynamic variables of the Van der Waals fluid and those of charged AdS black holes \cite{Kubiznak:2012wp}.
         }
    \label{tableanalogies}
   \end{table}

In  Section \ref{sec:canonical} we concluded that for certain parts of the parameter space, the hyperscaling violating black holes behave analogously to a Van der Waals liquid-gas system. The precise analogy, however, depends on the identification of the physical quantities associated to the black hole with the thermodynamic variables describing the liquid-gas system. In \cite{Kubiznak:2012wp} it was pointed out that there exist   three different options for this identification, and hence three different analogies between the Van der Waals   fluid and charged black holes. The three analogies are  shown in Table \ref{tableanalogies}. The first two analogies, which were already suggested in \cite{Chamblin:1999tk,Chamblin:1999hg}, involve standard thermodynamic quantities of the black hole, but they identify the ``wrong'' quantities with each other (e.g.    pressure with inverse temperature $\beta$ or   electric charge  $Q$). The   third analogy, which was   studied for the first time in \cite{Dolan:2011xt},  is based on   the extended version of black hole thermodynamics, where the cosmological constant is interpreted as a pressure, and it compares  the ``right'' physical quantities in the black hole system and the liquid-gas system. Although  the third analogy seems more physical, the first two analogies are nevertheless interesting in their own right.

In the current section we will investigate, through one particular analogy, the values of the critical exponents for hyperscaling violating black holes. Critical exponents describe the physical behaviour near critical points and are believed to be universal, in the sense that they do not depend on the microscopic details of the system. For charged AdS black holes the critical exponents were computed for the second analogy in \cite{Niu:2011tb} and for the third analogy in \cite{Kubiznak:2012wp}. It turns out that they exactly   coincide with the critical exponents of the Van der Waals system, suggesting that charged AdS black holes and the Van der Waals fluid belong to the same universality class.

For charged Lifshitz    black holes, however, it is rather   difficult to analytically compute the critical exponents  for the second and third analogy. This is because one needs an analytic expression for the equation of state to do so, e.g. $Q(\beta, \Phi)$ in the second analogy, and it is impossible to   invert the formulas (\ref{temperature}) and (\ref{eqntemp})  for the temperature for generic values of $z$.
  One could study the critical exponents numerically for the second and third analogy, but instead here we  focus on the first analogy, for which it is possible to find the equation of state.  In this section we closely follow the work and notation of \cite{Kubiznak:2012wp,Gunasekaran:2012dq}, and we refer to these papers for further details on the Van der Waals fluid and its critical exponents.

In order to compute the critical exponents we first need the equation of state $P(V,T)$.  In the first analogy in Table \ref{tableanalogies} the equation of state takes the   form
\begin{equation}
\beta (r_h, q) = \frac{4 \pi \ell^{z+1}}{r_h^{z}}   \left [(d-\theta +z)
 +k \frac{(d-1)^2  }{  (d-\theta +z-2)}  \frac{\ell^2}{r_h^2}
 -  \frac{ (d-\theta + z -2)  q^2 }{  r_h^{2 ( d- \theta + z-1)} }
   \right]^{-1}.
\end{equation}
Notice that the inverse temperature $\beta$ corresponds to pressure in this analogy with the Van der Waals system, and not to temperature. In this analogy the charge parameter $q$  is related to the temperature of the Van der Waals fluid and the horizon  radius $r_h$ is equivalent to the volume.  In order to obtain critical exponents we need to expand the equation of state around the critical point. We will do this using the following suitable choice of parameters
\bqn
p = \frac{\beta}{\beta_{crit}} \,, \qquad \omega = \frac{r_h}{r_{crit}} -1 \, , \qquad t = \frac{q}{q_{crit}} -1
\,,
\eqn
where the subscript $crit$ denotes the fact that we expand around the critical points $t=0$ and $\omega=0$. When we expand $p$ around the critical points, we obtain
\bqn
p = 1 + A t- B t \omega - C \omega^3 + O(t \omega^2, \omega^4) \, ,
\eqn
with
\bqn
A &=&\frac{z(2-z) }{2 (d-\theta +z-2) (d-\theta +z-1)}\,, \qquad B =\frac{z(2-z)  (2 d-2 (\theta +1)+z)}{2 (d-\theta +z-2) (d-\theta +z-1)} \,,\\
 C &=& \frac{1}{6} z(2-z)  (2 d-2 (\theta +1)+z)\,.   \nonumber
\eqn
Note that   the coefficients defined in this way are only positive for $z<2$. Interestingly, this seems to also hold for higher order coefficients, defined with a minus sign in front, since they are all proportional to $z(2-z)$. For $z=2$ the expansion turns out to be trivial and we do not have critical exponents, since there is no critical point in this case.

The four critical exponents, $\alpha, \beta, \gamma$ and $\delta$, which bear our interest, are defined as   \cite{Kubiznak:2012wp,Gunasekaran:2012dq}
\begin{equation}
	C_{r_{h}}\sim |t|^{-\alpha}
	\,,
	\quad
	\eta = r_{crit} (\omega_l - \omega_s)
	\sim
	|t|^{\beta}
	\,,
	\quad
		\kappa_Q = - \frac{1}{r_h} \left ( \frac{\partial r_h}{\partial \beta} \right)_{Q}
	\sim
	|t|^{-\gamma}
	\,,
	\quad
	p-1
	\overset{t=0}{\sim}
	|\omega|^{\delta}
	\,.
\end{equation}
Here $C_{r_{h}}$ is the analog of the specific heat at constant volume, $\eta$ is the order parameter on an isotherm (describing the difference $\omega_l - \omega_s$ between the ``volume'' of large and small black holes), $\kappa_{Q}$ is the equivalent of isothermal compressibility, and the exponent $\delta$ is a property of the critical isotherm $t=0$. The exponent $\beta$, introduced above, should not be confused with the inverse temperature.


First, we compute the critical exponent $\alpha$. It is easy to see that
\bqn
C_{r_h} = T \left ( \frac{\partial S}{\partial T} \right )_{r_h} =  0\,, \qquad \text{so} \qquad \alpha=0
\,,
\eqn
since the entropy does not vary with the temperature if the horizon radius $r_{h}$ is held fixed. Next we turn to exponent $\beta$. For a fixed $t<0$ we obtain
\bqn
dp = - (B t + 3 C \omega^2) d \omega
\,.
\eqn
We need the above expression in combination with   Maxwell's equal area law  to establish
\bqn
p &=& 1 + A t - B t \omega_l - C \omega_l^3 = 1 + A t - B t \omega_s - C \omega_s^3\,, \\
0&=& \int_{\omega_l}^{\omega_s} \omega dp
\,.
\eqn
We can now read off the   value for the exponent $\beta$
 \bqn
 \omega_s = - \omega_l = \sqrt{\frac{- B t}{ C}}
 \quad
 \Rightarrow
 \quad
  \eta = 2 r_{crit} \omega_l \sim \sqrt{-t}\,, \qquad \text{so} \qquad \beta = 1/2\,.
 \eqn
  Furthermore, we can obtain the critical exponent $\gamma$ by computing
 \bqn
 \kappa_Q = - \frac{1}{r_h} \left ( \frac{\partial r_h}{\partial \beta} \right)_{Q} \sim \frac{T_{crit}}{  B \, t}\,, \qquad \text{so} \qquad \gamma=1\,.
 \eqn
Finally, at the critical isotherm $t=0$, the defining equation for exponent $\delta$ takes the form
\bqn
p-1 \overset{t=0}{=} - C \omega^3\,, \qquad \text{so} \qquad \delta = 3\,.
\eqn
To summarize, in the first analogy the critical exponents for hyperscaling violating black holes  are
\bqn
\alpha = 0\,, \quad \beta = 1/2\,, \quad  \gamma = 1\,, \quad \delta =3\,.
\eqn
These critical exponents are precisely the same as those of the Van der Waals fluid. They do not depend on the dynamical exponent $z$ or the hyperscaling exponent $\theta$, so the physical behaviour of charged black holes near the critical point  is really universal.  This universality is   to be expected from a mean field theory point of view, since critical exponents  typically do not depend on   the details of the microscopic model. All   black hole solutions under consideration in this paper are solutions to Einstein gravity (plus matter fields), and this  macroscopic model seems to universally fix the value of the critical exponents.      For higher curvature corrections to Einstein gravity one might find different critical exponents (see e.g. \cite{Dolan:2014vba}). In conclusion,  the first analogy in the canonical ensemble between charged hyperscaling violating black holes and the Van der Waals system is quite  robust, since the critical exponents coincide for all $d$, $\theta$ and $1\leq z <2$.


%

\section{Extended thermodynamics}
\label{secextendedthermo}

In the previous section we observed that for   $1\leq z<2$  hyperscaling black hole solutions in the canonical ensemble undergo a phase transition similar to that of a Van der Waals fluid. This behavior is reminiscent of the more standard Reissner-Nordstr\"om-AdS black hole, which was observed for the first time in the seminal works \cite{Chamblin:1999tk}. However, contrary to the Hawking-Page transition (present in the grand canonical ensemble), the field theoretical interpretation of the  phase transition in the canonical ensemble is less understood.
Moreover, the comparison between    this phase transition and the liquid-gas phase transition seems to be rather unphysical, since it identifies intensive   with extensive thermodynamic quantities. For example, in the first analogy in Table \ref{tableanalogies}  the temperature of the fluid, which is an intensive quantity, is identified with the electric charge  of the black hole, which is an extensive quantity.

The analogy between Reissner-Nordstr\"om-AdS  black holes and  the Van der Waals system was reconsidered in the work of \cite{Dolan:2011xt,Kubiznak:2012wp}, where they used the framework of extended black hole thermodynamics (also known as black hole chemistry)  \cite{Kastor:2009wy,Dolan:2010ha,Kubiznak:2014zwa,Johnson:2014yja,Caceres:2015vsa,Karch:2015rpa,Armas:2015qsv}.\footnote{For a more comprehensive review of black hole chemistry  see \cite{Kubiznak:2016qmn}.} The advantage of this   analogy (see last column in  Table \ref{tableanalogies}) is that   the same physical quantities are identified with each other, e.g. the pressure of the Van der Waals fluid is compared with the pressure of the black hole system. Moreover, the authors of \cite{Kubiznak:2012wp} showed   there exists a critical point in the extended phase diagram of the canonical ensemble, and   the critical exponents associated to this point  coincide with those of the Van der Waals system. 
In the present section we will perform a similar analysis for the hyperscaling violating black holes discussed in this paper. At the end of this section we show that also for the third analogy in Table \ref{tableanalogies} the critical exponents are the same as for the van der Waals fluid. The beginning of this section    is devoted to  the more general  study of  extended thermodynamics, and to finding the thermodynamic volume and pressure  for our   black hole solutions.






\subsection{Quick review of thermodynamics with $\Lambda$}

In the framework of extended thermodynamics of AdS black holes, the cosmological constant is identified as a pressure\footnote{This pressure $P$ should not be confused with the pressure of the dual CFT. The latter one can be computed from the expectation value of the stress-energy tensor $\langle T^i\;\!\!_i\rangle$ and does not coincide with (\ref{pressure}).}
\bqn\label{pressure}
P=-\frac{\Lambda}{8\pi G} \qquad \text{with} \qquad \Lambda=-\frac{d(d+1)}{2\ell^2}\,,
\eqn
where $\ell$ is the AdS radius. The identification of the cosmological constant with the pressure of the black hole system
is very natural because the (bulk) stress-energy tensor associated to $\Lambda$ can be written precisely in the form of a perfect fluid with pressure $P$ given by (\ref{pressure}) and energy density by $\rho = - P$. The conjugate variable associated to $P$ is dubbed   the thermodynamic volume $V$ and enters into (an extended version of) the first law of black hole thermodynamics as
\bqn \label{extendedfirstlaw}
dM=TdS+ \Phi dQ +VdP\,.
\eqn
This form of the first law implies that the ADM mass has to be identified with the enthalpy of the system, i.e.  $M\equiv H=E+PV$.
From the first law one can derive the generalized Smarr relation using a scaling argument \cite{Kastor:2009wy}. For charged AdS black holes the Smarr formula is found to be
\bqn\label{Smarr}
(d-1)M=d  \, TS+(d-1) \Phi Q-2PV\,.
\eqn
Remarkably, for asymptotically AdS black holes the cosmological constant or $P V$ term \emph{has} to be included for the Smarr formula to hold.




Let us discuss  in more detail the physical interpretation of the thermodynamic volume $V$ and the pressure $P$. In simple cases
such as the Reissner-Nordstr\"om-AdS solution, the thermodynamical volume coincides with the naive volume of the black hole interior (in the flat space limit)
\bqn \label{AdSvolume}
V\equiv \left(\frac{\partial M}{\partial P}\right)\bigg|_{S,Q} = \frac{\omega_{k,d}r_h^{d+1}}{d+1}\,.
\eqn
However, such a statement needs to be taken with some caution, since the volume of a black
hole is not a coordinate independent quantity and, in particular, depends on the slicing of the spacetime. 
 In cases with a scalar potential, the thermodynamic volume is  given instead by the integral of the scalar potential \cite{Cvetic:2010jb}. For more general bulk fields, it is given by an integral of a certain combination of matter potentials. This was proven rigourously in \cite{Caceres:2016xjz,Couch:2016exn} by implementing a version of the Iyer-Wald formalism that allows variations of the cosmological constant.

Regarding the pressure $P$, it was noticed in \cite{Armas:2015qsv} that there is a certain degree of ambiguity in defining a thermodynamic variable that encodes variations of the extra scale $\ell$. In that paper the authors explored alternative variables, in particular, they considered $\mathbb{L}_\alpha=\lambda \ell^\alpha$, for some constants $\lambda$ and $\alpha$, and their thermodynamic conjugates $B_\alpha$. Only the $\alpha=-2$ case leads to a pair of variables with intrinsic units of pressure and volume, however, other values of $\alpha$ proved to be useful depending on the context. For instance, the authors argued that in the theory of membranes a more natural variable is $\mathbb{L}_{1}$, for which the conjugate variable $B_1$ can be interpreted as a tension. Regardless the choice of $\mathbb{L}_\alpha$, however, it was shown that the product $\mathbb{L}_\alpha B_\alpha$ is invariant under the mentioned scaling argument. This can be seen both by direct calculation, or simply by requiring consistency of their corresponding Smarr relations. Therefore, different definitions of the thermodynamic variable associated to variations of $\ell$ lead to the same physics since they all have the same free energy, $G=M-TS - \Phi Q$,   and the same first law, $dM=TdS+ \Phi dQ +B_\alpha d\mathbb{L}_\alpha$.

\subsection{Pressure, volume and the Smarr relation}

When the gravitational description contains extra scales, as is the case for the hyperscaling violating black holes studied in this paper, the story is a little more involved. In principle one could distinguish between variations of the different scales. However, for simplicity, we will define a single thermodynamic variable with units of pressure as follows
\begin{equation}\label{ourpressure}
P = \frac{V_0}{16 \pi G}   = \frac{1}{16 \pi G} (d -\theta  +z - 1) (d- \theta  + z ) \ell^{-2} r_F^{-2 \theta/d} e^{-\lambda_0 \phi_0 }\,.
\end{equation}
This is a natural generalization from the gravity perspective and it reduces to the AdS case (\ref{pressure})  for $z=1$ and $\theta=0$. For vanishing $\theta$, but generic values of the dynamical exponent $z$, this pressure also  agrees with the pressure defined for pure Lifshitz black holes in \cite{Brenna:2015pqa}. In that case the potential of the scalar is constant $V(\phi)=V_0=-2\Lambda$ and it can therefore be identified with   the pressure of a perfect fluid. For general values of $\theta$ the variable  \eqref{ourpressure} cannot be interpreted in a straightforward way as the pressure of a perfect fluid, because of nonlinear terms in the Einstein equations, but it can still be used as a pressure in the framework of extended thermodynamics.

Another possible definition of pressure would be to consider $\tilde{P}=d(d+1)\ell^{-2}/16\pi G$ as in the AdS case, or more generally $\tilde{P}=\lambda \ell^{-2}$ for any $\lambda$. With these alternative choices one would be varying only the curvature scale $\ell$ while keeping other scales fixed ($r_F$ and $\phi_0$). Conversely, one can choose to vary another scale, e.g.  by defining $\tilde{P}=\lambda r_F^{-2}$. However, following the logic and the discussion above, one can argue that all these definitions would give rise to the same $PV$ term, and hence to the same bulk physics (although, the interpretation in the boundary theory would be different depending on the specific variation that is considered).

In the remaining part of this section we will derive the conjugate thermodynamic volume associated to (\ref{ourpressure}) and write down another version of the Smarr relation  for   hyperscaling violating black branes.
We follow the method of \cite{Brenna:2015pqa} for obtaining the thermodynamic volume in the sense that we assume the extended first law (\ref{extendedfirstlaw}) and Smarr relation (\ref{Smarr}) to hold, and derive an expression for the volume that is consistent with both relations. Obviously, the   volume  is uniquely fixed by the Smarr relation (\ref{Smarr}), our definition of the pressure in (\ref{ourpressure}), and the expressions for all other thermodynamic quantities found in Section \ref{firstlaw}.
 By combining all these  expressions, we find  the following thermodynamic volume  for our    hyperscaling violating black hole solutions
 \begin{equation}
\begin{aligned}  \label{thermovolume}
V \, \, &= \, \,  \omega_{k,d}r_h^{d-\theta+z} \ell^{-z+1}  r_F^{\theta + 2 \theta/d} e^{ \lambda_0 \phi_0  } \frac{1}{2 (d- \theta +z-1)(d-\theta+z)} \times \\
 &\qquad   \times \left ( d( z+1)-\theta + k \frac{(d-1)^2 (d(z-1)-\theta)}{(d - \theta+ z-2)^2} \frac{\ell^2}{r_h^2}   - ( d(z -1)-\theta) \frac{q^2}{r_h^{2 (d-\theta + z - 1)}} \right)  \, .
\end{aligned}
\end{equation}
This result is valid for $k=0,1$.\footnote{For   hyperbolic black holes we defined the mass with respect to the ground state  given by $q=0$ and $m=m_{ground}$, i.e. $M = M - M_{ground}$. The thermodynamic volume would change accordingly  to $V_{k=-1}  = V - V_{ground}$, where $V_{ground} = -(d-1) M_{ground} / (2P).$ The absolute value of the thermodynamic volume (in contrast to the background subtracted value) can be found using the counterterm method \cite{Henningson:1998gx,Balasubramanian:1999re}.}
Note that for  the special value~$\theta  = d(z-1)$ the expression simplifies significantly, since the last two terms in the second line vanish.
 As a consistency check, we note that for $z= 1$ and $\theta = 0$ this formula  reduces to the standard expression (\ref{AdSvolume}) for AdS black holes. Moreover, for $\theta =0 $ our result     agrees with the thermodynamic volume for  pure Lifshitz black holes stated in equation~(74) of   \cite{Brenna:2015pqa}. It would be interesting, though, to check our expression for general $\theta$      using   different methods,  such as the Komar integral relation   \cite{Kastor:2009wy} or   the extension of the Iyer-Wald formalism  \cite{Caceres:2016xjz,Couch:2016exn}.

Lastly, we mention that for  planar black holes ($k=0$) the standard Smarr relation (\ref{Smarr}) can be split up into two separate equations
 \begin{align}
 (d-\theta+z)M  &= (d-\theta) T S +  (d-\theta+z-1)  \Phi Q   \, ,  \label{smarrkis0}  \\
2   (d-\theta+z) PV  &=   (d(z+1)-\theta)TS + (d -1) \Phi Q       \, .
\end{align}
This follows from the explicit expressions for all the thermodynamic quantities involved, such as the mass and the temperature.
It is straightforward to verify that these equations can be combined into the single formula (\ref{Smarr}).
 For pure Lifshitz black branes ($\theta =0$)  the first equation is often called the Smarr relation, e.g. \cite{Liu:2014dva}, but it was pointed out in \cite{Brenna:2015pqa} that the standard Smarr relation  (\ref{Smarr}) is more  general, since it applies to all horizon topologies and asymptotics. This is already apparent from the fact that the equations above  explicitly depend on $z$ and $\theta$, whereas these parameters do not enter into the standard Smarr relation.

\subsection{$P-V$ criticality}

Until now we have considered the extended thermodynamics in the grand canonical ensemble. For charged AdS black holes in the canonical ensemble there exists a phase transition in the ($P,V$) phase diagram, analogous to the liquid-gas phase transition for the Van der Waals system \cite{Kubiznak:2012wp}. Furthermore, it was shown in this analogy  (which corresponds to the third analogy in Table \ref{tableanalogies}) that the critical exponents coincide with those of the Van der Waals fluid.
 In this section we study these phase transitions in the canonical ensemble and    compute the critical exponents for hyperscaling violating black holes. Contrary to the AdS case, we cannot define the equation of state $P(V,T)$ explicitly, since   the expressions for the temperature and thermodynamic volume cannot be inverted for all values of $z$ and $\theta$. Therefore, we proceed numerically and   work with the full thermodynamic volume instead of the specific volume, defined in \cite{Kubiznak:2012wp} through the equation of state. Note that this section focuses solely on   spherical black holes $k=1$, since this is the only case that exhibits phase transitions.

We   first compute the thermodynamic volume in the fixed charge ensemble. In this ensemble we have defined the mass   with respect to the extremal black hole, and therefore the thermodynamic volume   changes accordingly. The Smarr relation in the canonical ensemble   is given by
\bqn\label{Smarr2}
(d-1) \tilde M=d  \, TS+(d-1) \tilde \Phi Q-2P\tilde V\,,
\eqn
where $\tilde M$ and $\tilde \Phi$ are the  mass and   potential compared to the extremal case, given in (\ref{masscanonical}) and (\ref{potentialcanonical}) respectively.  This identity follows from subtracting the  Smarr formula for the extremal black hole, $(d-1) M_{ext} = (d-1) \Phi_{ext} Q_{ext} - 2 P V_{ext}$, from the   Smarr relation    (\ref{Smarr}) itself. The thermodynamic volume in the canonical ensemble can be obtained by solving  (\ref{Smarr2}) for $\tilde V$. As a  result,  the volume is   $\tilde V = V - V_{ext}$, where $V$ is  given by  (\ref{thermovolume}), and $V_{ext}$ can be computed to be
\begin{equation}
 V_{ext}    
   = \omega_{k,d}  r_{ext}^{d-\theta  +z} \ell^{-z+1}   r_F^{\theta + 2 \theta/d} e^{ \lambda_0 \phi_0  }  \frac{(d-1) (d-\theta )}{(d-\theta +z-2) (d-\theta +z-1) (d-\theta
   +z)} \, .
\end{equation}
  The extremal horizon radius $r_{ext}$ is an implicit function of    the extremal charge $q_{ext}$,  through  (\ref{extremalcharge}).

  In the extended phase space, the on-shell Euclidean action  in the canonical ensemble is associated with the Gibbs free energy, instead of the Helmholtz free energy, i.e. $ G\equiv I /\beta =   \tilde M-TS\,$.
The Helmholtz free energy, on the other hand, is defined as $F = G - P \tilde V $.
The Gibbs free energy $G=G(T)$ at constant $P$ inherits  the ``swallow tail" behaviour of the  canonical ensemble in the standard thermodynamic framework, since the on-shell Euclidean action is the same \cite{Karch:2015rpa}. So we expect a line of first order phase transitions for our black holes (similar to the liquid-gas phase transition in the Van der Waals system), with a critical point at the end of the line.

The value of the critical point can be easily determined from the following argument. The specific heat at constant pressure $C_P = T \frac{\partial S}{\partial T} \big |_P$
  is the same as the heat capacity at constant charge $C_Q$ in the standard canonical ensemble, given in \eqref{heatcapQ}. The critical point occurs at the point of divergence of the specific heat at constant pressure, which is thus the same as the critical point in the standard canonical ensemble (since $C_Q$ also diverges at the critical point). The critical horizon radius and electric charge are given by \eqref{critical}, which yield the critical temperature  \eqref{criticaltemp}.
Using \eqref{critical}  we can also compute  the   critical value for the pressure
\begin{equation}
P_{crit}  = \frac{k (d-1)^2(2-z)}{ z  r_{crit}^2  } \frac{r_F^{-2 \theta/d} e^{-\lambda_0 \phi_0}  }{16 \pi G} \, ,
\end{equation}
and for the thermodynamic volume
\begin{eqnarray}
\begin{aligned} \label{criticalvolume}
V_{crit} \,\, &= \,\,  \omega_{k,d} \ell^{-z+1} r_{crit}^{d-\theta +z} r_F^{\theta + 2 \theta/d} e^{ \lambda_0 \phi_0  } \frac{1}{2 (d - \theta +z -1)(d-\theta +z)} \times \\
 &\qquad   \times \left ( d (z+1)-\theta-\frac{z (2 d-2 \theta +z) (d (z-1)-\theta ) (d-\theta +z)}{(z-2) (d-\theta +z-2) (2 d-2\theta +z-2)}\right)   \, .
\end{aligned}
\end{eqnarray}
This is the first expression in the literature for the critical value of the thermodynamic volume and pressure for general $z$ and $\theta$.   For the full background subtracted volume $\tilde V$ we also need  to know the extremal value $V_{ext}$ at the critical point, but there is no  analytic expression for this quantity for general $z$ and $\theta$. It could   be determined numerically, however,   by writing $\ell$ in terms of $r_{crit}$  and solving $ q_{ext} (r_{ext}, \ell)$ in (\ref{extremalcharge}) for $r_{ext}$.

\begin{figure}[t!]
\begin{center}
\begin{overpic}[width=.45\textwidth]{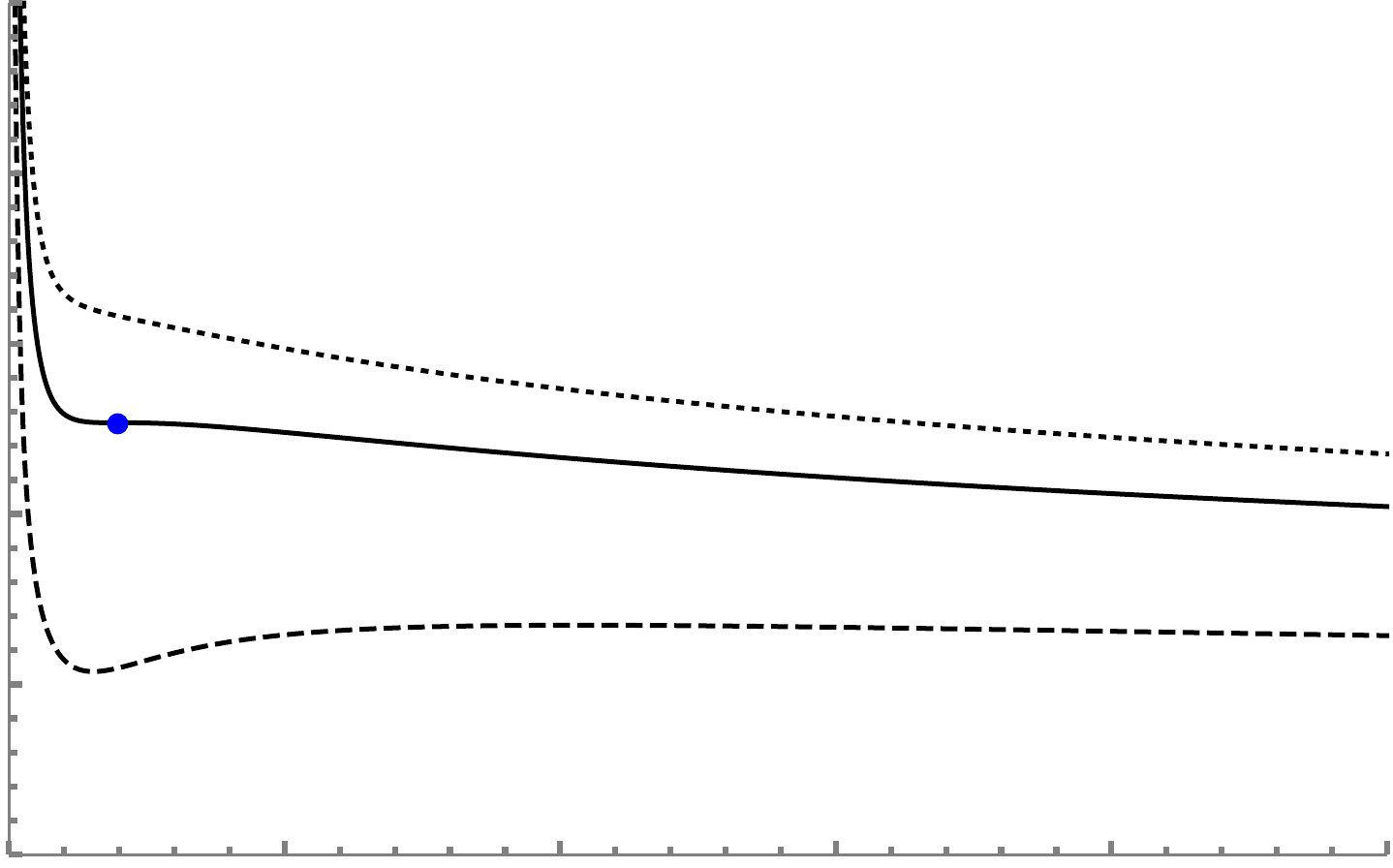}

		\put (-8,47.75) {\footnotesize{0.20}}
		\put (-8,36) {\footnotesize{0.15}}
		\put (-8,24.25) {\footnotesize{0.10}}
		\put (-8,12.5) {\footnotesize{0.05}}
		\put (-4,-3.5) {\footnotesize{0}}
		
		\put (19.5,-4) {\footnotesize{1}}
		\put (39.25,-4) {\footnotesize{2}}
		\put (59.5,-4) {\footnotesize{3}}
		\put (79,-4) {\footnotesize{4}}

		\put (33,64) {$\underline{1\leq z <2,\,\theta=1/2}$}
		
		\put (2.4,26.9) {\footnotesize{$(P_{crit},\tilde{V}_{crit})$}}
		
		\put (-4.5,58) {$P$}
		\put (95,-4.5) {$\tilde{V}$}
		
		\put (55,34) {\footnotesize{$T>T_{crit}$}}
		\put (32,25) {\footnotesize{$T=T_{crit}$}}
		\put (3,10) {\footnotesize{$T<T_{crit}$}}
		
\end{overpic}
\hspace{0.35cm}
\begin{overpic}[width=.45\textwidth]{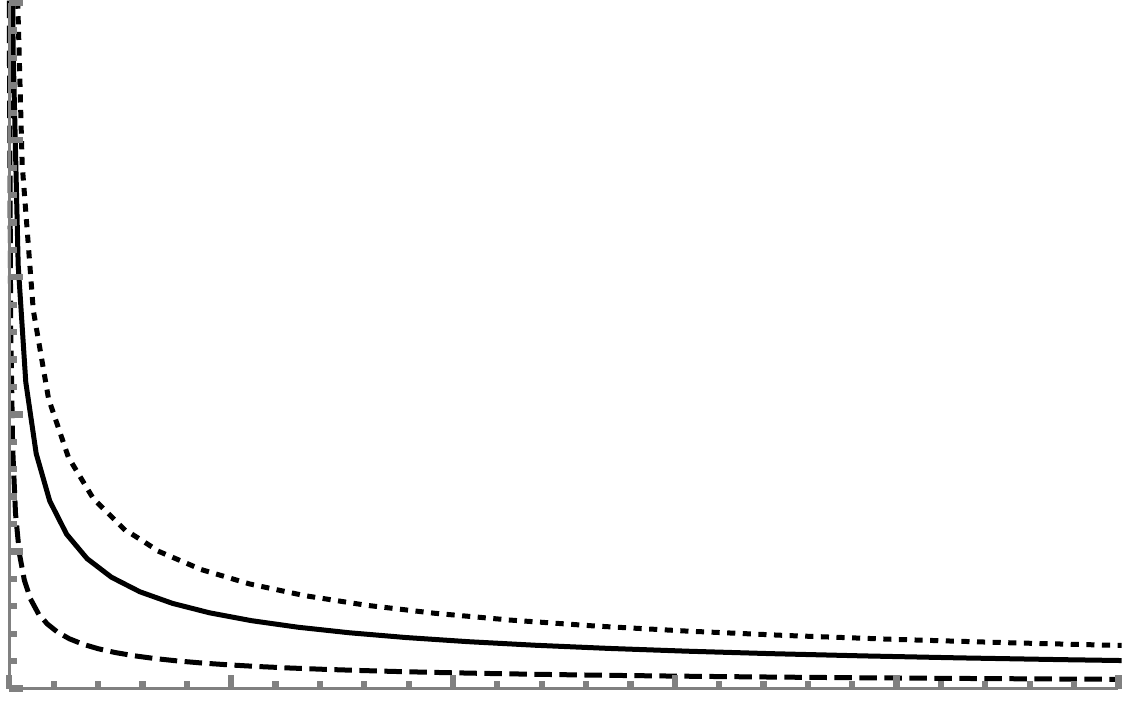}

		\put (-8,47.75) {\footnotesize{0.20}}
		\put (-8,36) {\footnotesize{0.15}}
		\put (-8,24.25) {\footnotesize{0.10}}
		\put (-8,12.5) {\footnotesize{0.05}}
		\put (-4,-3.5) {\footnotesize{0}}
		
		\put (19,-4) {\footnotesize{1}}
		\put (39,-4) {\footnotesize{2}}
		\put (59,-4) {\footnotesize{3}}
		\put (79,-4) {\footnotesize{4}}

		\put (33,64) {$\underline{z\geq 2,\,\theta=1/2}$}

		\put (-4.5,58) {$P$}
		\put (95,-4.5) {$\tilde{V}$}
		
		
\end{overpic}
\end{center}
\vspace{-0.2cm}
\caption{
Two $(P,\tilde{V})$ diagrams in which different isotherms are plotted for two representative values of $z$, i.e. $z=3/2$ and $z=3$, respectively. Other parameters are fixed at: $k=\phi_{0}=Z_{0}=r_{F}=G=1$, $\theta=1/2$ and $d=3$. In the left panel, there exists a first order phase transition (similar to a liquid-gas phase transition) below a  critical temperature. In the right panel,  there is no critical temperature and hence no phase transitions.
}\label{fig:vanderwaals}
\end{figure}
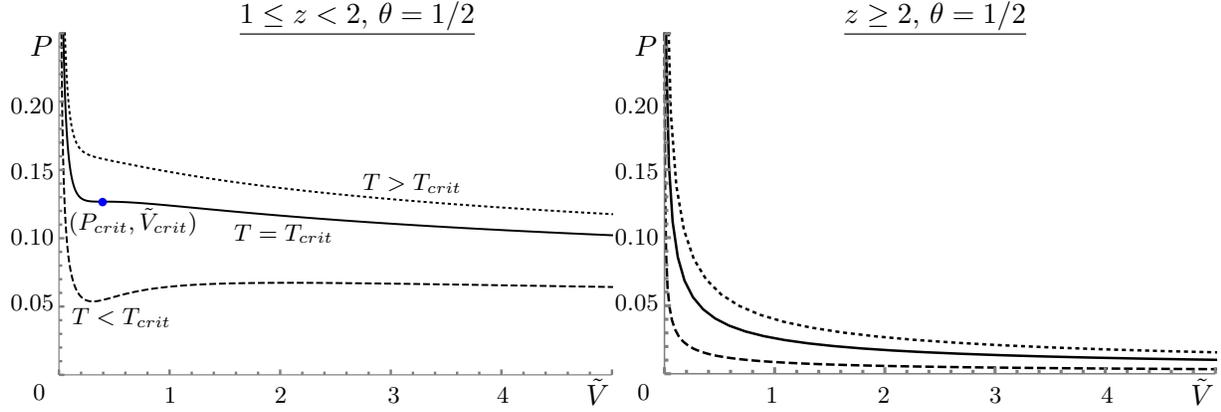

In practice, the full $P(\tilde{V},T)$ diagram can be obtained by fixing $Q$, which gives an implicit relation for $q=q(\ell)$, and then making a parametric plot by varying $r_h$ and $\ell$.  In Figure \ref{fig:vanderwaals} we give examples of isotherms in a ($P,\tilde V$) diagram for two representative values of $z$ and a fixed value of $\theta$. For $1 \le z < 2$ we find a critical point at  the critical temperature $T=T_{crit}$ and first order liquid-gas like phase transitions below this temperature, whereas for $z\ge 2$ there are no phase transitions. A comment on the phase transitions in the extended $PV$ space is in order here. As explained in \cite{Caceres:2015vsa}, the statement about mechanical stability in the bulk and, in particular, the positivity of the isothermal compressibility, is mapped holographically into a monotonicity constraint in the space of QFTs, analogous to a $c$-theorem. Therefore, the existence of a Van der Waals transition translates into a non-trivial prediction (the discontinuity of this central function) for non-relativistic theories with $1\le z<2$ (in a suitable large-$N$ limit) defined on a sphere.
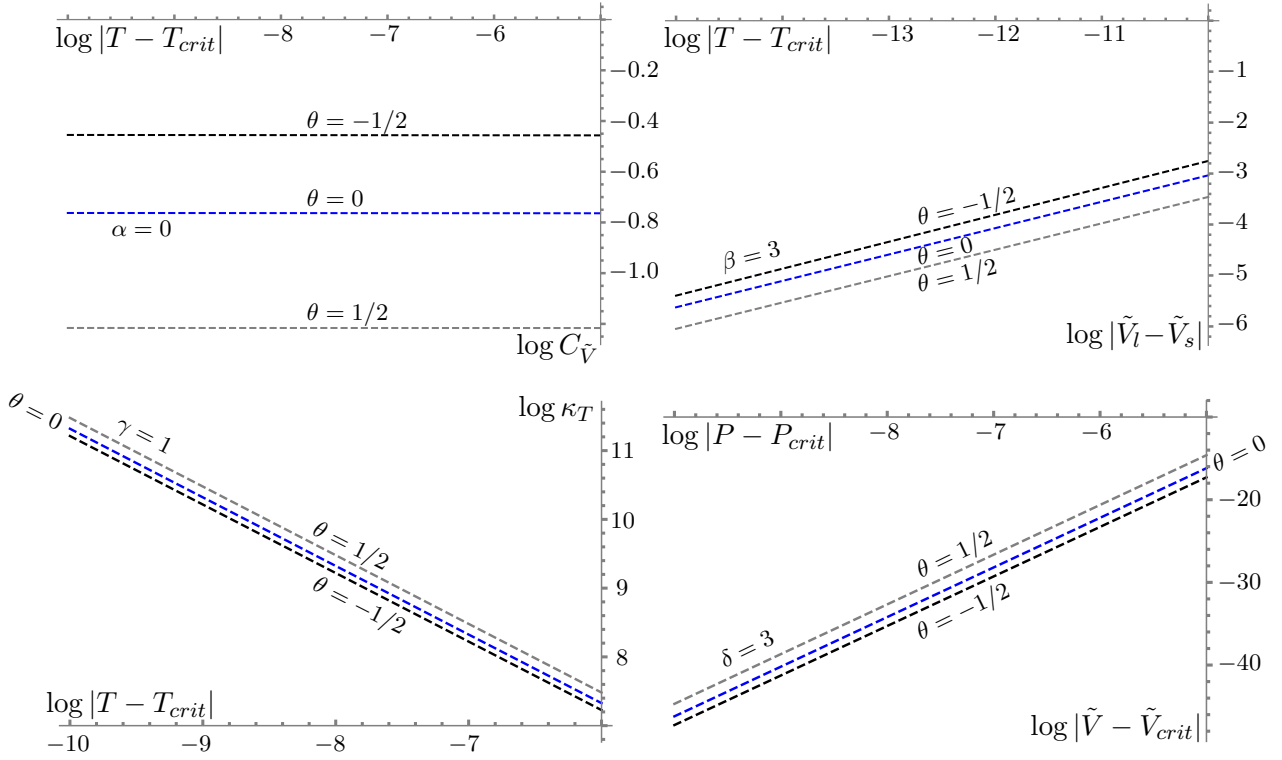
\begin{figure}[t!]
\begin{center}
\begin{overpic}[width=.45\textwidth,grid=false]{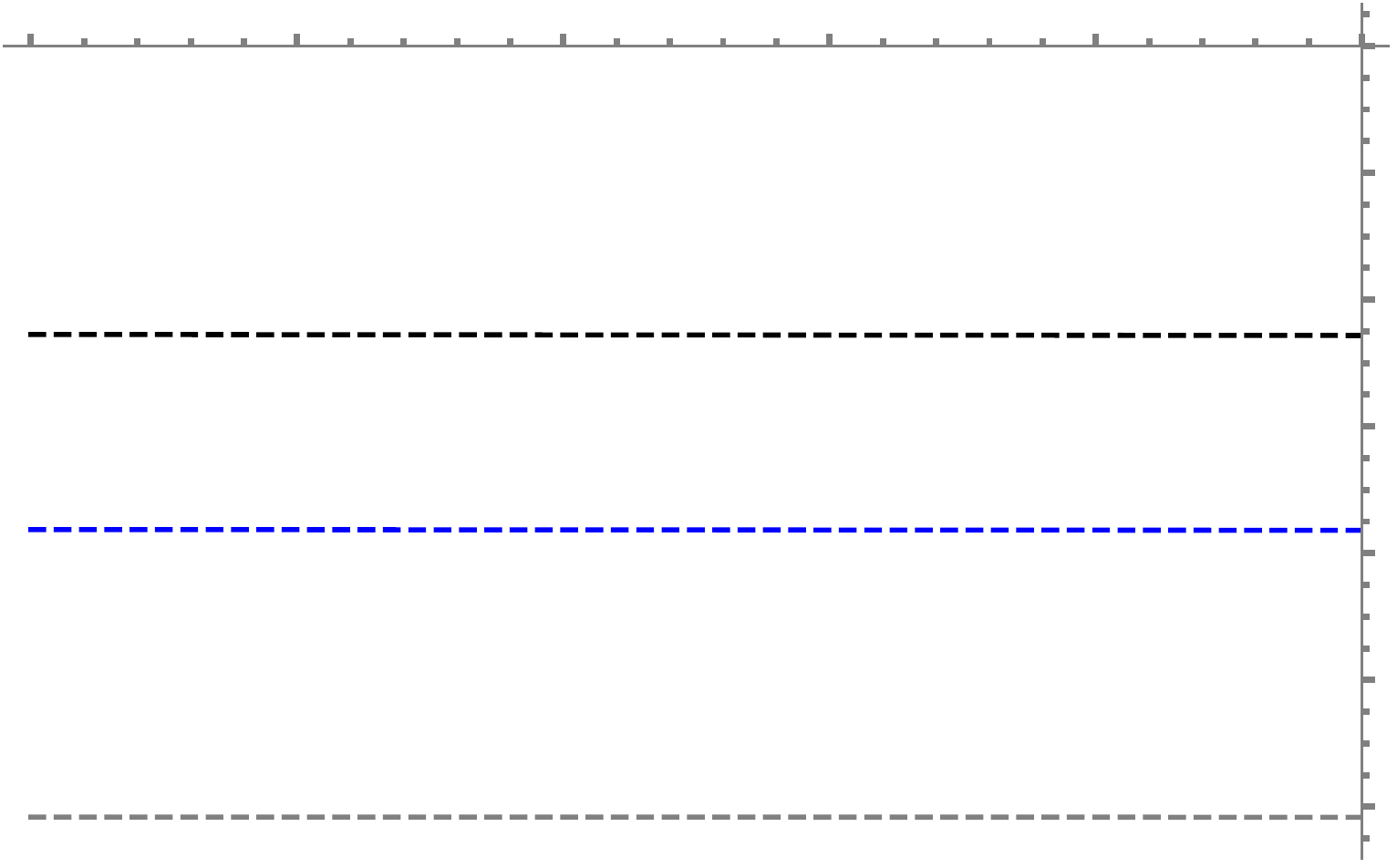}

		\put (99,48.5) {\footnotesize{$-0.2$}}
		\put (99,39.5) {\footnotesize{$-0.4$}}
		\put (99,30.5) {\footnotesize{$-0.6$}}
		\put (99,21.5) {\footnotesize{$-0.8$}}
		\put (99,12.5) {\footnotesize{$-1.0$}}
		\put (82.5,-1) {$\log C_{\tilde{V}}$}
		
		\put (75,55) {\footnotesize{$-6$}}
		\put (56,55) {\footnotesize{$-7$}}
		\put (37,55) {\footnotesize{$-8$}}
		\put (0,54) {$\log|T-T_{crit}|$}
		
		\put (10,20) {\footnotesize{$\alpha=0$}}
		
		\put (45,5) {\footnotesize{$\theta=1/2$}}
		
		\put (45,25.5) {\footnotesize{$\theta=0$}}
		
		\put (45,39.5) {\footnotesize{$\theta=-1/2$}}

\end{overpic}
\hspace{0.4cm}
\vspace{0.65cm}
\begin{overpic}[width=.45\textwidth,grid=false]{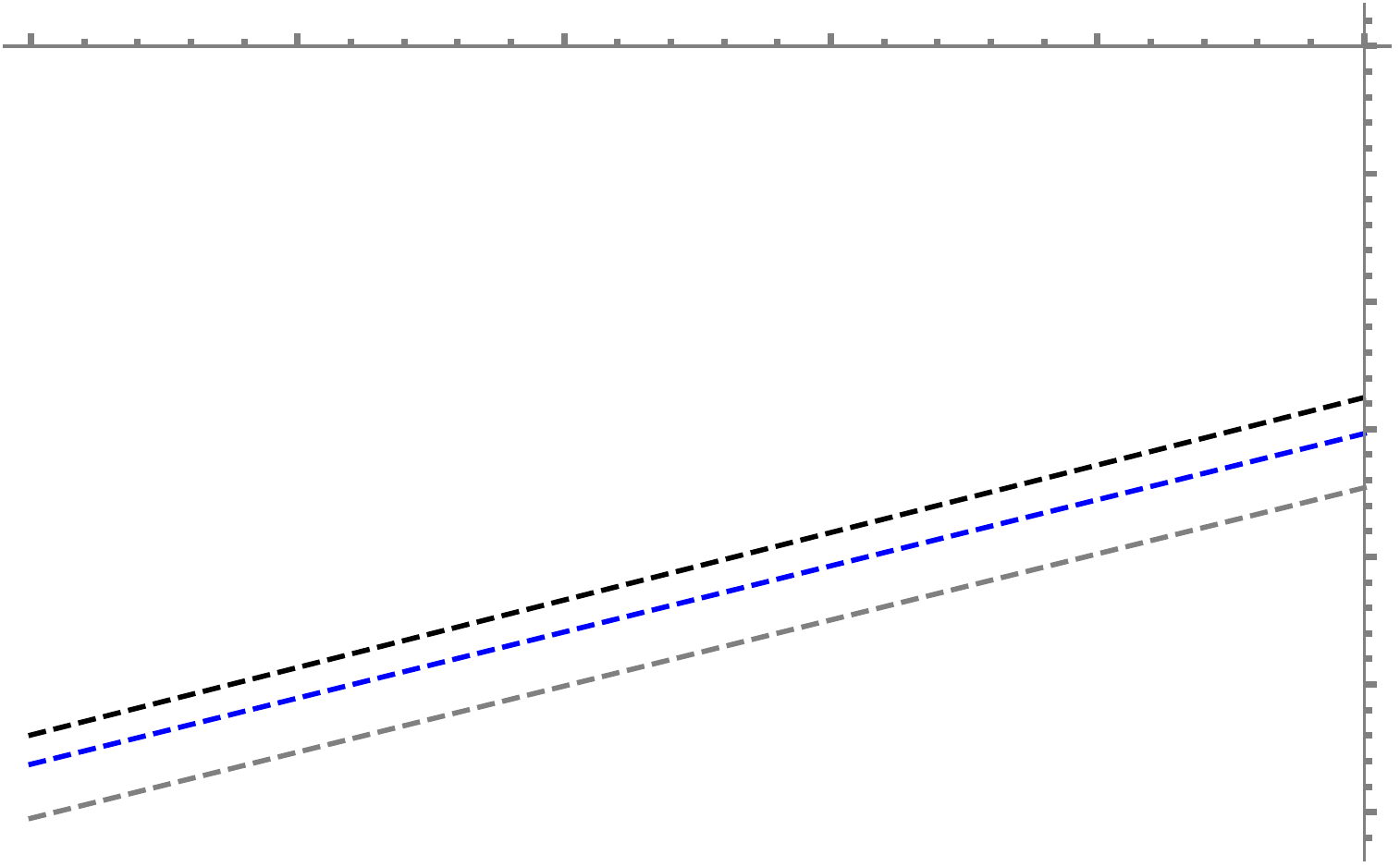}

		\put (99,48.5) {\footnotesize{$-1$}}
		\put (99,39.5) {\footnotesize{$-2$}}
		\put (99,30.5) {\footnotesize{$-3$}}
		\put (99,21.25) {\footnotesize{$-4$}}
		\put (99,12.25) {\footnotesize{$-5$}}
		\put (99,3.25) {\footnotesize{$-6$}}
		\put (72,1) {$\log|\tilde{V}_{l}\!-\!\tilde{V}_{s}|$}
		
		\put (75,55) {\footnotesize{$-11$}}
		\put (56,55) {\footnotesize{$-12$}}
		\put (37,55) {\footnotesize{$-13$}}
		\put (0,54) {$\log|T-T_{crit}|$}

		\put (10,14) {\rotatebox{14}{\footnotesize{$\beta=3$}}}
		
		\put (45,10) {\rotatebox{14}{\footnotesize{$\theta=1/2$}}}
		
		\put (45,14.9) {\rotatebox{14}{\footnotesize{$\theta=0$}}}
		
		\put (45,22) {\rotatebox{14}{\footnotesize{$\theta=-1/2$}}}

\end{overpic}
\begin{overpic}[width=.45\textwidth,grid=false]{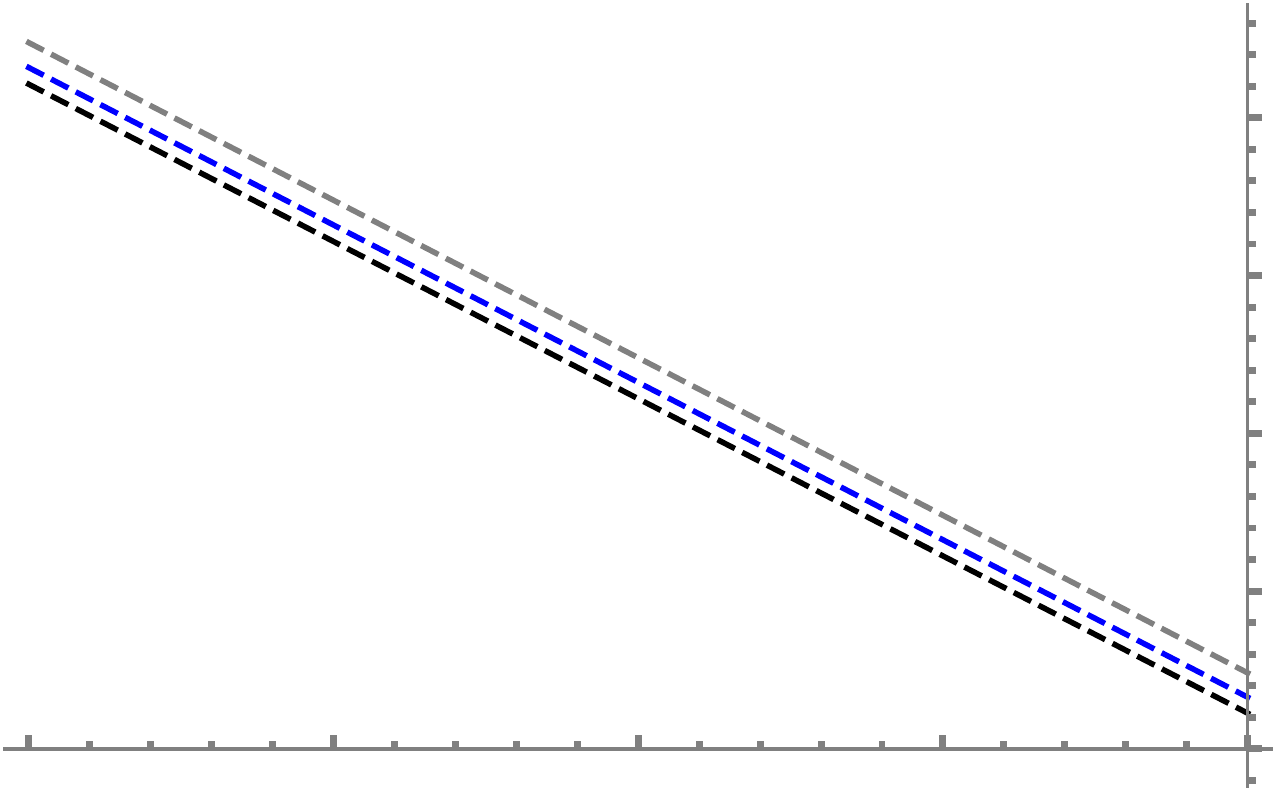}

		\put (99,51.5) {\footnotesize{$11$}}
		\put (99,39) {\footnotesize{$10$}}
		\put (100,27) {\footnotesize{$9$}}
		\put (100,14.5) {\footnotesize{$8$}}
		\put (82.5,59) {$\log \kappa_{T}$}

		\put (70,-1) {\footnotesize{$-7$}}
		\put (46,-1) {\footnotesize{$-8$}}
		\put (22,-1) {\footnotesize{$-9$}}
		\put (-2,-1) {\footnotesize{$-10$}}
		\put (-2,6) {$\log|T-T_{crit}|$}
		
		\put (10,57) {\rotatebox{-27}{\footnotesize{$\gamma=1$}}}
		
		\put (45,38) {\rotatebox{-27}{\footnotesize{$\theta=1/2$}}}
		
		\put (-9,61) {\rotatebox{-27}{\footnotesize{$\theta=0$}}}
		
		\put (45,28) {\rotatebox{-27}{\footnotesize{$\theta=-1/2$}}}

\end{overpic}
\hspace{0.35cm}
\begin{overpic}[width=.45\textwidth,grid=false]{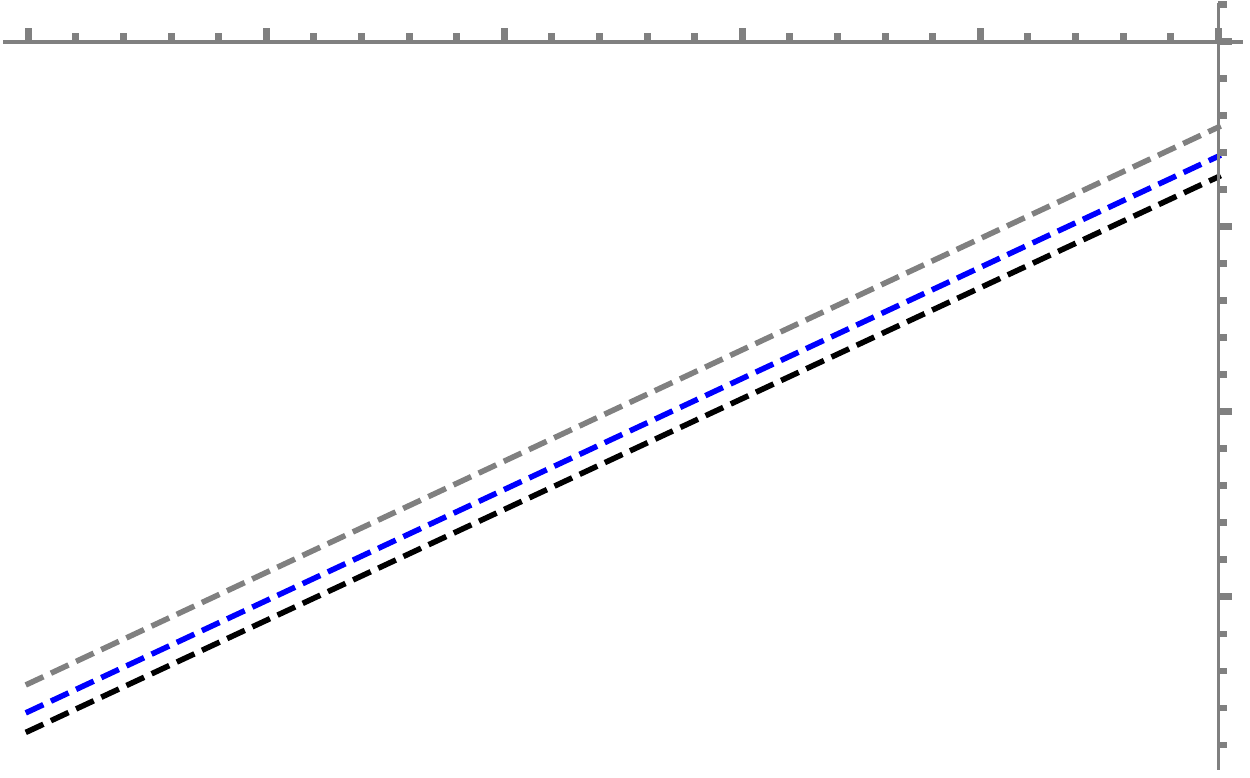}

		\put (99.5,43) {\footnotesize{$-20$}}
		\put (99.5,28) {\footnotesize{$-30$}}
		\put (99.5,13.5) {\footnotesize{$-40$}}
		\put (66,2) {$\log|\tilde{V}-\tilde{V}_{crit}|$}
		
		\put (75,55) {\footnotesize{$-6$}}
		\put (56,55) {\footnotesize{$-7$}}
		\put (37,55) {\footnotesize{$-8$}}
		\put (0,54) {$\log|P-P_{crit}|$}

		\put (10,14) {\rotatebox{26}{\footnotesize{$\delta=3$}}}
		
		\put (45,30) {\rotatebox{26}{\footnotesize{$\theta=1/2$}}}
		
		\put (98,49) {\rotatebox{26}{\footnotesize{$\theta=0$}}}
		
		\put (45,19) {\rotatebox{26}{\footnotesize{$\theta=-1/2$}}}

\end{overpic}
\end{center}
\vspace{-0.2cm}
\caption{
In these four diagrams, we show $\log-\log$ plots for the quantities defined in (\ref{criticalquantities}) near the critical point, for three different values of $\theta=-1/2,0,1/2$.   Other parameters are fixed at: $k=\phi_{0}=Z_{0}=r_{F}=G=1$,  $d=3$, and $z=3/2$. From the slopes of these $\log-\log$ plots one can easily read off the critical exponents $\alpha, \beta, \gamma$ and $\delta$, which turn out to be consistent with the critical exponents for the Van der Waals system.
}\label{fig:critical}
\end{figure}

In cases where the Van der Waal transition is present,  we can compute the critical exponents  numerically. They are defined  through the following relations:
\begin{equation}
\begin{aligned} \label{criticalquantities}
	&& C_{\tilde V} = T \frac{\partial S}{\partial T} \bigg |_{\tilde V}\sim |T-T_{crit}|^{-\alpha}
	\,,
	\qquad  \quad
	\eta = \tilde V_l - \tilde V_s
	\sim
	|T-T_{crit}|^{\beta}
	\,, \\
	&&	\kappa_T = - \frac{1}{\tilde V} \frac{\partial  \tilde V}{\partial P} \bigg |_T
	\sim
	|T-T_{crit}|^{-\gamma}
	\,,
	\qquad
	|P- P_{crit}|
	\overset{ T=T_{crit}}{\sim}
	|\tilde V- \tilde V_{crit}|^{\delta}
	\,.
\end{aligned}
\end{equation}
Here, $\tilde V_l$ and $\tilde V_s$ denote the thermodynamic volume of large and small black holes, respectively.
From  the $P(\tilde{V},T)$ diagram the exponents can be read off by zooming in around the critical point $(\tilde{V}_{crit}, T_{crit}, P_{crit})$. In Figure \ref{fig:critical} we plot examples of the above quantities for various values of the parameters. Similarly to the first analogy in Table \ref{tableanalogies}, we find that the critical exponents in the third analogy are the same as those of the Van der Waals system
\bqn
\alpha = 0\,, \quad \beta = 1/2\,, \quad  \gamma = 1\,, \quad \delta =3\,,
\eqn
for $1\leq z<2$ and any value of $\theta$.  This agrees with the expectation from mean field theory.

\section{Summary and discussion}
In this paper we constructed two different, but related, families of charged hyperscaling violating black hole solutions and studied the standard and extended thermodynamics of these black holes. The first set of backgrounds is a family of charged hyperscaling violating black holes with spherical horizons, which are solution to a generalized EMD theory. These black holes can be viewed as a natural extension of the planar hyperscaling violating black branes \cite{Alishahiha:2012qu,Gath:2012pg,Gouteraux:2012yr}, but also a generalization of the pure Lifshitz spherical black holes \cite{Tarrio:2011de} to arbitrary hyperscaling violating parameter $\theta$. The second family is obtained by a subtle limit of the dilaton field in the aforementioned model. This limit yields a solution for spherical, planar and hyperbolic charged black holes with $1\leq z<2$, as long as we set $\theta=d(z-1)$. The value $\theta=d(z-1)$ seems singular in the former model, but we showed that the limit is physically well behaved.

From the perspective of thermodynamics, we found that only the spherical black hole exhibits phase transitions. Curiously, the phase transitions only take place for $1\leq z\leq2$, as was already established for $\theta=0$ \cite{Tarrio:2011de}, but now extended to other values of $\theta$. The qualitative behavior of the phase transitions does not depend on $d$ and $\theta$, although most of the thermodynamic quantities depend on all parameters. This resonates with the intuition that $\theta$ effectively only reduces the dimensionality of the system, i.e. $d_{\text{eff}}=d-\theta$. However, looking at the detailed computations, this seems a fairly non-trivial outcome.

Another result from our thermodynamic studies is the computation of four critical exponents for hyperscaling violating black holes, determined analytically for the first analogy and numerically for the third analogy in Table \ref{tableanalogies}. The exponents associated to the critical point  are shown to be   the same for both analogies as the critical exponents for a Van der Waals liquid. This result is consistent with previous findings for   charged AdS black holes for the second \cite{Niu:2011tb} and third analogy \cite{Chamblin:1999tk}. Moreover, it agrees with the expectation from mean field theory that critical exponents are universal and do not depend on the details of the microscopic model.


For future work, we remark that in order to study the thermodynamics  we used the background subtraction method. Although it seems reasonable to apply this method for our purposes, it could be insightful to use the counterterm method \cite{Papadimitriou:2005ii} to compute the Casimir energies and rederive the various thermodynamic quantities, especially in the case of the hyperbolic black hole. The general algorithm for constructing the counterterms for asymptotically hyperscaling violating
Lifshitz backgrounds was given in \cite{Chemissany:2014xsa,Cremonini:2018jrx}. Such an analysis would also improve our   Smarr relation in the canonical ensemble for extended thermodynamics, since we used the background subtraction method  in that context.  Furthermore, it would   be useful to check our rather complex expression for the thermodynamic volume (\ref{thermovolume}) using other methods, such as the Komar integral relation   \cite{Kastor:2009wy} or   the extension of the Iyer-Wald formalism  \cite{Caceres:2016xjz,Couch:2016exn}.

A puzzle is provided by the $\theta=d(z-1)$ background, for which we find that there are no singularities or diverging tidal forces at $r=0$. This raises the question whether this space has extra symmetries as opposed to $\theta\neq d(z-1)$ or whether one can construct  global coordinates as was done,  e.g., in\cite{Blau:2009gd}. Furthermore, we note that especially the massless, uncharged hyperbolic case   is curious, since it has finite entropy and temperature, but no curvature singularities at $r=0$. This seems to imply that this solution is actually the Rindler wedge of some spacetime, similar to how the massless hyperbolic black hole in AdS is identical to the AdS-Rindler wedge. A natural question for future work is: which spacetime is this? Due to all these properties we note that it seems that the $\theta=d(z-1)$ background has, in fact, a lot in common with AdS.

Finally, another avenue for future research is studying observables in these background geometries, e.g., two-point correlators, quasinormal modes and transport coefficients. There are various important questions that might be of interest. How do the observables depend on $z$, $\theta$ and the horizon topology $k$ (see also \cite{Mukherjee:2017ynv,Arav:2012ud} for similar work)? Are these solutions stable under perturbations and can they become overdamped (see e.g. \cite{Sybesma:2015oha,Puletti:2017gym,Neupane:2003vz})? Are there extra cases with hidden Fermi surfaces besides the well-known $\theta=d-1$, $k=0$ case \cite{Ogawa:2011bz,Gouteraux:2011ce,Huijse:2011ef}? And what happens with the zero sound mode? Does the Ohmic vs. sub-Ohmic dissipative transition at $z=2(1-\theta/d)$ prevail for the different topologies \cite{Edalati:2012tc,Edalati:2013tma}? Is the pattern of entanglement short-range or long-range? Can there be new violations to the area law? How does the entanglement entropy depend on the various parameters \cite{Alishahiha:2012cm,Fischler:2012ca}? And can one construct holographic heat engines for hyperscaling violating black holes \cite{Johnson:2014yja}?
More generally, it would be interesting to study holography for the new spherical and hyperbolic hyperscaling violating black holes, and to interpret the phase transitions found in this paper in terms of  a dual boundary system.  \\

\noindent \textbf{Note added:} The   paper \cite{KastorRayTraschen2018}  appeared simultaneously with ours and also studies extended thermodynamics for hyperscaling violating black branes, which are solutions to Einstein-Dilaton theory ($z=1, \, k=q=0$ in our notation). They derive the ADM mass, spatial tensions, and the first law  in   detail using the Hamiltonian formalism and  prove the Smarr formula using Komar integrals.

\subsection*{Acknowledgments}
It is a pleasure to thank J. Armas, J. Cresswell, J. Hartong, A. Janssen, I. Jardine, E. Kiritsis, I. Papadimitriou, A. Peet, N. Poovuttikul, and S. Vandoren for discussions and D. Cohen-Maldonado for collaborating in the initial stage of the project. JFP is supported by the Netherlands Organization for Scientific Research (NWO) under the VENI scheme. The research of WS is funded by the Icelandic Research Fund (IRF) via a Postdoctoral Fellowship Grant (number 185371-051). WS acknowledges that he was employed at Utrecht University when this project was initiated and he  thanks the Universities of Amsterdam and Toronto for their hospitality while this project was being completed.   MRV is supported by the Spinoza Grant,   financed by NWO.

\appendix

\section{On-shell Euclidean action}
\label{euclideanaction}

In this appendix we   compute the (background subtracted) on-shell Euclidean action for the black hole solutions of the generalized  Einstein-Maxwell-Dilation theory. This quantity is related to the free energy of the black holes \cite{Gibbons:1976ue} or, through the holographic dictionary, to the free energy of the dual theory coupled to a thermal ensemble \cite{Witten:1998zw}. From the free energy (and temperature) of the solutions it is then straightforward to compute the entropy and energy  of the black holes. Hence, the results of this appendix can be considered as an independent check of the expressions presented in Section \ref{firstlaw}.
Although we will focus on solutions to the action (\ref{action1}), a similar analysis can be performed for the action (\ref{action2}). However, these analyzes should  lead to the same result after taking the limit $\theta\to d(z-1)$ (or $\gamma\to0$), since (\ref{action2}) can be obtained as a limiting case of (\ref{action1}). We perform the relevant computations both in the grand canonical and the canonical ensemble, following closely the original works on AdS black hole thermodynamics \cite{Hawking:1982dh,Witten:1998zw,Chamblin:1999tk}.


\subsection{Grand canonical ensemble}

In the grand canonical ensemble there are three contributions to the on-shell action: the Euclidean version of (\ref{action1}) evaluated on-shell,
the Gibbons-Hawking boundary term and extra boundary terms from matter fields.\\

\noindent \textbf{Einstein-Maxwell-Dilation term}
We start with the Euclidean form of the action (\ref{action1}). Replacing $t\to i \tau$ and inserting the explicit solutions (\ref{ansatz})-(\ref{ansatz2}) with (\ref{phispherical})-(\ref{blackeningspherical}) we arrive at:
\begin{equation}
\begin{aligned}
I_{EMD}
&=  \frac{1}{8 \pi G d} \int d^{d+2} x \, \, \ell^{1-z} r_F^{ \theta} r^{d- \theta  +z -1 }    \bigg [ (d-\theta ) (d-\theta +z)   \ell^{-2 }              \\
&
- (d-\theta)(d - \theta + z - 2)  q^2  \ell^{-2}        r^{-2 (d - \theta + z -1)}
-    k \frac{(d-1)(d(z-1)-\theta)}{d-\theta+z-2} r^{-2}     \bigg]          \\
&\equiv \frac{1}{8\pi G d}  \int d^{d+2} x  \,\,  v(r)\,,     \label{SEMDos}
\end{aligned}
 \end{equation}
where we used $\sqrt{g} = \ell^{1-z} r_F^{\theta + 2 \theta/d} r^{-2 \theta/d - \theta + d + z -1}$. As usual, the Euclidean time
direction shrinks to zero size at $r=r_h$, so we must require $\tau$ to be periodically identified $\tau\sim \tau+\beta$, with temperature $T=\beta^{-1}$ given by (\ref{HawkT}).
However, the integration in (\ref{SEMDos}) is divergent because the volume is infinite near the boundary. In order to regularize it we use the background subtraction method as in \cite{Gibbons:1976ue,Witten:1998zw,Chamblin:1999tk}. In the grand canonical ensemble, the natural reference background is the thermal state (i.e. the vacuum state heated up to a temperature $T$) with $m=0$, $q=0$ for $k=0,1$, or $m=m_{ground}$, $q=0$ for $k=-1$. In the following, we will present explicit results for the cases $k=0,1$, but we will come back to the $k=-1$ case at the end of the section.

Putting an upper cutoff $R$ on the radial integrals, the regularized volume integral of the thermal state can be expressed as follows
\begin{equation}\label{V1int}
V_1(R) =  \int_0^{\beta'} d\tau \int_0^R dr \int_{S_{k,d}} d \Omega_{k,d} \, v_0(r)\,,
\end{equation}
where $v_0(r)$ is the same integrand as in (\ref{SEMDos}) evaluated at $m=0$, $q=0$.
Similarly, the regularized volume integral for the black hole spacetime is
\begin{equation}\label{V2int}
V_2 (R) =  \int_0^{\beta} d\tau \int_{r_h}^R dr \int_{S_{k,d}} d \Omega_{k,d} \, v(r)\,.
\end{equation}
The temperature  $\beta'$  of the thermal state can in principle take any value. However, in  order to compare the black hole spacetime to the thermal spacetime, we need to choose $\beta'$ such that the geometry of the hypersurface $r=R$ is the same in these solutions, i.e.
\begin{eqnarray}
\beta'  \sqrt{1 + k \frac{(d-1)^2}{(d-\theta + z - 2)^2} \frac{\ell^2}{R^2}  }   =
 \beta   \sqrt{1 - \frac{m}{ R^{d-\theta+z}}    + k \frac{(d-1)^2}{(d-\theta + z - 2)^2} \frac{\ell^2}{R^2}
+ \frac{q^2  }{ R^{2 (d-\theta  + z - 1)}} }\,.
\end{eqnarray}
At large $R$ this condition becomes
\begin{equation}
\beta' = \beta \left (  1- \frac{m}{2 R^{d-\theta+z}}  + \dots \right)\,.
\end{equation}
Finally, by evaluating the integrals (\ref{V1int})-(\ref{V2int}) we obtain that the difference of on-shell actions is
\begin{eqnarray}
\Delta I_{EMD} &\!\!\!\!=\!\!\!\!&  \frac{1}{8\pi G d}  \lim_{R\to \infty} \left (  V_2 (R) - V_1 (R)  \right)  \\
&\!\!\!\!=\!\!\!\!&
-\frac{\omega_{k,d}}{16\pi G} \frac{\beta    r_F^{\theta } r_h^{d-\theta +z-2}}{ \ell^{1+z}} \frac{d-\theta}{d}\bigg[ r_h^2  +    \frac{q^2}{r_h^{2(d-\theta+z-2)}}  \!- \frac{d(d-1)   (  d+2
   z-3-(d+1) \theta/d )}{ (d-\theta)(d-\theta +z-2)^2} k  \ell^2  \bigg].    \nonumber
\end{eqnarray}
For $z=1$ and $\theta=0$ this result agrees with the on-shell actions computed in \cite{Witten:1998zw,Chamblin:1999tk}.\\

\noindent \textbf{Gibbons-Hawking boundary term} In general relativity, the Gibbons-Hawking boundary term  needs to be added to the Einstein-Hilbert action, if the underlying spacetime manifold has a boundary, so that the variational principle is well-defined \cite{Gibbons:1976ue}. If the spacetime is asymptotically AdS the Gibbons-Hawking boundary term gives a vanishing contribution to the on-shell Euclidean action \cite{Hawking:1982dh,Witten:1998zw,Chamblin:1999tk}. However, as we show below, for asymptotically hyperscaling violating Lifshitz spacetimes the Gibbons-Hawking term does yield a non-zero contribution.

The Gibbons-Hawking boundary term for the Einstein-Hilbert action takes the form
\begin{equation}
I_{GH} = - \frac{ 1}{8\pi G} \int  d^{d+1} x \sqrt{h} K\,,
\end{equation}
where   $\sqrt{h} =\sqrt{f(r)}  \ell^{-z} r_F^{\theta +  \theta/d} r^{ -  \theta/d +d- \theta + z }$ and $K$ is the trace of the extrinsic curvature of the timelike surface at $r=R\to\infty$. For our solution (\ref{ansatz}) the latter is given by
\begin{eqnarray}
K =  \frac{1}{2 \ell \sqrt{ f(r)}} \left (  \frac{r}{r_F} \right)^{\theta/d} \bigg [ 2 (d+z - (d+1) \theta/d) f(r) + r f' (r) \bigg]_{r=R}
\end{eqnarray}
where $f(r)$ is the blackening factor (\ref{blackeningspherical}).
Again, there are two contributions that we must consider, one from the thermal state and the other from the black hole spacetime, and in order to get a finite result one should subtract the first contribution from the second.
After some algebra we get
\begin{eqnarray}
\Delta I_{GH} 
=  - \frac{\omega_{k,d}}{16 \pi G}   \frac{\beta     r_F^{\theta}  r_h^{d-\theta+z -2 } }{\ell^{z+1}}   \frac{\theta  }{d}\left [ r_h^2  +   \frac{q^2}{r_h^{2 (d-\theta +z - 2)}}             +  \frac{ (d-1)^2  }{ (d-\theta +z-2)^2} k \ell^2 \right]     .
\end{eqnarray}
Notice that the   result for the Gibbons-Hawking term is proportional to the hyperscaling violating parameter $\theta$, so it indeed vanishes for asymptotically AdS and asymptotically Lifshitz spacetimes.\\

\noindent \textbf{Boundary terms from matter fields}
We also need to consider appropriate boundary terms coming from the variation of the   matter fields of the EMD theory, specifically, from the gauge fields $A$, $B$ and $C$ (\ref{gaugefield1})-(\ref{gaugefield3}). These boundary terms vanish if we keep the gauge potentials fixed at infinity,  but not otherwise. In  the grand canonical ensemble we need to keep  the  gauge potential $\Phi = C_t(\infty)$ fixed that supports the electric charge of the black hole, and hence in this ensemble the boundary term associated to this gauge field vanishes.   However, for the other two gauge fields $A$ and $B$ we need keep the charges fixed, instead of the potentials, since otherwise these gauge fields would spoil the  asymptotic Lifshitz geometry and the  non-trivial horizon topology of the solution. In order to keep these two charges fixed, (\ref{chargeF}) and (\ref{chargeH}) respectively, we must add the following two boundary terms to the Euclidean action (see for example \cite{Tarrio:2013tta,Chamblin:1999tk})
 \begin{equation}
I_M  =  - \frac{1}{16 \pi G} \int d^{d+1} x \sqrt{h}  X(\phi) F^{\mu \nu} n_\mu A_\nu  - \frac{1}{16 \pi G} \int d^{d+1} x \sqrt{h}  Y(\phi) H^{\mu \nu} n_\mu B_\nu\,,
\end{equation}
\noindent where  $\sqrt{h} =\sqrt{f(r)}  \ell^{-z} r_F^{\theta +  \theta/d} r^{ -  \theta/d - \theta + d + z }$  and $n_\mu$ is the radial unit vector pointing outwards
\begin{equation}
n_\mu =  \left ( \frac{r}{r_F}  \right)^{-\theta/d} \frac{\ell}{r} \frac{1}{\sqrt{f(r)}}   \delta_\mu^r\,. 
\end{equation}
Now, by evaluating the matter boundary terms  on-shell we arrive at
\begin{equation}
\begin{aligned}
I_M  
 &= \frac{1}{16\pi G} \int d^{d+1} x \,  2\ell^{-1-z}  r_F^{\theta }     \bigg [   (z-1)  ( r^{d-\theta +z} -  r_h^{d-\theta +z}  )   \\
 &\qquad\qquad\qquad\qquad\qquad\quad + k  \ell^2     (r^{d-\theta +z -2} - r_h^{d-\theta +z -2}    ) \frac{  (d-1)     (d (z-1)-\theta ) }{(d-\theta +z-2)^2}    \bigg]\, .     
\end{aligned}
\end{equation}
Next, we need to subtract the contribution from the thermal spacetime to obtain a finite result.
By doing so, we obtain the following difference in actions
\begin{align}
  \Delta I_M 
   \!=\! -\frac{\omega_{k,d}}{16\pi G} \beta  \ell^{-z-1} r_F^{\theta } r_h^{d-\theta +z-2}(z-1)\left [r_h^2  - \frac{q^2}{r_h^{2(d-\theta + z-2)}} + \frac{  (d-1) ((d+1) (z-1)-2 \theta )  }{(z-1) (d-\theta +z-2)^2}k \ell^2     \right].
\end{align}
Note that the first two terms vanish for $z=1$, but the last term survives. This is because     gauge field $A$ is absent in this limit, while   field $B$ only vanishes if we impose in addition $k=0$ or $\theta=0$.\\



\noindent \textbf{Final result}
  By adding all three contributions to the on-shell Euclidean action we find
  \begin{equation}
 \begin{aligned}
 I & = \Delta I_{EMD} + \Delta I_{GH} + \Delta I_M   \\
 & =  - \frac{\omega_{k,d}}{16\pi G} \beta  \ell^{-z-1} r_F^{\theta } r_h^{d-\theta +z-2} \left [ z r_h^2     + (2-z)  \frac{q^2}{r_h^{2 (d - \theta +z -2)}}  - \frac{(d-1)^2  (2-z)}{ (d-\theta +z-2)^2}k \ell^2 \right].\label{Igce}
 \end{aligned}
 \end{equation}
The grand canonical (Gibbs) potential is then
\begin{equation}
W = \frac{I}{\beta} = M - TS - \Phi Q\,.
\end{equation}
From this free energy we can derive the entropy
\begin{eqnarray}
S = \beta \left ( \frac{ \partial I }{\partial \beta}\right)_\Phi - I = \frac{ \omega_{k,d}}{4G}  r_h^{d-\theta} r_F^\theta\,,
\end{eqnarray}
the energy
\begin{eqnarray}
M = \left (  \frac{\partial I }{\partial \beta}    \right)_\Phi - \frac{\Phi}{\beta} \left (   \frac{\partial I }{\partial \Phi}   \right)_\beta
=	\frac{\omega_{k,d}  }{16\pi G}  (d - \theta) m  \ell^{-z-1 }   r_F^\theta\,,  
   \end{eqnarray}
and the charge
\begin{eqnarray}
Q = - \frac{1}{\beta} \left ( \frac{\partial I}{\partial \Phi}  \right)_\beta 
=  \frac{\omega_{k,d}}{16 \pi G}  \sqrt{2 Z_0 (d-\theta)(d-\theta + z-2)} \, q  \, \ell^{-1} r_F^{\theta -  \theta/d} e^{\lambda_3 \phi_0 /2}\,.
\end{eqnarray}
These expressions coincide with the ones derived in Section \ref{firstlaw}, and together they satisfy
the first law of thermodynamics, given in (\ref{first_law}).

Finally, notice that (\ref{Igce}) only holds for the $k=0,1$ cases. For $k=-1$ we need to subtract the action of the true vacuum (solution with $m=m_{ground}$, $q=0$) and match the corresponding asymptotic geometries. The computation is very similar to the one presented above, so we will omit the details of the derivation. The final result for the on-shell action for $k=-1$ and $\theta=d(z-1)$ reads:
\begin{eqnarray}
I =  \frac{ \omega_{-1,d}}{16 \pi G} \frac{r_F^{d(z-1) } \beta}{\ell^{z+1}}
    \left[ -r_h^{d (2 - z)+z} \left (  z     + \frac{1}{2-z} \frac{\ell^2}{r_h^2}  \right)   - (2-z)  q^2    -  d(2-z) m_{ground} \right]\,. \nonumber
\end{eqnarray}
Again, one can check that the entropy, mass and charge  derived from this on-shell action agree with the expressions   in Section \ref{firstlaw}, and together they satisfy the first law of thermodynamics, given in (\ref{first_law}).

\subsection{Canonical ensemble}

In the canonical ensemble we keep the charge of the black hole $Q$ fixed, instead of the electric potential $\Phi$. There are three essential differences in the computation of the on-shell Euclidean action with respect to the grand canonical ensemble. First, we must add another boundary term to the action, in order to have a well-defined variational principle in a fixed charge ensemble
\begin{equation}
 \begin{aligned}
S_C &= - \frac{1}{16 \pi G} \int d^{d+1} x \sqrt{h} Z(\phi) K^{\mu\nu} n_\mu C_\nu   \\
 &=  \frac{1}{16 \pi G} \int d^{d+1} x \, 2 q^2 (d-\theta ) \ell^{-z-1}  r_F^{\theta
   }  \left(  r_h^{-d+\theta -z
   +2}-r^{-d+\theta-z +2}  \right)\,.
    \end{aligned}
   \end{equation}
Second, we need to consider a different reference background, which in this case is given by the extremal black hole with $m=m_{ext}$, $q=q_{ext}$ (this applies to all $k$). Finally, in order to match the geometries of the black hole and reference spacetime at $r=R$ we impose the following relation between the temperatures of the two spacetimes
\begin{equation}
 \beta'  = \beta \left ( 1- \frac{m-m_{ext}}{2R^{d-\theta+z}} + ... \right)\,.
\end{equation}
We will omit the details of the calculation, since they are very similar to the ones presented for the grand canonical ensemble. The final result for the on-shell Euclidean action in this ensemble yields
\begin{equation}
  \begin{aligned}
 \tilde I &= - \frac{ \omega_{k,d} }{16 \pi G } \beta \ell^{-z-1}r_F^\theta\Big[  z ( r_h^{d-\theta +z} -  r_{ext}^{d-\theta +z} ) +  k \ell^2 \frac{(d-1)^2 (z-2)  }{(d-\theta +z-2)^2} ( r_h^{d-\theta +z-2} -r_{ext}^{d-\theta +z-2})   \\
   &\quad  \qquad \qquad \qquad  \qquad - (2 d-2    \theta +z-2) q^2  (r_h^{-d+\theta -z+2}  - r_{ext}^{-d+\theta -z+2}   )
  \Big]  \, .
  \end{aligned}
  \end{equation}
We can eliminate $r_{ext}$ from the final expression by using  (\ref{extremalcharge}) and (\ref{extremalmass})
\begin{equation}
 \begin{aligned}
 \tilde I &= 
  -\frac{\omega_{k,d}}{16 \pi G} \beta \ell^{-z-1} r_F^\theta r_h^{d+z-\theta-2}\bigg [ z r_h^2  - (2 d -2\theta +z -2 )  \frac{q^2}{r_h^{2(d-\theta+z-2)}}   \\
 & \qquad\qquad\qquad\qquad\qquad\qquad   - \frac{(d-1)^2   (2-z)}{  (d-\theta +z-2)^2}k \ell^2 + (d-\theta) \frac{m_{ext}}{r_h^{d-\theta + z -2}}  \bigg]\, .
 \end{aligned}
 \end{equation}
In the canonical ensemble the (Helmholtz) free energy is given by
 \begin{eqnarray}
 F = \tilde I /\beta = \tilde M - T S\,.
 \end{eqnarray}
From the free energy we can read off the entropy
  \begin{equation}
S = \beta \left (  \frac{\partial \tilde I}{\partial \beta} \right)_{Q}  - \tilde I = \frac{\omega_{k,d}}{4 G} r_h^{d-\theta} r_F^\theta\,, 
 \end{equation}
the energy
 \begin{equation}
 \tilde M = \left (  \frac{\partial \tilde I}{\partial \beta} \right)_{Q} =  \frac{\omega_{k,d}}{16 \pi G} (d-\theta) (m - m_{ext}) \ell^{-z-1} r_F^\theta\,,
 \end{equation}
and the electric potential
   \begin{equation}
\tilde{\Phi}= \frac{1}{\beta} \left (  \frac{\partial \tilde I}{\partial Q} \right)_{\beta}  =  \sqrt{ \frac{2 (d-\theta)}{Z_0 (d-\theta + z-2)} } \ell^{-z} r_F^{\theta/d} e^{- \lambda_3 \phi_0 /2}     \left ( \frac{q}{ {r_h^{d-\theta+z-2}}} - \frac{q}{r_{ext}^{d-\theta+z-2} } \right)\,.
 \end{equation}
These quantities agree with the ones derived in Section \ref{firstlaw} and together they satisfy the first law of thermodynamics in the canonical ensemble, given in (\ref{first_lawB}).


\bibliographystyle{JHEP-2}
\bibliography{ref}

\end{document}